\documentclass[11pt]{article}

\usepackage{amsmath,amsthm,latexsym,amssymb,amsfonts,epsfig}

\usepackage[utf8]{inputenc}

\usepackage{graphicx} 
\usepackage{physics}
\usepackage{amssymb}
\usepackage{color}

\usepackage[hidelinks]{hyperref}
\usepackage{booktabs}
\usepackage[compress]{cite}

\usepackage{bm}

\numberwithin{equation}{section}
\allowdisplaybreaks[1] 

\title{{\sf Quantum Field Theory of}\\
{\sf Black Hole Perturbations with Backreaction}\\
{\sf V. Beyond Second Order Perturbations}}

\author{
{\sf J. Neuser}$^1$\thanks{{\sf jonas.neuser@fau.de}},
{\sf T. Thiemann}$^1$\thanks{{\sf
thomas.thiemann@gravity.fau.de}}\\
\\
{\sf $^1$ Institute for Quantum Gravity, FAU Erlangen -- N\"urnberg,}\\
{\sf Staudtstr. 7, 91058 Erlangen, Germany}\\
}
\date{{\small\sf \today}}

\begin{document}

\maketitle

\begin{abstract}
    Black hole perturbation theory beyond second order is not well understood because typically one defines the meaning of gauge invariance order by order
    which is ambiguous. In this series of works we therefore developed a new approach which disentangles the meaning of gauge invariance from the perturbative 
    order. It is based on the reduced phase space approach to the Hamiltonian formulation of General Relativity and constructs a non-perturbative, albeit
    implicit, formulation of the dynamics of only observables that are gauge invariant to all orders. To obtain explicit expressions, perturbation theory 
    is then employed, but now only perturbations are considered that are gauge invariant to all orders. There are both spherically symmetric and  non-symmetric 
    observables and the formulation takes the (perturbative) backreaction between those fully into account. The formulation has access to both the exterior and 
    interior of the dynamical horizon.
    
    In previous papers of this series we have introduced the general formalism and applied it to Einstein-Maxwell theory. We developed perturbation theory 
    to second order and checked that our formalism can be matched to previous works on second order black hole perturbation theory without backreaction when we 
    neglect backreaction terms in our formulation and restrict to the exterior of the fixed event horizon. 
    
    This served as a consistency check. The real virtue of our approach starts emerging at higher than second order where we expect differences from previous 
    works both due to backreaction effects and because we work with observables that are gauge invariant to all orders, not only up to a given order. 
    In this paper, we consider the third order. Also new to our approach is that we start from a non-perturbative, namely polynomial, version of the constraints
    which therefore are finite polynomials in all degrees of freedom before reducing, rather than an infinite series. This allows for an exact and 
    non-perturbative, while implicit, solution of the constraints which does not need to truncate the series and thus is of tremendous technical advantage. 
\end{abstract}

\section{Introduction}

General relativity only admits few closed, analytic solutions for highly symmetric spacetimes.
In particular, for black holes in four dimensions, one has the non-rotating Schwarzschild solution which is spherically symmetric and the 
rotating Kerr solution
which is axially symmetric around the axis of rotation. 
In order to explore more general spacetimes and for the study of dynamical processes such as black hole mergers, accretion of matter 
or the emission of gravitational waves one needs to go beyond exact symmetry.
There are two options to include non-symmetric degrees of freedom: numerical relativity and perturbation theory.
In the present work we will focus on the perturbative approach.

The field of black hole perturbation theory was introduced in the seminal works by Regge, Wheeler and Zerilli \cite{ReggeWheeler, Zerilli}.
Later it was generalized to arbitrary spherically symmetric backgrounds and matter fields \cite{Gerlach,Poisson}.
In all of these works the perturbation theory is truncated after the first order and higher order corrections are neglected.
Since the first direct detection of gravitational waves by the LIGO and Virgo collaborations \cite{LIGO}, perturbation theory became 
increasingly important for our understanding of black holes.
In the ring-down phase of the merger, quasi-normal modes are used to describe how the horizon of the black hole settles down to the Schwarzschild / Kerr solution.
Recently, non-linear perturbation theory gained increasing attention in the literature \cite{Madrid1,Madrid2,Madrid3,Italy1,UK1,Japan1}.
It provides corrections to the quasi-normal mode spectrum and allows for a more accurate test of the predictions of general relativity in black hole mergers.

In perturbation theory, one assumes spacetime to be highly symmetric and then adds ``small'' perturbations.
There are two main approaches to black hole perturbation theory:
(i) In the Lagrangian approach, one considers the equations of motion and expands them around a known solution of the symmetry reduced Einstein equations.
(ii) In the Hamiltonian approach, one considers general relativity formulated in the ADM formulation \cite{ADM} and performs a perturbative expansion of 
the constraints. The result is a Hamiltonian describing the dynamics and ideally the Hamiltonian equations of motion then agree with what is obtained in the Lagrangian approach.
In this series of papers, we follow the Hamiltonian approach.
For black holes, the Hamiltonian perturbation theory was pioneered by Moncrief \cite{Moncrief} and later extended in \cite{Brizuela1,Brizuela2} to 
arbitrary backgrounds.

In this series of papers, we follow a recent proposal for perturbation theory based on the reduced phase space formulation for constrained Hamiltonian systems 
\cite{I}.
The idea is based on the approach of relational observables \cite{BrownKuchar, Rovelli1, Rovelli2, Dittrich1, Dittrich2, Dittrich3, ThiemannReduced, Giesel1}:
One identifies a subset of the degrees of freedom as `clocks' or reference fields and uses them to define physical observables for the remaining degrees of freedom. 
The observables are then defined to be the value of a given field when the reference field takes a certain value.
In practice, two possibilities arise:
(i) We add additional matter fields (e.g. dust \cite{BrownKuchar}) that serve as reference fields to deparametrize the theory.
(ii) We use some of the existing gravitational degrees of freedom as reference fields with respect to which we define observables for the remaining 
gravitational degrees of freedom.
While the second approach seems more natural as no additional matter fields are required, one faces the problem of solving the constraints which 
is highly non-trivial in general because it requires to solver partial differential equations rather than algebraic equations.
In \cite{I}, it was shown that the second approach can be performed in a perturbative way.
The strategy is as follows: 
First, we split all the degrees of freedom into spherically symmetric degrees of freedom and non-symmetric degrees of freedom. 
Then, within each sector we designate a subset of the degrees of freedom to be the `gauge', i.e. redundant degrees of freedom, and the 
remaining degrees of freedom to be the "true" or observable ones. One imposes gauge fixing conditions on the configuration degrees of 
freedom of the gauge sector and solves the constraints for the momentum degrees of freedom of the gauge sector.
The number of canonical pairs in the gauge sector must equal the number of constraints in order that this can be done unambiguously.  
Lapse and shift function are determined by requiring that the gauge fixing condition be invariant with respect to the time evolution generated by the 
constrained primary Hamiltonian. The reduced or true Hamiltonian is obtained as the effective Hamiltonian depending only on the true degrees of freedom 
which generates the same time evolution as the primary one on those true degrees of freedom when the constraints, the gauge conditions and their preservation
are installed. This results in a non-perturbative, albeit implicit, expression for the reduced Hamiltonian. To obtain explicit expressions one 
uses perturbation theory and it is at this point that the split into symmetric and non-symmetric sector becomes important: The
non-symmetric degrees of freedom are defined as linear or first order in the perturbations while the symmetric degrees of freedom are defined 
as zeroth order. A novelty of our approach is that we use a polynomial formulation of the constraints which therefore are finite order polynomials 
in all degrees of freedom and in particular in the perturbations rather than an infinite series and thus are non-perturbative. 
These are split into symmetric and non-symmetric ones and are also finite order polynomials in all degrees of freedom and we can identify 
its contributions of a given perturbative order. By construction, the symmetric constraints have no first order contribution while the 
non-symmetric constraints have no zeroth order contribution. Given that the symmetric constraints allow for an exact zeroth order solution, 
by a kind of ping-pong 
mechanism, one can successively solve the constraints order by order in the perturbations: Denote by $(b,n)$ the $n-th$ order contribution to 
the symmetric or non-symmetric constraints respectively when $b=0,1$ respectively. Then the constraint solution scheme is 
$(0,0)\to (1,1)\to (0,2)\to (1,2)\to (0,3)\to (1,3)...(0,n)\to (1,n)\to (0,n+1)...$ where each new solution step uses all the solutions of the 
previous step. That series never terminates because the constraints are not linear polynomials. These solutions are then substituted for into 
the reduced Hamiltonian which can be then expanded by itself. The end result, to any given order, is an explicit expression which depends only on the
symmetric and non-symmetric true degrees of freedom and which includes interaction (i.e. backreaction) terms between those.   

This is the description of the general scheme. Concretely we work with Gullstrand-Painlevé gauge (GPG) conditions which allow the exploration of 
the black hole interior and exterior simultaneously. Moreover, careful attention is paid to the decay behaviour of all degrees of freedom at spatial infinity 
which have a major impact on the reduced Hamiltonian which in the asymptotically flat context is a boundary term as the radial coordinate approaches 
infinity. In \cite{I} we considered as matter content electromagnetic fields and charged scalar fields (Abelian Higgs model). The backreaction
between symmetric and non-symmetric true degrees of freedom concerns then the symmetric scalar field modes and the non-symmetric electromagnetic and 
gravitational modes. In \cite{II,III,IV} we focussed on the electromagnetic fields only for which there is not much backreaction as the only symmetric 
true degrees of freedom consist of mass and charge and their conjugate momenta. We developed second order perturbation theory and 
checked that in this case we recover the Regge-Wheeler-Zerilli results when we restrict to the black hole exterior and after translating between 
the Gullstrand-Painleve and Schwarzschild coordinates.  
  
In this paper we develop third order perturbation theory within our formalism for pure, i.e. vacuum, General Relativity without matter. 
This includes, as a preparation for future general order perturbation 
theory, a complete and non-perturbative expansion of the polynomial version of the constraints to all orders. Quite astonishingly, even the polynomial 
version of the Hamiltonian constraint is only a sixth order polynomial. This is an accident that happens in the GP gauge, in a general gauge, the polynomial 
degree is ten in four spacetime dimensions. 
Obviously, this is a tremendous advantage of these gauge conditions as compared to others as the number of algebraically independent 
terms in perturbative expansions grows factorially. The decomposition of the $n-th$ order contribution to the constraints with $n=0,1,..,6$ into 
symmetric and non-symmetric is then just an exercise in angular momentum recoupling theory by decomposing degrees of freddom into scalar, vector and 
tensor harmonics. We then follow the above sketched iterative scheme to third order.\\

This work is organized as follows:

In section \ref{sec:ExpansionConstraints}, we fully expand the Hamiltonian and spatial diffeomorphism vacuum 
constraints to all orders in terms of the non-symmetric
true degrees of freedom and the non-symmetric gauge momenta. In our gauge, the non-symmetric gauge configuration degrees of freedom
gauged to zero. These are then expanded in terms of spherical scalar, vector and tensor harmonics. We briefly review their theory and
their recoupling theory in terms of Clebsch-Gordan coefficients.

In section \ref{sec:SolConstraints}, we explicitly solve the vacuum constraints up to third order in perturbation theory for the symmetric and 
non-symmetric gauge momenta. We include a discussion of the decay behaviour of the fields because the constraints include different contributions 
in terms of inverse powers of the radial coordinate and in order to being able to solve for the gauge momenta only, their decay behaviour must be able 
to capture the dominant contributions with respect to the radial expansion. This requires a modification of the decay behaviour as compared to 
earlier works where we solved the constraints also partly for the configuration degrees of freedom. This is based on a reduced phase space 
induced decay analysis \cite{TTnew}. Alternatively one keep the previous decay behaviour by 
modifying the GP gauge.
We show that both has no effect on the reduced Hamiltonian and focus on the first alternative in the remainder of the paper in order to be able to recycle 
previous results unchanged.   
 
In section \ref{sec:PhysHam} the reduced Hamiltonian is then computed in general terms and more concretely for the pure axial perturbations. 
We discuss its relation to the Lagrangian perturbation theory based on perturbing the equations of motion obtained in the Regge-Wheeler-Zerilli (RWZ)
gauge different from the GP gauge. We see that the results do not match at third order even after performing the canonical transformation between 
true degrees of freedom found in previous which brought the two frameworks to match. This difference is expected: As discussed in more detail in \cite{TTNew1},
the description of a physical system in terms of different relational Dirac observables with which the true degrees of freedom corresponding to 
different gauges are identified, generically results in different physical Hamiltonians which are identified with the reduced Hamiltonians in the 
respective gauge fixed frameworks. This is true even after performing a natural transformation between the two different reductions. 
Nevertheless, the frameworks are physically equivalent. As a simple example, the Hamiltonian of a Poincare invariant free quantum field depends on the 
inertial frame because the Hamiltonian is not a Lorentz scalar but the zero component, with respect to a given intertial frame, of the 4-momentum. 
   
In section \ref{sec:Conclusion}, we give a summary and an outlook for future work.

In appendix \ref{sec:RWG}, the pure axial perturbations are discussed in the Regge-Wheeler gauge which is more often used in the literature.

In appendix \ref{sec:BTModifiedGauge}, the boundary term of the second order physical Hamiltonian for the modified gauge fixing considered 
in section \ref{sec:SolConstraints} is recorded.

\section{Expansion of the Constraints to All Orders}
\label{sec:ExpansionConstraints}

In the previous papers of the series \cite{II,III,IV}, we investigated non-spherically symmetric degrees of freedom around spherically symmetric spacetimes up to second order in perturbation theory. 
Working in the ADM formulation of general relativity, the spatial metric $m_{\mu \nu}$ and the conjugate momentum $W^{\mu \nu}$ ($\mu,\nu,\dots=1,2,3$) are decomposed as follows
\begin{equation}
\label{eq:DefVariablesFull}
    \begin{aligned}
        m_{33} &= q_v + x_v, \quad m_{3A} = x_A, \quad ~~~ m_{AB} = (q_h + x_h)\Omega_{AB} + X_{AB}\\
        W^{33} &= \sqrt{\Omega}(p_v + y_v), \quad W^{3A} = \frac{\sqrt{\Omega}}{2} y^A, \quad W^{AB} =  \frac{\sqrt\Omega}{2} (p_h  + y_h) \Omega^{AB} + \sqrt{\Omega} Y^{AB}
    \end{aligned}
\end{equation}
In this decomposition, we assume the three dimensional hypersurfaces $\Sigma$ of the foliation underlying the ADM formulation to have the topology $\mathbb{R}^+ \times S^2$, 
where $\mathbb{R}^+$ is the radial direction $3$ and $S^2$ is the two sphere.
On the sphere we choose spherical coordinates $z^A = (\theta, \phi)$, the standard metric $\Omega_{AB}$ and the determinant of the metric $\Omega := \det(\Omega)$. 
The symmetric degrees of freedom are $(q_v,p_v)$ and $(q_h,p_h)$ which are assumed to be spherically symmetric. They only depend on time and the radial coordinate. 
The variables $(x_{v/h/A},y^{v/h/A})$ and $(X_{AB},Y^{AB})$ are non-symmetric, i.e. they additionally have angular dependence.
The tensors $(X_{AB}, Y^{AB})$ are symmetric and trace free with respect to the metric $\Omega_{AB}$, i.e. $\Omega_{AB} X^{AB} = \Omega^{AB} Y_{AB} = 0$.
We also assume that all the symmetric modes are already contained in the spherically symmetric variables $(q,p)$ and hence the non-symmetric degrees of freedom do not contain any zero modes.
That is, the average over the sphere of $x_{v/h}$ and $y^{v/h}$ vanishes.

For the computation of the physical Hamiltonian, we have to single out a subset of the degrees of freedom that we declare to be the true (physical) degrees of freedom.
The remaining degrees of freedom are then called gauge (redundant) degrees of freedom and can be chosen freely through some gauge condition usually restricting the form of the metric $m_{\mu \nu}$.
The Gullstrand-Painlevé gauge (GPG) inspired by the Gullstrand-Painlevé coordinates turned out to be a good choice.
These coordinates are physically motivated by an observer freely falling along a radial geodesic in a Schwarzschild spacetime
and the resulting spherically symmetric metric is regular across the black hole horizon.
Hence, they allow for a simultaneous description of both the interior and exterior of the black hole spacetime.
In Gullstrand-Painlevé gauge, we impose $m_{33} = 1$, $m_{3A} = 0$ and $m_{AB}\Omega^{AB} = 2 r^2$, where $r$ is the radial coordinate.
In terms of the variables introduced in equation \eqref{eq:DefVariablesFull}, we find $q_v = 1$, $q_h = r^2$ and $x= 0$. 
Therefore, the metric in Gullstrand-Painlevé gauge reduces to
\begin{equation}
    \begin{aligned}
        m_{33} &= 1, \quad m_{3A} = 0, \quad m_{AB} = r^2 \Omega_{AB} + X_{AB}\\
    \end{aligned}
    \label{eq:DefVariables}
\end{equation}
In \cite{II,III,IV}, we used a different notation for the spherically symmetric degrees of freedom in terms of $(\mu, \pi_\mu)$ and $(\lambda,\pi_\lambda)$. 
The variables are related to the ones used here and in \cite{II} by $\pi_\mu = 2 p_v$ and $\pi_\lambda = 2 r^2 p_h$.

For the derivation of the reduced Hamiltonian it is important to take special care of boundary terms and the behaviour of the fields at the boundary. 
For black holes the relevant boundary is the asymptotic boundary $r\to\infty$ and one typically assumes asymptotic flatness there.
In \cite{I}, this approach was used to derive the fall-off conditions of the canonical variables in the present GPG frame by transcribing the 
usual Cartesian frame decay conditions.  
However, as it turns out, those standard asymptotic flatness conditions are incompatible with the solution of the constraint equations
{\it for the momenta} at higher than second order because their contribution to the constraints at that order 
decays faster than the contribution of the configuration degrees of freedom. This effect did not show up in the computation
of the reduced Hamiltonian to second order in our previous works because there we partly solved 
the constraints for configuration degrees of freedom rather than momentum degrees of freedom.
While not a problem in principle, it is inconvenient in practice because one can solve the constraints easier for the momenta than 
configuration degrees of freedom because one just has to solve algebraic or first order partial differential equations rather than 
second order partial differential equations, especially at higher order. Fortunately, as observed in \cite{TTnew},
the standard decay behaviour of the gauge degrees of freedom is unnecessarily weak: as these are gauged to zero anyway one can 
require their decay arbitrarily strong. By strengthening
their decay one can relax the decay of the momenta in order avoid the problem mentioned above.  
We discuss a possible choice of, in this sense self-consistent decay conditions, for the canonical variables in section \ref{sec:SelfConDecay}. 
These are $X_{AB}=O(r^{3/4})$, $Y^{AB} = O(r^{-7/4})$, where $X_{AB}$ is an even function and $Y_{AB}$ is an odd function on the sphere.
The constraints then imply that the gauge momenta behave as $y_v = O(r^{1/4})$, $y_e = O(r^{-3/4})$ and $y_h = O(r^{-7/4})$ as $r$ 
approaches infinity. These decay conditions are tailored to have the property, that the  
implicit formula for the physical Hamiltonian derived in \cite{I} remains intact also with the modified fall-off conditions.

After these introductory remarks, we explicitly perform the perturbative analysis of the constraints. 
First, we introduce the spherical scalar, vector and tensor harmonics and discuss their recoupling theory in terms of spin-weighted spherical harmonics.
Then, we discuss how self-consistent decay conditions are obtained and we motivate the choice mentioned above.
Next, we study the radial and angular diffeomorphism constraints.
They are quadratic in the canonical variables and therefore, their perturbative series naturally truncates after second order. 
Next, the Hamiltonian constraint is analysed. 
It is a non-polynomial function of the canonical variables due to the square root of the determinant of the metric which is involved in its definition.
Hence, the expansion would never terminate. 
However, it turns out that multiplying the constraint with a suitable power of the determinant of the metric, the constraint becomes polynomial and the
expansion terminates after finite order.
We use this and obtain a finite list of constraints that need to be solved in section \ref{sec:SolConstraints}.

\subsection{Spherical Harmonics and Spin-Weighted Spherical Harmonics}
\label{sec:Spin-Weighted-Harmonics}
For the computation, it is convenient to use the spherical symmetry of the background to expand the non-symmetric degrees of freedom into spherical harmonics. 
In the following, we briefly discuss their definition and show how we can use the recoupling theory of angular momenta to reduce the products of harmonics to a sum over single harmonics.

\subsubsection{Spherical Scalar, Vector and Tensor Harmonics}

The scalar spherical harmonics $Y_{lm}$ are complex eigenfunction of the Laplace operator on the sphere. 
Let $\Omega_{AB}$ be the metric on the sphere and $D_A$ the corresponding covariant derivative.
Then, the harmonics satisfy $\Omega^{AB}D_A D_B Y_{lm} = - l(l+1) Y_{lm}$. 
The harmonics are defined by
\begin{equation}
    Y_{lm}(\theta,\phi) = \sqrt{\frac{2l+1}{4\pi}\frac{(l-m)!}{(l+m)!}}e^{im\phi}P^l_m(\cos \theta)
\end{equation}
where $P^l_m$ are the associated Legendre polynomials. Under complex conjugation, the harmonics transform as $(Y_{lm})^* = (-1)^m Y_{l,-m}$. 
It is convenient to introduce real-valued spherical harmonics by
\begin{equation}
    L_{lm} = \begin{cases}
        \frac{1}{\sqrt{2}}((Y_{lm})^* +Y_{lm}), & m>0\\
        Y_{l,0}, &m=0\\
        \frac{i}{\sqrt{2}}((Y_{lm})^* - Y_{lm}), &m>0
    \end{cases}
\end{equation}

For later convenience we introduce an operator $U^{(l)}_{m m'}$ that transforms between the real and complex valued spherical harmonics, i.e.
\begin{equation}
    L_{lm} = \sum_{m'} U^{(l)}_{m m'} Y_{l m'}
\end{equation}
The transformation $U$ is explicitly given by
\begin{align}
    \begin{aligned}
    U^{(l)}_{0 m'} &= \delta_{m'0}\\
    U^{(l)}_{m m'} &=\begin{cases}
    \frac{1}{\sqrt{2}} \delta_{m, m'}, &(m, m'>0)\\
    \frac{1}{\sqrt{2}} (-1)^m \delta_{m, - m'}, &(m>0, m'<0)\\
    \frac{i}{\sqrt{2}} (-1)^m \delta_{m, - m'}, &(m<0, m'>0)\\
    - \frac{i}{\sqrt{2}} \delta_{m, m'}, &(m, m'<0)
    \end{cases}
    \end{aligned}
\end{align}
We introduce the inner product on the space $L^2(S^2,\dd \Omega)$ for scalar functions $f,g$ by
\begin{equation}
\label{eq:InnerPrductSphere}
    \langle f, g\rangle := \int_{S^2}\dd\Omega \sqrt{\Omega} f^* g
\end{equation}
The spherical harmonics are orthonormal with respect to that inner product, i.e. $\langle L_{l_1 m_1}, L_{l_2 m_2}\rangle = \delta_{l_1,l_2} \delta_{m_1, m_2}$. 

The vector harmonics are defined by
\begin{equation}
    [L^e_{lm}]_A := \frac{1}{\sqrt{l(l+1)}} D_A L_{lm}, \quad [L^o_{lm}]_A:= \frac{1}{\sqrt{l(l+1)}} \eta_{AB} \Omega^{BC} D_C L_{lm}
\end{equation}
where $\eta_{AB} = \sqrt{\Omega} \epsilon_{AB}$.
The vector harmonics are orthonormal, i.e. 
\begin{equation}
    \int_{S^2} \Omega^{AB} [L^{I_1}_{l_1 m_1}]_A [L^{I_2}_{l_2 m_2}]_B \sqrt{\Omega} \dd \Omega = \delta^{I_1, I_2}\delta_{l_1, l_2}\delta_{m_1, m_2}
\end{equation}
The label $I=e,o$ stands for the harmonics with even (e) and odd (o) "polarity". Perhaps a better label would be 
$p,a$ for polar and and axial taking into accound that these are vectors and pseudo-vectors on the sphere respectively.
This should not be confused with even and odd "parity" i.e. their change by a sign under reflection on the sphere.

The tensor harmonics are given by
\begin{equation}
    \begin{aligned}
        \relax[L^{\mathrm{tr}}_{lm}]_{AB} &:= \frac{1}{\sqrt{2}}\Omega_{AB} L_{lm}\\
        [L^{e}_{lm}]_{AB} &:= \sqrt{\frac{(l-2)!}{2(l+2)!}}\qty(D_A D_B + \frac{1}{2} l(l+1) \Omega_{AB}) L_{lm}\\
        [L^{o}_{lm}]_{AB} &:= \sqrt{\frac{(l-2)!}{2(l+2)!}}D_{(A}\qty(\eta_{B) C}\Omega^{CD} D_D L_{lm})
    \end{aligned}
\end{equation}
They satisfy the following orthonormality relations
\begin{equation}
     \int_{S^2} \Omega^{AC} \Omega^{BD} [L^{I_1}_{l_1 m_1}]_{AB} [L^{I_2}_{l_2 m_2}]_{CD} \sqrt{\Omega} \dd \Omega = \delta^{I_1, I_2}\delta_{l_1, l_2}\delta_{m_1, m_2}
\end{equation}
where $I$ takes the values $\mathrm{tr}, e, o$. A list of useful identities satisfied by the harmonics can be found in \cite{I,II}.

\subsubsection{Spin-Weighted Spherical Harmonics}

In the computation of the mode expansion for the perturbed constraints, integrals over products of scalar, vector and tensor spherical harmonics arise. 
For their computation, we use spin-weighted spherical harmonics \cite{SPSH}. 
They are an alternative to the scalar, vector and tensor spherical harmonics defined in the previous section and allow for an easier computation.
After choosing a basis of the tangent space on the sphere, one contracts the basis with the vector and tensor harmonics.
The resulting expressions are said to have spin-weight $0,\pm 1, \pm 2$ for scalar, vector and tensor harmonics respectively.

For their definition, we introduce a complex null basis $m^A$, $\overline m^A$ on the sphere. 
A good choice is
\begin{equation}
	m^A = \frac{1}{\sqrt{2}}\qty(1, \frac{i}{\sin \theta})\,.
\end{equation}
By definition, we have $\Omega_{AB} m^A m^B =\Omega_{AB} \overline m^A \overline m^B = 0$ and $\Omega_{AB} m^A \overline m^B = 1$. 
In terms of this basis, the metric and the Levi-Civita pseudo-tensor read
\begin{align}
    \Omega^{AB} &= m^A \overline m^B + \overline m^A m^B\,,\\
    \eta_{AB} &= \frac{1}{i}\qty(m_A \overline m_B - \overline m_A m_B)\,.
\end{align}

Consider now, a transformation of the basis $m \to e^{i \psi} m$, which leaves the inner products between $m$ and $\overline m$ invariant. 
A quantity $A$ is said to have spin-weight $s$, if it transforms according to  $A \to e^{i s \psi}A$. 
Thus, by definition, $m$ has spin weight $+1$ while $\overline m$ has spin weight $-1$. 
By construction, the metric $\Omega^{AB}$ and the Levi-Civita pseudo-tensor both have spin-weight 0 as expected. 

For the construction of the spin-weighted spherical harmonics, we introduce the differential operators $\eth$ and $\overline \eth$.
On quantities of spin-weight $s$ they act as
\begin{align}
	\eth &= \sqrt{2} m^A D_A - s \cot \theta\\
	\overline \eth &= \sqrt{2} \overline m^A D_A + s \cot \theta\,.
\end{align}
This equations differ by a minus sign from the ones in \cite{SPSH}.
The operator $\eth$ raises the spin weight by one and $\overline \eth$ lowers the spin weight by one.
Furthermore, we have $\overline \eth \eth - \eth \overline \eth = - 2 s$. 
In addition, the operators are constructed such that $\eth m_A= \eth \overline m_A = \overline \eth m_A = \overline \eth \overline m_A = 0$.

Consider the real valued spherical harmonics $L_{lm}$ defined before. 
A direct computation shows that
\begin{align}
    \eth \overline \eth L_{lm} = \overline \eth \eth L_{lm} = - l(l+1) L_{lm}\,.
\end{align}
The covariant derivative on the sphere acting on spin-wight zero functions is written as
\begin{align}
	D_A = \frac{1}{\sqrt{2}}\qty(m_A \overline \eth + \overline m_A \eth)
\end{align}

From the spherical harmonics $L_{lm}$, we can construct the spin-weighted spherical harmonics with $s \neq 0$. 
For $|s| \leq l$, they are defined as
\begin{equation}
    L^s_{lm} := \frac{1}{\lambda_{l,|s|}} \begin{cases}
        (-1)^s \eth^s L_{lm}, \quad & 0\leq s \leq l\\
        \overline \eth^{|s|}L_{lm}, & -l \leq s \leq 0
    \end{cases}\,,
\end{equation}
where 
\begin{equation}
    \lambda_{l,s} := \sqrt{\frac{(l+ s)!}{(l- s)!}}\,.
\end{equation}
Note that the spin-weighted spherical harmonics  for $s\neq0$ are complex valued because the operator $\eth$ involves the complex dyad $m^A$. 
We also have that $(L^s_{lm})^* = (-1)^s L^{-s}_{lm}$. 

The action of the derivative operators $\eth$ and $\overline \eth$ on the spin-weighted spherical harmonics is
\begin{align}
    \begin{split}
	\eth L^s_{lm} &=  - \sqrt{(l+s+1)(l-s)} L^{s+1}_{lm}\\
    \overline \eth L^s_{lm} &= \sqrt{(l+s)(l-s+1)} L^{s-1}_{lm}\\
    \overline \eth \eth L^s_{lm}&=-(l+s+1)(l-s) L^s_{lm}
    \end{split}
\end{align}

Consider the space of spin-weighted spherical harmonics for a given spin-weight $s$: $L^s_{lm}$ ($l \geq |s|$, $-l \leq m \leq l$). 
The spin-weighted spherical harmonics for \emph{fixed} $s$ form an orthonormal basis for functions on the sphere of spin-weight $s$.
First, we prove orthogonality. 
For functions $f,g$ with spin weight $s$, we have
\begin{equation}
    \langle \eth f, g\rangle = - \langle f, \overline \eth g\rangle, \quad \quad \langle \overline \eth f, g\rangle = - \langle f, \eth g\rangle
\end{equation}
where $\langle \cdot, \cdot \rangle$ is the $L^2$ inner product on the sphere defined in \eqref{eq:InnerPrductSphere}.
Then, the inner product of two spin-weighted spherical harmonics for $s>0$ is
\begin{align}
    \langle L^s_{l_1 m_1}, L^s_{l_2 m_2}\rangle &= \frac{(-1)^s}{\lambda_{l_1,s}} \langle\eth^s L_{l_1m_1}, L^s_{l_2 m_2}\rangle\nonumber\\
    &= \frac{1}{\lambda_{l_1,s}} \langle L_{l_1m_1}, \overline \eth{} ^s L^s_{l_2 m_2}\rangle\rangle\\
    &= \frac{\lambda_{l_2,s}}{\lambda_{l_1,s} } \langle L_{l_1m_1}, L_{l_2 m_2}\rangle\nonumber\\
    &= \delta_{l_1l_2} \delta_{m_1 m_2}\nonumber
\end{align}
In the last step we used the orthonormality of the ordinary spherical harmonics. 
A similar calculation shows the same result for $s<0$.
The integral over a function on the sphere is only well defined, if the integrand has total spin weight zero. 
Otherwise, the result of the integral would depend on the choice of the complex null basis and is ill-defined.
Thus, there is no meaning in computing $\langle L^s_{l m}, L^{s'}_{l' m'}\rangle$ for $s \neq s'$.

Second, we use the fact that the spin-weighted spherical harmonics with fixed $s$ are complete:
\begin{equation}
    \sum_{lm} (L^s_{lm}(\theta,\phi))^* L^s_{l m}(\theta',\phi') = \delta(\phi- \phi')\delta(\cos\theta-\cos \theta')
\end{equation}
This is proven in \cite{SPSH}. 

Therefore, let $f$ be any function of spin weight $s$.
Then, it is decomposed as
\begin{equation}
    f = \sum_{lm} \langle L^s_{lm}, f\rangle L^s_{lm}
\end{equation}
into the basis of spin-weighted spherical harmonics.
Notice, that for a well-defined inner product we have to first determine the spin-weight $s$ of $f$ and then choose the correct set of spin-weighted spherical harmonics
for that spin weight.


We are now ready to rewrite the vector and tensor spherical harmonics in terms of spin-weighted ones and obtain
\begin{align}
    \begin{split}
	(L^e_{lm})^A &= \frac{1}{\sqrt{l(l+1)}} D^A L_{lm} = \frac{1}{\sqrt{2}}\qty(m^A  L^{-1}_{lm} - \overline m^A L^{1}_{lm})\\
	(L^o_{lm})^A &= \frac{1}{\sqrt{l(l+1)}}\epsilon^{AB} D_B L_{lm} = \frac{i}{\sqrt{2}}\qty(m^A L^{-1}_{lm} + \overline m^A L^{1}_{lm})\\
(L^e_{lm})^{AB} &= \sqrt{\frac{2}{(l-1)l(l+1)(l+2)}}\qty(D^A D^B + \frac{1}{2} l(l+1) \Omega^{AB})L_{lm} = \frac{1}{\sqrt{2}}\qty(m^A m^B L^{-2}_{lm} + \overline m^A \overline m^B L^2_{lm})\\
(L^o_{lm})^{AB} &= \sqrt{\frac{2}{(l-1)(l+2)}}D^{(A}(L^o_{lm})^{B)} = \frac{i}{\sqrt{2}} \qty(m^A m^B L^{-2}_{lm} - \overline m^A \overline m^B L^{2}_{lm})
    \end{split}
\end{align}
It is convenient to condense the notation by considering the label $I={e,o}$:
\begin{equation}
\begin{aligned}
\relax
    [L^I_{lm}]^A &= \frac{i^I}{\sqrt{2}}\qty(m^A L^{-1}_{lm} - (-1)^I \overline m^A L^1_{lm})\\
    [L^I_{lm}]^{AB} &= \frac{i^I}{\sqrt{2}}\qty(m^A m^B L^{-2}_{lm} + (-1)^I \overline m^A \overline m^B L^2_{lm})
\end{aligned}
\end{equation}
In the equations we use the convention that $I=e=0$ and $I=o=1$.

\subsubsection{Recoupling Theory of Spin-Weighted Spherical Harmonics}

For the computation of the perturbed constraints, we have to express products of spin-weighted spherical harmonics in terms of the sum of spin-weighted spherical harmonics. 
Consider two harmonics $L^{s_1}_{l_1 m_1}$ and $L^{s_2}_{l_2 m_2}$ of spin weight $s_1$ and $s_2$ respectively. 
Their product has spin-weight $s = s_1 + s_2$ and we can decompose the product as
\begin{equation}
    L^{s_1}_{l_1 m_1} L^{s_2}_{l_2 m_2} = (-1)^s \sum_{lm} C^{l m, - s}_{l_1 m_1 s_1, l_2 m_2 s_2} L^{s}_{l m}
\end{equation}
where the coefficients $C$ are the Clebsch-Gordan coefficients.
By construction the coefficients are given by
\begin{equation}
    (-1)^s C^{lm -s}_{l_1 m_1 s_1, l_2 m_2 s_2} = \int_{S^2}\dd\Omega\sqrt\Omega (L^s_{lm})^* L^{s_1}_{l_1 m_1} L^{s_2}_{l_2 m_2}
\end{equation}
We observe that
\begin{equation}
    C^{l m s}_{l_1 m_1 s_1, l_2 m_2 s_2} := 
    \int_{S^2}\dd\Omega  \sqrt\Omega L^s_{l m} L^{s_1}_{l_1 m_1} L^{s_2}_{l_2 m_2}
\end{equation}
This integral is only well-defined provided that $s+s_1+s_2 = 0$, i.e. no total spin-weight.

In order to find an explicit expression for the triple integral, we use the fact that starting from the comlex-valued spherical harmonics $Y_{lm}$, we can define a different set of spin-weighted spherical harmonics $Y^s_{lm}$, which satisfy the well-known identity
\begin{equation}
    \int_{S^2}\dd\Omega \sqrt\Omega Y^{s_1}_{l_1 m_1} Y^{s_2}_{l_2 m_2} Y^{s_3}_{l_3 m_3} = \sqrt{\frac{(2l_1 + 1)(2l_2+1)(2l_3+1)}{4\pi}} \mqty(l_1 & l_2 & l_3\\m_1 & m_2 & m_3)\mqty(l_1 & l_2 & l_3\\- s_1 & -s_2 & -s_3)
\end{equation}
In order to find the expression for our situation, we define the transformation $U^{(l)}_{m m'}$ relating the real and complex valued harmonics.
Then, the coefficients $C$ are given by
\begin{equation}
    C^{lms}_{l_1 m_1 s_1, l_2 m_2 s_2} = \sum_{m' m_1' m_2'}U^{l}_{m m'}U^{l_1}_{m_1 m_1'}U^{l_2}_{m_2 m_2'}\sqrt{\frac{(2l + 1)(2l_1+1)(2l_2+1)}{4\pi}} \mqty(l & l_1 & l_2\\m' & m_1' & m_2')\mqty(l & l_1 & l_2\\- s & - s_1 & -s_2)
\end{equation}

The Clebsch-Gordan coefficients have many symmetries. 
First of all, it follows from the symmetries of the 3j-symbols that it is fully symmetric under exchanging the tuples $(lms)$, $(l_1m_1s_1)$ and $(l_2m_2s_2)$ in any order.
Then, changing the signs of $s$ gives
\begin{equation}
    C^{l m -s}_{l_1 m_1 -s_1, l_2 m_2 -s_2} = (-1)^{l + l_1 + l_2}C^{lms}_{l_1 m_1 s_1, l_2 m_2 s_2}
\end{equation}
Setting $l = m = s = 0$, we find using properties of the $3j$-symbols
\begin{equation}
\begin{aligned}
    C^{000}_{l_1 m_1 s, l_2 m_2 -s} &= \sum_{m_1' m_2'}U^{l_1}_{m_1 m_1'}U^{l_2}_{m_2 m_2'}\sqrt{\frac{1}{4\pi}} \delta_{l_1,l_2}\delta_{m_1', -m_2'}(-1)^{m_1' + s}\\
    &= \frac{(-1)^s}{\sqrt{4\pi}} \delta_{l_1 l_2} \delta_{m_1 m_2}
\end{aligned}
\end{equation}
where we performed the product of the transformations $U$ explicitly. 

Another useful identity follows from the recursion relation for the $3j$-symbols. We have
\begin{equation}
\begin{split}
    \sqrt{(l_1 \mp s_1)(l_1 \pm s_1 +1)}&C^{l_1 m_1 s_1 \pm1}_{l_2 m_2 s_2, l_3 m_3 s_3} + \sqrt{(l_2 \mp s_2)(l_2 \pm s_2 +1)}C^{l_1 m_1 s_1}_{l_2 m_2 s_2 \pm 1, l_3 m_3 s_3}\\
    &+ \sqrt{(l_3 \mp s_3)(l_3 \pm s_3 +1)}C^{l_1 m_1 s_1}_{l_2 m_2 s_2, l_3 m_3 s_3 \pm 1} = 0
\end{split}
\end{equation}

Finally, we compute the triple integrals of vector and tensor spherical harmonics.
We transform the vector and tensor spherical harmonics into spin-weighted spherial harmonics and evaluate the integral using
the triple integral formula:
\begin{align}
        \int_{S^2} L_{l_1 m_1} \Omega_{AB} [L^I_{l_2m_2}]^A [L^J_{l_3m_3}]^B  \sqrt\Omega \dd \Omega &= - \frac{i^{I+J}}{2} \int_{S^2} L_{l_1 m_1}\qty((-1)^I L^1_{l_2 m_2} L^{-1}_{l_3 m_3} + (-1)^J L^{-1}_{l_2 m_2} L^1_{l_3 m_3}) \sqrt\Omega \dd\Omega\nonumber\\
        & = - \frac{i^{I+J}}{2} \qty((-1)^I C^{l_1 m_1 0}_{l_2 m_2 1, l_3 m_3 -1} + (-1)^J C^{l_1 m_1 0}_{l_2 m_2 -1, l_3 m_3 1})\\
        &= - C^{l_1 m_1 0}_{l_2 m_2 1, l_3 m_3 -1} q^{IJ}\nonumber
\end{align}
where $q^{ee} = q^{oo} = \sigma_+$ and $q^{eo} = -q^{oe} = i \sigma_-$. Notice that $(q^{JI})^* = q^{IJ}$. With the projector
\begin{align}
    \sigma_{\pm} = \frac{1 \pm (-1)^{l_1 + l_2 + l_3}}{2}
\end{align}
It satisfies $\sigma_\pm^2 = \sigma_\pm$ and $\sigma_+ \sigma_- = 0$ and $\sigma_+ + \sigma_- = 1$ as expected.
For the tensor modes, we have
\begin{align}
    \begin{aligned}
         \int_{S^2} L_{l_1 m_1} [L^I_{l_2 m_2}]^{AB} [L^J_{l_3 m_3}]_{AB}  \sqrt\Omega \dd\Omega &= \frac{i^{I+J}}{2} \int_{S^2} L_{l_1 m_1} \qty((-1)^I L^{2}_{l_2 m_2} L^{-2}_{l_3 m_3} + (-1)^J L^{-2}_{l_2 m_2} L^2_{l_3 m_3}) \sqrt\Omega \dd \Omega\\
         &= C^{l_1 m_1 0}_{l_2 m_2 2, l_3 m_3 -2} q^{IJ}
    \end{aligned}
\end{align}
From this computation, it is not difficult to see that
\begin{align}
    [L^I_{l_1 m_1}]_{AB} [L^J_{l_2 m_2}]^{AB} &= \sum_{lm} C^{lm 0}_{l_1 m_1 2, l_2 m_2 -2} q^{IJ} L_{lm}
\end{align}

\subsection{Modified Decay Conditions for the Canonical Variables}
\label{sec:SelfConDecay}

In the previous section, we briefly reviewed the definition of the variables and their splitting into symmetric and non-symmetric degrees of freedom. 
The derivation of the physical Hamiltonian in \cite{I} relies on a careful analysis of the Hamiltonian system including fall-off conditions and boundary terms.
The following asymptotic behaviour were derived by transcribing the Cartesian frame decay condition to the GPG frame:
\begin{align}
    \begin{split}
    x^v &= \frac{x^v_+}{r} + \frac{x^v_-}{r^2}, \quad x_A = x^+_A + \frac{x^-_A}{r}, \quad x^h = x^h_+ r + x^h_-, \quad X_{AB} = X^+_{AB} r + X^-_{AB},\\
    \frac{y_v}{\omega} &= y_v^- + \frac{y_v^+}{r}, \quad \frac{y_A}{\omega} = \frac{y^A_-}{r} + \frac{y^A_+}{r^2}, \quad \frac{y_h}{\omega} = \frac{y_h^-}{r^2} + \frac{y_h^+}{r^3}, \quad \frac{Y^{AB}}{\omega} = \frac{Y^{AB}_-}{r^2} + \frac{Y^{AB}_+}{r^3}
    \label{eq:FallOffConds}
    \end{split}
\end{align}
Here, $\omega = \sqrt{\det \Omega}$ and the variables on the right-hand side are independent of $r$.
The sub-/superscript $+/-$ indicates whether the leading order contribution is positive (even) or negative (odd) under the parity operation $x \rightarrow -x$.
Using these conditions, one can show that the symplectic form is well defined and the boundary terms are finite at infinity (see \cite{I}). 
However, if one wants to follow the systematic strategy to solve the constraints only for the momenta $y$ but not for the $x$, then there is an additional 
condition that needs to be satisfied by the decay condition, namely that the solution $y=y_\ast(X,Y)$ in terms of the true degrees of freedom $X,Y$ is such 
that the decay of $y$ matches that of $y_\ast$. In our previous works we did not check this additional condition because at second order it is 
in fact possible to solve the constraints partly for the $x$ which is what we did there. However, especially at higher order we want to follow 
the systematic approach and solve the constraints purely for the $y$.

In order to better understand the issue, we first consider the full Hamiltonian constraint of general relativity in ADM variables:
\begin{equation}
    V_0 = \frac{1}{\sqrt{\det m}} \qty(m_{\mu \rho} m_{\nu \sigma} - \frac{1}{2}m_{\mu \nu} m_{\rho \sigma})W^{\mu \nu} W^{\rho \sigma}- \sqrt{\det m} R
\end{equation}
For simplicity, we work in Cartesian coordinates and assume the standard asymptotic flatness conditions on the metric and its conjugate momentum.
At infinity, we find that the perturbation of the spatial metric decays like $r^{-1}$ and therefore, the Christoffel symbols decay as $r^{-2}$.
Then, the Ricci scalar, which is a contraction of derivatives of Christoffel symbols and terms quadratic in the Christoffel symbol, will decay as $r^{-3}$.
Hence, in summary, the Ricci scalar term will decay as $r^{-3}$.
Looking at the term involving the momentum $W^{\mu \nu}$, we see that its leading order needs to decay like $r^{-3/2}$ in order to consistently solve the equation asymptotically.
This is not the standard decay behaviour of the momentum which is $r^{-2}$.
Therefore, if one wants to keep the standard decay conditions one is forced to solve the constraints at least in part for some of the $m_{\mu\nu}$ rather 
than the $W^{\mu\nu}$. If on the other hand one wants to solve the constraints for the momenta only one either has to weaken the 
decay of some of the $W$ and strengthen the decay of the corresponding conjugate $m$ or 
one imposes a decay condition which avoids the $r^{-3}$ contributions to the constraints while keeping the conditions of having the symplectic structure, constraints and 
Hamiltonian vector fields of the constraints well defined. For example one can impose the symmetric transverse traceless (STT) gauge 
used in the theory of gravitational waves on a flat Minkowski background. 
In this gauge one assumes the perturbations to satisfy $\delta^{\mu \nu} h_{\mu \nu} = 0$ and $\delta^{\rho\mu}\partial_\rho h_{\mu \nu} = 0$. 
The contribution of order $r^{-3}$ to the Ricci scalar are given by
\begin{equation}
    R \sim \partial^\mu \partial^\nu h_{\mu \nu} - \partial^\rho \partial_\rho (\delta^{\mu \nu} h_{\mu \nu}) + O(r^{-4})
\end{equation}
Obviously, imposing the STT gauge, we find that the $r^{-3}$ contribution vanishes. 

The discussion reveals that it would be desirable to have a systematic 
procedure at one's disposable to actually {\it derive} decay conditions on the kinematical phase space that are tailored to a strategy to construct 
the reduced phase space and explore the freedom involved. Note that one cannot dispose of the decay behaviour of the kinematical phase space and directly 
work on the reduced phase space because the constraints, that are functions on the kinematical phase space and their Hamiltonian flow on it, is what defines 
the reduced phase space. Such a systematic procedure is developed in \cite{TTnew}. The key idea is that if one solves the constraints for $y$ and gauges 
the conjugate variables $x$ to zero, one can assume arbitrarily strong decay on $x$ (they become even exactly zero when the gauge is installed). Then 
one can weaken the decay of $y$ without making the symplectic structure ill-defined. Of course it is not as simple as this because there are additional 
conditions to be obeyed which have to do with boundary terms. Details can be found in \cite{TTnew}. Then the decay of $y$ is dictated by that of $X,Y$ via 
the constraints $y=y_\ast(X,Y)$

In more detail, in our concrete application to the split into spherically symmetric and non-symmetric degrees of freedom employed in this series of papers
we have the following: 
With the choice of fall-off conditions in \eqref{eq:FallOffConds}, the leading order contribution to the Ricci tensor comes from the first-order 
contributions of the non-symmetric constraints. 
In \cite{II}, we computed them explicitly (in any gauge) and found 
\begin{align}
    \label{eq:Zvlm1}
    {}^{(1)}Z^v_{lm} &= \frac{1}{2r^2}(\pi_\mu - \pi_\lambda)y_v - \frac{1}{2} \pi_\mu y_h + 2 \qty(\partial_r^2 - \frac{1}{r} \partial_r - \frac{(l+2)(l-1)}{2 r^2} - \frac{r_s}{r^3}) x^h\\
    &- \qty(2 r \partial_r + l(l+1) + 2 - 2 \frac{r_s}{r}) x^v + 2 \qty(\partial_r + \frac{1}{r}) \sqrt{l(l+1)} x^e - \sqrt{\frac{(l+2)(l+1)l(l-1)}{2}} \frac{1}{r^2} X^e = 0\nonumber
\end{align}
For the positive polarity contributions, one finds that the momentum contributions behave as $r^{-5/2}$ while the remaining terms behave as $r^{-1}$. 
For the negative polarity, the momentum terms fall-off like $r^{-3/2}$ and the other terms like $r^{-2}$.
Solving for the momenta $y$ (here written as $\pi_\mu,\pi_\nu$) we observe that for the negative polarity harmonics, 
one can find solutions to the constraints in terms of a power series in $\sqrt{r}$ 
for the momenta with above decay conditions. 
For the positive polarity, this is not possible because we do not have corresponding momentum contributions to cancel the leading order terms. 

Thus we have the just mentioned two options in order to be able to solve the constraints just for the $y$: Either we weaken the decay of the $y$ and 
strengthen that of the $x$ or we impose a gauge different from the GPG condition in order to cancel unwanted contributions or a combination of both.
In the former case, we also have to modify the decay of some of $X,Y$ in order that the physical Hamiltonian be unchanged as compared 
to previous analysis while in the latter case the decay can be kept $X,Y$ unchanged. 
We will explore both options below. In general, in the latter case we would need to perform a different 
split of the phase space into gauge, true, symmetric and non-symmetric degrees of freedom which would lead to major 
rewriting of our previous results. Thus to keep as close as possible to the previous analysis, we impose a modified gauge condition still on $x$ 
only which however changes the GPG. We will therefore not follow this path further in the main text and rather 
modify the decay behaviour while keeping the GPG intact. In both cases these modifications potentially affect the form of the reduced  
Hamiltonian so that we have to revisit the analysis in \cite{I}. 
We will tailor the freedom in modifying the decay conditions or gauge fixing conditions such that the previously derived physical
Hamiltonian remains unchanged.

\subsubsection{Modified Fall-Off Conditions}

First, we explore the modification of the fall-off conditions on the variables. 
The goal is to find a self-consistent collection of fall-off conditions.
We follow the strategy of reference \cite{TTNew1}.

The true degrees of freedom are $(X,Y)$ and in order that $\int Y \delta X$ remains finite, we need to have $X = O(r^\alpha)$ and $Y = O(r^\beta)$ with $\alpha + \beta \leq -1$ if $X$ and $Y$ are of opposite parity. 
Since, the leading order contribution to $m_{AB}$ is $r^2 \Omega_{AB}$ we assume that $\alpha \leq 1$.
The largest number of modes are achievable when the bound is saturated, i.e. $\alpha + \beta = -1$.
Now, we consider the constraint equations to determine the fall-off behaviour of the gauge momenta $y$ depending on the fall-off conditions of $X,Y$.
For this, we use the already known expressions that $p_v = O(\sqrt{r})$ and $p_h = O(r^{-3/2})$ (see \cite{II}).
This is the decay behaviour of the corresponding functions for the Schwarzschild black hole in Gullstrand-Painlevé coordinates.

In the odd polarity sector, we find that $y_o$ scales like $r Y_o + r^{-5/2} X^o$ and for $\alpha \leq 1$ the first term dominates and we have $y_o = O(r^{\beta +1})$.
From the first order even polarity constraints in \cite{II}, we observe that $y_h$ behaves as $r^{-2}$ times $y_v$ and $y_e$ behaves as $r^{-1}$ times $y_v$.
Then, the constraints imply that for $\alpha< 3/4$ the $Y_e$ contribution dominates the solution for $y_v$ at infinity and for $\alpha > 3/4$ the $X^e$ terms dominate.
One good choice is the value $\alpha = 3/4$ and for this value we have $y_v = O(r^{1/4})$, $y_h = O(r^{-7/4})$ and $y_e = O(r^{-3/4})$ as well as
$y_o = O(r^{-1})$ as before.
Thus $(X_o,Y^o)=(O(r),O(r^{-2})$ as before and $(X_e,Y^e)=(O(r^{3/4}),O(r^{-7/4})$. Thus we have achieved that $p_v$ asymptotically dominates 
$y_v$. As before all momenta have opposite parity as compared to the configuration degrees of freedom.

For the next step, we consider the GP gauge fixing condition and evaluate its stability under Hamiltonian evolution. 
This gives us expressions $S = S_*$ for the lapse and shift and we obtain asymptotic expressions based on the fall-off conditions of $y, X, Y$ above.
The computation requires the full Hamiltonian constraint $V_0$ and the full diffeomorphism constraint $V_\mu$ which are given by
\begin{gather}
    V_0 = \frac{1}{\sqrt{\det m}}\qty(m_{\mu \rho} m_{\nu \sigma}- \frac{1}{2} m_{\mu \nu} m_{\rho \sigma}) W^{\mu \nu} W^{\rho \sigma} - \sqrt{\det m} R\\
    V_\mu = W^{\rho \sigma} \partial_\mu m_{\rho \sigma} - 2 \partial_\rho(W^{\rho \sigma} m_{\mu \sigma}) 
\end{gather}
We smear these constraints with suitable test functions $S^\mu_\parallel$, $S_\bot$.
$V_{\parallel}[S_\parallel]$ is defined by $\int V_\mu S^\mu_\parallel$ and $V_{\bot}[S_\bot]$ is given by $\int V_0 S_\bot$.
In \cite{I} we discussed that in order to make the variational principle well defined, we have to add suitable boundary terms.
The improved constraints are denoted by $H_\parallel[S_\parallel] := V_\parallel[S_\parallel] + B_\parallel[S_\parallel]$ and $H_\bot[S_\bot] := V_\bot[S_\bot] + B_\bot[S_\bot]$, where $B_\parallel, B_\bot$ are the suitably chosen boundary terms.

The GP gauge fixing conditions are $G_1 = m_{33}-1=0$, $G_2=m_{3A}=0$ and $G_3 = \Omega^{AB} m_{AB} - 2 r^2=0$.
They involve only the metric $m_{\mu \nu}$ and therefore we calculate: 
\begin{align}
    \qty{m_{\alpha\beta}, H_\bot[S_\bot]} = \tilde S_\bot \qty(2 m_{\alpha \rho} m_{\beta \sigma} W^{\rho \sigma} - m_{\alpha \beta} m_{\rho \sigma} W^{\rho \sigma}),
\end{align}
where $\tilde S_\bot = S_\bot/\sqrt{m}$. The diffeomorphism constraint gives
\begin{align}
    \qty{m_{\alpha\beta}, H_\parallel[S_\parallel]} = S^\mu_\parallel \partial_\mu m_{\alpha \beta} + 2 m_{\mu (\alpha} \partial_{\beta)} S^\mu_\parallel
\end{align}

We now specialise the variables to the GP gauge. The metric components are $m_{33}=1$, $m_{3 A}=0$ and $m_{AB} = r^2 \Omega_{AB} + X_{AB}$, where $X_{AB}$ is a symmetric, traceless tensor. 
The conjugate momenta are $W^{33}=\sqrt{\Omega}(p_v + y_v)$, $W^{3A} = \sqrt{\Omega} y^A/2$ and $W^{AB} = \sqrt{\Omega}(p_h + y_h)\Omega^{AB}/2 + \sqrt{\Omega} Y^{AB}$ as before.
For the test functions we use the convention $S^3_\parallel = f_h + g_h$ and $S^A_\parallel = g^A$ and $\tilde S_\bot = \sqrt{\Omega}^{-1}(f_v + g_v)$. 

The stability condition for the GP gauge ($G_1 = G_2 = G_3 = 0$) is given by the following Poisson brackets: 
\begin{align}
    \{G_1,H_\bot[S_\bot] +  H_\parallel[S_\parallel]\} &= 2 \partial_3 (f_h + g_h) +(f_v+g_v)((p_v + y_v) - r^2 (p_h + y_h) - X_{AB} Y^{AB}) = 0\nonumber\\
    \label{eq:GPfixingEqn1}
    \{G_2,H_\bot[S_\bot] +  H_\parallel[S_\parallel]\} &= \partial_A g_h + (r^2 \Omega_{AB} + X_{AB})\partial_3 g^B + (f_v+g_v) (r^2 \Omega_{AB} + X_{AB}) y^B =0\\
    \{G_3,H_\bot[S_\bot] +  H_\parallel[S_\parallel]\} &= 4 r (f_h + g_h) + 2 r^2 D_A g^A + 2 \Omega^{AB} X_{AC} D_B g^C + (f_v+g_v)\Big[(X^{AB}X_{AB}) (p_h + y_h)\nonumber\\
    &~~~~~~~~~~+ 2\Omega^{AB}X_{AC}X_{BD} Y^{CD} - 2 r^2 (p_v + y_v - X_{AB} Y^{AB})\Big]=0\nonumber
\end{align}
These are three coupled differential equations for the unknowns $f_h,g_h,f_v,g_v,g^A$.

Since $p_v = O(\sqrt{r})$, we can neglect to leading order $X_{AB}Y^{AB}$ in the first and last equations.
In addition, using that the requirement $\alpha \leq 1$, we can neglect more terms in the second and last equation involving $X_{AB}$.
We end up with the equations
\begin{align}
    \begin{split}
    &2 \partial_3 (f_h + g_h) +(f_v+g_v)((p_v + y_v) - r^2 (p_h + y_h)) = 0\\
    & \partial_A g_h + r^2 \partial_3 g_A + (f_v+g_v) r^2 y_A =0\\
    & 4 r (f_h + g_h) + 2 r^2 D_A g^A - 2 r^2 (f_v+g_v)(p_v + y_v) = 0
    \end{split}
\end{align}
Extracting the symmetric components of these equations, we have
\begin{gather}
    \begin{split}
    2 \partial_r f_h + f_v (p_v - r^2 p_h) + g_v \cdot (y_v - r^2 y_h) = 0\\
    4 r f_h - 2 r^2 f_v p_v - 2 r^2 g_v \cdot y_v = 0
    \end{split}
\end{gather}
In the above equations, we can drop the terms involving $g_v$ provided that $g_v$ fall-off faster than $r^{-7/4}$.
Then, we solve the equations for $f_h$ and $f_v$.
The last equation implies that
\begin{equation}
    f_v = \frac{2}{r p_v} f_h
    \label{eq:Solfv}
\end{equation}
Then, the first equation becomes
\begin{equation}
    r \partial_r f_h + (1 - r^2 \frac{p_h}{p_v}) f_h = 0
\end{equation}
The radial diffeomorphism constraint is given by 
\begin{equation}
    V_3 = \sqrt{\Omega}[ - 2 \partial_r (p_v +y_v) + 2 r (p_h + y_h) - \partial_A y^A + \partial_r X_{AB} Y^{AB}]
\end{equation}
To leading order, we derive the asymptotic relation $p_v \sim 2 r^2 p_h$ as $r$ approaches infinity.
Therefore, we find that $f_h \sim r^{-1/2}$ and through \eqref{eq:Solfv} we have $f_v \sim r^{-2}$. 
Fixing that $f_v$ asymptotes to $r^{-2}$ as suggested by the Schwarzschild solution in GP coordiantes, we find that $f_h$ asymptotes to $p_v/(2r)$.

In the following, we assume that the perturbations $g_v$ and $g_h$ fall off faster than the spherically symmetric lapse and shift $f_v$ and $f_h$. 
Extracting the non-symmetric components of the stability conditions we have the leading order equations
\begin{align}
    \begin{split}
    &2 \partial_3 g_h^{lm} + f_v (y_v^{lm} - r^2 y_h^{lm}) + g_v^{lm} (p_v - r^2 p_h ) = 0\\
    &r^2 \partial_3 g_o^{lm} + f_v r^2 y_o^{lm} =0\\
    &\sqrt{l(l+1)} g_h^{lm} + r^2 \partial_3 g_e^{lm} + f_v r^2 y_e^{lm} =0\\
    &4 r g_h^{lm} - 2 \sqrt{l(l+1)} r^2 g_e^{lm} - 2 r^2 f_v y_v^{lm} - 2 r^2 g_v^{lm} p_v = 0
    \end{split}
\end{align}
The second equation implies that $g_o^{lm} = O(r^{-2})$ when we assume $y_o = O(r^{-1})$. 
Based on the above equations a consistent choice for the other test functions is $g_h^{lm} = O(r^{-3/4})$, $g_e^{lm} = O(r^{-7/4})$ and $g_v^{lm} = O(r^{-9/4})$.

With this choice of variables it remains to show that the boundary terms are finite.
The boundary terms were worked out in \cite{I} and read
\begin{gather}
    B_\parallel[S_\parallel] = 2 \int \dd \Omega S^\nu W^{3 \mu} m_{\mu \nu}\\
    \begin{split}
    B_\bot[S_\bot] &= - \int \dd \Sigma_{\rho} \sqrt{\det m} \qty[\qty(m^{\mu \rho} m^{\nu \sigma} - m^{\mu \nu} m^{\rho \sigma})\qty(S^0 \nabla_\sigma (m_{\mu \nu} - m^{\mathrm{ND}}_{\mu \nu}) -  \nabla_\sigma S^0 (m_{\mu \nu} - m^{\mathrm{ND}}_{\mu \nu}))]
    \end{split}
\end{gather}
From the first boundary term, we observe that it reduces to
\begin{equation}
    B_\parallel[S^*_\parallel] = 2 \lim_{r \to \infty} \int \dd \Omega f_h^* p_v
\end{equation}
This is the same result as before.
For the boundary term of the Hamiltonian constraint, we obtain
\begin{align}
    \begin{split}
    B_\bot[S_\bot] &= - \lim_{r\to\infty} \int \dd \Omega \sqrt{\det m} m^{AB} \qty(S^0 \nabla_3 X_{AB} -  \nabla_\sigma S^0 X_{AB})\\
    &= - \lim_{r\to\infty} \int \dd \Omega \sqrt{\det \Omega}(r^2 + O(1)) (r^{-2} \Omega^{AB} + O(r^{-3})) \qty(S^0 \nabla_3 X_{AB} -  \nabla_\sigma S^0 X_{AB})
    \end{split}
\end{align}
Since $X_{AB}$ is trace-free, we observe that the leading order contribution of order $1$ vanishes and we are left with decaying contributions.
This shows that the boundary term to the Hamiltonian constraint vanishes as before.

Combining the boundary terms, we see that the physical Hamiltonian for one asymptotic end with the modified fall-off conditions reads
\begin{equation}
    H_{\mathrm{phys}} = \lim_{r\to \infty} \int_{S^2} \dd \Omega \frac{p_v^2}{2 \kappa r}
\end{equation}
where $\kappa$ is the gravitational coupling constant.

\subsubsection{Modified Gauge Fixing Condition}

There are several choices to cancel the leading order contributions  in \eqref{eq:Zvlm1} by modifying the gauge conditions.
For the simplicity of the computation, we would like to stay as close as possible to the original GP gauge ($x=0$).
One option is
\begin{align}
    x^h_{lm} = - \sqrt{\frac{(l+2)(l-1)}{2 l(l+1)}} r \lim_{r \to \infty} \frac{X^e_{lm}}{r} 
    \label{eq:NewGauge}
\end{align}
rather than $x_h=0$ while keeping $x^v= x^e = x^o=0$. As we will not follow this path in the remaining main text,
in this subsection, we restrict attention to second order perturbation theory and neglect any higher order corrections in order 
to illustrate the changes that would become necessary.

Due to the change of gauge, the first and second order constraints are modified and we find the following corrections:
\begin{align}
    \begin{split}
    {}^{(1)} \tilde Z^h_{lm} &= {}^{(1)} Z^h_{lm} - \frac{\pi_\lambda}{2 r^2} \sqrt{\frac{(l+2)(l-1)}{2 l(l+1)}} X^e_\infty\\
    {}^{(1)} \tilde Z^{e/o}_{lm} &= {}^{(1)} Z^{e/o}_{lm}\\
    {}^{(1)} \tilde Z^v_{lm} &= {}^{(1)} Z^v_{lm} + \frac{l(l+1)r + 2 r_s}{r^2}\sqrt{\frac{(l+2)(l-1)}{2 l(l+1)}} X^e_\infty\\
    {}^{(2)} \tilde C^h_{lm} &= {}^{(2)} C^h_{lm}\\
    {}^{(2)} \tilde C^v_{lm} &= {}^{(2)} C^v_{lm} - \frac{(l+2)(l-1)}{4 l(l+1) r^2}\qty(1 - 4 \frac{r_s}{r}) X^e_\infty \cdot X^e_\infty + \frac{\pi_\mu}{2 r^3}  \sqrt{\frac{(l+2)(l-1)}{2 l(l+1)}}  y_v \cdot X^e_\infty
    \end{split}
\end{align}
We abbreviated $\lim_{r \to \infty} \frac{X^e}{r}$ as $X^e_\infty$ which is an $l,m$ dependent constant with respect to $r$.
The constraints $Z,C$ on the right-hand side are the original constraints (see \cite{II}) imposing the GP gauge and the constraints $\tilde Z, \tilde C$ on the left-hand side are the modified constraints in the new gauge.
Notice that when solving the constraints, we have to insert the solution of the first-order constraints into the second-order ones and this will give additional contributions that we will compute in the following.

For the reduced phase space formulation, we solve the constraints order by order for the gauge momenta $p$ and $y$. 
At zeroth order there is no corrections and the previous results in \cite{II} are still valid. 
Next, for the first-order, even-polarity constraints we start by solving ${}^{(1)} Z^v_{lm} = 0$ for $y_h$ and find the solution
\begin{align}
    y_h^{(1)} &= \frac{y_v^{(1)}}{2 r^2} - \frac{\sqrt{2(l+2)(l+1)l(l-1)}}{4 r^2 \sqrt{r r_s} } X^e + \frac{l(l+1)r + 2 r_s}{2 r^2 \sqrt{r r_s}}\sqrt{\frac{(l+2)(l-1)}{2 l(l+1)}} X^e_\infty
\end{align}
Then, the constraint ${}^{(1)} Z^h_{lm} = 0$ gives
\begin{align}
    \begin{split}
    \sqrt{l(l+1)} y_e^{(1)} &= 2 \partial_r y_v^{(1)} - 2 r y_h^{(1)} + \frac{\sqrt{r r_s}}{r^2} \sqrt{\frac{(l+2)(l-1)}{2 l(l+1)}} X^e_\infty\\
    &= 2 \partial_r y_v^{(1)} - \frac{1}{r}y_v^{(1)} + \frac{\sqrt{2(l+2)(l+1)l(l-1)}}{2 r \sqrt{r r_s} } X^e - \frac{l(l+1)r + r_s}{r \sqrt{r r_s}}\sqrt{\frac{(l+2)(l-1)}{2 l(l+1)}} X^e_\infty
    \end{split}
\end{align}
Without the modification, the solution for $y_v^{(1)}$ is given in terms of an integral of a source term (see \cite{II} and also \eqref{eq:yv1} in this paper).
Here, the source term $\tilde s(r)$ is modified and given in terms of the original source term $s(r)$ as
\begin{align}
    \tilde s(r) = s(r) + \sqrt{\frac{(l+2)(l-1)}{2 l (l+1)}} \frac{l^2(l+1)^2 r - 3 l(l+1) r + 2 l(l+1) r_s  - r_s}{2 \sqrt{r r_s}} X^e_\infty 
\end{align}

Next, we explicitly compute the change in the solution for $y_v^{(1)}$ due to this change in the source term.
We find for the new solution $\tilde y_v^{(1)}$ in terms of the old solution $y_v^{(1)}$:
\small
\begin{align}
    \tilde y_v^{(1)} &= y_v^{(1)} + \sqrt{\frac{(l+2)(l-1)}{2 l(l+1)(4 l(l+1) - 9)}} \frac{i}{\sqrt{r_s}} r^{\alpha_-} \int \dd r \qty(r^{\alpha_+}l(l+1)(l(l+1) -3) + r^{\alpha_+ -1}r_s (2 l(l+1) -1))X^e_\infty - \alpha_+ \leftrightarrow \alpha_-\nonumber\\
    &= y_v^{(1)} + \sqrt{\frac{(l+2)(l-1)}{2 l(l+1)(4 l(l+1) - 9)}} \frac{i}{\sqrt{r r_s}} \qty(\frac{r}{\alpha_+ + 1} l(l+1)(l(l+1) -3) + \frac{r_s}{\alpha_+} (2 l(l+1) -1))X^e_\infty - \alpha_+ \leftrightarrow \alpha_-\nonumber\\
    &= y_v^{(1)} + \sqrt{\frac{(l+2)(l-1)}{2l(l+1)}} \frac{2}{\sqrt{r r_s}} \qty(r (l(l+1) -3) + \frac{r_s}{(l+2)(l-1)}  (2 l(l+1) -1)) X^e_\infty
\end{align}
\normalsize
In the computation we used the relations $\alpha_+ + \alpha_- = - 1/2$, $\alpha_+ - \alpha_- = \frac{i}{2}\sqrt{4l(l+1) - 9}$ and $\alpha_+ \alpha_- = \frac{1}{4} (l+2)(l-1)$.
For the other two momenta, we find
\begin{align}
    \tilde y_h^{(1)} = y_h^{(1)} &+ \sqrt{\frac{(l+2)(l-1)}{2 l(l+1)}} \frac{1}{2 r^2 \sqrt{r r_s}} \qty(3 r (l+2)(l-1) + \frac{2 r_s}{(l+2)(l-1)}  (3 l(l+1) - 3 )) X^e_\infty
\end{align}
\begin{align}
    \sqrt{l(l+1)} \tilde y_e^{(1)} &= \sqrt{l(l+1)} y_e^{(1)} + \sqrt{\frac{(l+2)(l-1)}{2 l(l+1)}} \frac{1}{r \sqrt{r r_s}} \qty( - r l(l+1) - \frac{r_s}{(l+2)(l-1)}  (9 l(l+1) - 6)) X^e_\infty\nonumber
\end{align}

This completes the solution of the first-order constraints in the new gauge. 
We now need to verify the fall-off conditions of the solutions $y$ in terms of the physical degrees of freedom $X,Y$.
As it turns out we have to multiply the fall-off conditions for $y_{e/v/h}$ by a factor of $\sqrt{r}$ as compared with \eqref{eq:FallOffConds}.
Comparing with \cite{I} shows that such a strengthening of the fall-off conditions does not violate the conclusions in this paper.

Therefore, with a weakening of the fall-off conditions for the momenta $y$  by a factor of $\sqrt{r}$ compared to the original proposal, 
everything is consistent. 
Next, we need to insert these solutions into the second order constraints. 
Fortunately, in \cite{II}, we showed that the physical Hamiltonian to second order is weakly gauge invariant and gauge-variant terms only amount to boundary terms at infinity. 
For completeness, we recorded the change in the boundary term in appendix \ref{sec:BTModifiedGauge}.
Thus, we will obtain the same physical Hamiltonian as before but with a stronger fall-off behaviour of the $y$ variables.

\subsubsection{Summary}

In the previous two subsections, 
we explored two different strategies to obtain a well-defined set of fall-off conditions in the reduced phase space formulation. 
Both have some advantages and disadvantages:

The first approach is based on a modification of the fall-off conditions on the true degrees of freedom.
The new fall-off conditions were derived in a systematic way by noting that the fall-off conditions of the physical degrees of freedom $X,Y$ determine the 
fall-off conditions
of the gauge momenta $y$ and the fixed lapse and shift.
Then, tuning the fall-off conditions of $X,Y$ we found a self-consistent set of decay conditions such that the boundary terms are finite and the symplectic
form is well-defined.
The advantage of this approach is that we keep the original GP gauge fixing condition which has a nice physical interpretation and is easy to implement.
Furthermore, we derived the physical Hamiltonian in \cite{I,II} in this gauge and thus the results directly apply. 
On the other hand, the fall-off conditions for the even polarity sector are unusual and it is unclear whether they capture all the interesting physical solutions 
of the theory.

In the second approach, we modified the gauge fixing condition so that the problematic leading order terms in the Hamiltonian constraint cancel.
This gives rise to new constraint equations and different solutions of the constraints that need to be calculated explicitly.
In the end, we find that the physical Hamiltonian is unchanged because the new contributions only modify the boundary terms that were shown to vanish 
in \cite{II}. 

For the remaining discussion in this paper, we stick to the first approach because in it we can use the results of \cite{I}
in order to determine the physical Hamiltonian without too many modifications as compared to the previous analysis.

\subsection{Expansion of the Diffeomorphism Constraint}

The diffeomorphism constraint of general relativity is given by
\begin{equation}
    V_\mu = - 2 \nabla_\rho \qty(m_{\mu \nu} W^{\nu \rho})
\end{equation}
Expressing the covariant derivative in terms of a partial derivative and Christoffel symbols we arrive at
\begin{equation}
    V_\mu = -2 \partial_\rho \qty(m_{\mu \nu} W^{\nu \rho}) + \partial_\mu m_{\nu \rho} W^{\nu \rho}
\end{equation}
This is a quadratic expression in the variables $(m_{\mu \nu}, W^{\mu \nu})$.
Thus, any perturbative expansion of the diffeomorphism constraint will naturally terminate at second order and the perturbation theory is exact for any higher truncations.

The radial component $\mu = 3$ of the diffeomorphism constraint is given by
\begin{equation}
\begin{aligned}
    V_3 &= -2 \partial_\rho(m_{33} W^{3 \rho}) + \partial_3 (m_{AB}) W^{AB}\\
    &= \sqrt\Omega \qty[- 2 \partial_r (p_v + y_v) + 2 r (p_h + y_h) - D_A y^A + Y^{AB} \partial_r X_{AB} ]
\end{aligned}
\end{equation}
The angular diffeomorphism constraint gives
\begin{align}
        V_A &= -2 \partial_\rho(m_{AB} W^{\rho B}) + \partial_A (m_{CD}) W^{CD}\nonumber\\
        &= - 2 \partial_r(m_{AB} W^{3B}) - 2 D_C (m_{AB} W^{BC}) + D_A X_{CD} W^{CD}\\
        &= \sqrt\Omega\qty[ - \partial_r (r^2 y_A + X_{AB} y^B) - \frac{2r^2}{\sqrt{\Omega}} D^B W_{AB} - 2 D_C (X_{AB} Y^{BC}) - D^B (X_{AB}(p_h + y_h)) + Y^{CD} D_A X_{CD}]\nonumber
\end{align}
We collect terms of equal order. At zeroth order we find
\begin{equation}
    {}^{(0)}V_3 = 2 \sqrt\Omega \qty(r p_h - \partial_r p_v), \quad {}^{(0)}V_A = 0
\end{equation}
At first order, we have
\begin{equation}
    {}^{(1)}V_3 =  \sqrt\Omega\qty(2 r y_h - 2 \partial_r y_v -  \partial_A y^A), \quad {}^{(1)}V_A = \sqrt\Omega\qty(- \partial_r (r^2 y_A) - \frac{2}{\sqrt{\Omega}} r^2 D^B W_{AB} - D^B X_{AB} p_h )
\end{equation}
At second order, we find
\begin{equation}
\begin{split}
    {}^{(2)}V_3 &= \sqrt\Omega Y^{AB} \partial_r X_{AB}\\
    {}^{(2)}V_A &= \sqrt\Omega \qty(- \partial_r (X_{AB} y^B)- 2 D_C (X_{AB} Y^{BC}) - D^B(X_{AB} y_h) + D_A X_{BC} Y^{BC})
\end{split}
\end{equation}

The constraints are now decomposed into the spherically symmetric zero modes and the non-symmetric modes with $l>0$. 
For the radial components, the zero modes are given by an average over the sphere
\begin{equation}
    C_h = \int \dd\Omega V_3 = 8 \pi (r p_h - \partial_r p_v) + \sum_{I l m} Y_{I lm} \partial_r X^{Ilm}
\end{equation}
The non-symmetric contributions to the constraint are
\begin{equation}
    Z^h_{lm} := \int_{S^2} \dd\Omega L_{lm} V_3 = 2 r y_h^{lm} - 2 \partial_r y_v^{lm}  + \sqrt{l(l+1)} y_e^{lm} + \sum_{I l_2 m_2, J l_3 m_3} C^{l m 0}_{l_2 m_2 2, l_3 m_3 -2} q_{IJ} \partial_r X^I_{l_2 m_2} Y^J_{l_3 m_3}
\end{equation}

For the angular components, we calculate the contributions order by order. 
Since there are no spherically symmetric vector fields on the sphere, there are no spherically symmetric contributions and we only need to compute the non-symmetric constraints.
We start with the first order contribution:
\begin{equation}
    {}^{(1)}Z^I_{lm} := \int_{S^2} \dd \Omega [L^I_{lm}]^A {}^{(1)} V_A = \sqrt{2(l+2)(l-1)}\qty(r^2 Y_I^{lm} + \frac{1}{2} p_h X^I_{lm}) - \partial_r (r^2 y_I^{lm}) - \delta^I_e r^2 \sqrt{l(l+1)} y_h^{lm}
\end{equation}
The second order contribution is the sum of three terms that we compute separately.
The first term is given by
\begin{align}
\label{eq:2ndAngDiffeo1}
    \int_{S^2} \dd \Omega [L^I_{lm}]^A \partial_r(X_{AB}y^B) &= \sum_{\{l_i,m_i\}, J K} C^{l m -1}_{l_2 m_2 2, l_3 m_3 -1} \frac{i^{I+J+K}}{2\sqrt{2}}(-1)^J (1+ (-1)^{I+J+K} (-1)^{l + l_2 + l_3}) \partial_r(X^J_{l_2 m_2} y_K^{l_3 m_3}) 
\end{align}
Next, we have
\begin{align*}
     &[L^I_{lm}]^A D^B (X_{AB}y_h) = \sqrt{\Omega} \sum_{\{l_i,m_i\},J}\frac{i^{I+J}}{2\sqrt{2}}X^J_{l_2 m_2} y_h^{l_3 m_3} ( (-1)^J L^{-1}_{lm} \overline \eth(L^2_{l_2 m_2}L_{l_3 m_3}) - (-1)^I L^1_{lm} \eth (L^{-2}_{l_2 m_2} L_{l_3 m_3})\Big]\\
     &=  \sum_{\{l_i,m_i\},J}\frac{i^{I+J}}{2\sqrt{2}}X^J_{l_2 m_2} y_h^{l_3 m_3} \Big[(-1)^J L^{-1}_{lm} (\sqrt{(l_2+2)(l_2-1)} L^1_{l_2 m_2} L_{l_3 m_3} + \sqrt{l_3(l_3+1)}L^2_{l_2 m_2} L^{-1}_{l_3 m_3})\\
     &~~~~~~~~~~~~~~~~~+ (-1)^I L^1_{lm} (\sqrt{(l_2+2)(l_2-1)} L^{-1}_{l_2 m_2} L_{l_3 m_3} + \sqrt{l_3(l_3+1)}L^{-2}_{l_2 m_2} L^1_{l_3 m_3})\Big]
\end{align*}
Integrating this expression, we find
\begin{align*}
     \int_{S^2} \dd \Omega [L^I_{lm}]^A D^B (X_{AB}y_h) = \sum_{\{l_i,m_i\},J}&\frac{i^{I+J}}{2\sqrt{2}}(-1)^I(1 + (-1)^{I+J} (-1)^{l+l_2 +l_3}) X^J_{l_2 m_2} y_h^{l_3 m_3} \times\\
     &\times\Big[\sqrt{(l_2+2)(l_2-1)}C^{l m 1}_{l_2 m_2 -1, l_3 m_3 0} + \sqrt{l_3 (l_3 + 1)} C^{lm1}_{l_2 m_2 -2,l_3 m_3 1}\Big]
\end{align*}
From the recursion relation of the Clebsh-Gordan coefficients $C$ we have the relation
\begin{equation}
    \sqrt{(l+2)(l-1)}C^{l m 2}_{l_2 m_2 -2, l_3 m_3 0} + \sqrt{(l_2+2)(l_2-1)}C^{l m 1}_{l_2 m_2 -1, l_3 m_3 0} + \sqrt{l_3 (l_3+1)}C^{l m 1}_{l_2 m_2 -2, l_3 m_3 1} = 0
\end{equation}
This simplifies the sum of Clebsh-Gordan coefficients and we have
\begin{align}
    \label{eq:2ndAngDiffeo2}
    \int_{S^2} \dd \Omega [L^I_{lm}]^A D^B (X_{AB}y_h) = - \sum_{\{l_i,m_i\},J}\sqrt{\frac{(l+2)(l-1)}{2}} q^{IJ}  X_J^{l_2 m_2} y_h^{l_3 m_3} C^{lm2}_{l_2 m_2 -2, l_3 m_3 0}
\end{align}
The last contribution to the second order non-symmetric constraint is
\begin{align*}
    &[L^I_{lm}]^A(- 2 D_C(X_{AB}Y^{BC}) + D_A X_{BC} Y^{BC})=\\
    &= \sqrt{\Omega} \sum_{\{l_im_i\},JK}\frac{i^{I+J+K}}{4}(L^{-1}_{lm} ((-1)^K \eth L^{-2}_{l_2 m_2} L^2_{l_3 m_3}- (-1)^J \eth L^{2}_{l_2 m_2} L^{-2}_{l_3 m_3} - 2 (-1)^J L^{2}_{l_2 m_2} \eth  L^{-2}_{l_3 m_3})\\
    &~~~~~~~~+ (-1)^I L^1_{lm} ((-1)^K \overline \eth L^{-2}_{l_2 m_2} L^2_{l_3 m_3} + 2 (-1)^K L^{-2}_{l_2 m_2} \overline \eth L^2_{l_3 m_3}- (-1)^J \overline \eth L^{2}_{l_2 m_2} L^{-2}_{l_3 m_3}) X^J_{l_2 m_2} Y_K^{l_3 m_3}
\end{align*}
Integrating over the sphere, we have
\begin{align*}
    &\int_{S^2} \dd\Omega [L^I_{lm}]^A(- 2 D_C(X_{AB}Y^{BC}) + D_A X_{BC} Y^{BC})=\\
    &= \sum_{\{l_im_i\},JK} \frac{i^{I+J+K}}{4} \Big[ \qty(2 \sqrt{(l_3 +2)(l_3 -1)}C^{l m 1}_{l_2 m_2 -2, l_3 m_3 1}+ \sqrt{(l_2+3)(l_2-2)} C^{lm1}_{l_2 m_2 -3, l_3 m_3 2})(-1)^{I+K}\\
    &~~~~~~~~- \sqrt{(l_2+2)(l_2-1)}C^{lm1}_{l_2 m_2 1, l_3 m_3 -2}(-1)^{I+J}\Big](1 + (-1)^{I+J+K}(-1)^{l+l_2+l_3})X^J_{l_2 m_2} Y_K^{l_3 m_3}
\end{align*}
From the recursion relation of the Clebsch-Gordan coefficients, it follows that
\begin{equation}
    \sqrt{(l_2+3)(l_2-2)}C^{l m 1}_{l_2 m_2 -3,l_3 m_3 2} + \sqrt{l(l+1)}C^{l m0}_{l_2 m_2 -2, l_3 m_3 2} + \sqrt{(l_3 + 2)(l_3 - 1)}C^{l m 1}_{l_2 m_2 -2, l_3 m_3 1} = 0
\end{equation}
This gives
\begin{align}
\label{eq:2ndAngDiffeo3}
    &\int_{S^2} \dd \Omega [L^I_{lm}]^A(- 2 D_C(X_{AB}Y^{BC}) + D_A X_{BC} Y^{BC})=\nonumber\\
    &= \sum_{\{l_im_i\},JK} \frac{i^{I+J+K}}{4} \Big[\qty(\sqrt{(l_3 +2)(l_3 -1)}C^{l m 1}_{l_2 m_2 -2, l_3 m_3 1} - \sqrt{l(l+1)}C^{l m0}_{l_2 m_2 -2, l_3 m_3 2})(-1)^{I+K}\\
    &~~~~~~~~- \sqrt{(l_2+2)(l_2-1)}C^{lm1}_{l_2 m_2 1, l_3 m_3 -2}(-1)^{I+J} \Big](1 + (-1)^{I+J+K} (-1)^{l+l_2 +l_3})X^J_{l_2 m_2} Y_K^{l_3 m_3}\nonumber
\end{align}

The second order contributions to the angular diffeomorphism constraint are then the three parts in equations \eqref{eq:2ndAngDiffeo1}, \eqref{eq:2ndAngDiffeo2} and \eqref{eq:2ndAngDiffeo3}.

\subsection{Expansion of the Hamiltonian Constraint}

In this section, we expand the Hamiltonian constraint in terms of the non-symmetric degrees of freedom. 
For the diffeomorphism constraint, we observed that it is a polynomial of degree two in the dynamical variables. 
The Hamiltonian constraint of general relativity in the ADM formulation is 
\begin{equation}
    V_0 = \frac{1}{\sqrt{\det(m)}}\qty(m_{\mu \rho} m_{\nu \sigma} - \frac{1}{2} m_{\mu \nu} m_{\rho \sigma})W^{\mu \nu}W^{\rho \sigma}- \sqrt{\det(m)} R
\end{equation}
We notice that the expression above is non-polynomial in the metric $m_{\mu \nu}$ due to the square root of the determinant of the metric and the Ricci scalar, which involves contractions with the inverse metric.

The determinant of the metric $\det m$ is quadratic in the perturbations, as can be seen by computing 
\begin{equation}
    \det m = \det( r^2 \Omega + X) = \frac{1}{2} \epsilon^{AC} \epsilon^{BD} (r^2 \Omega_{AB} + X_{AB}) (r^2 \Omega_{CD} + X_{CD}) = r^4 \det \Omega + \det X
\end{equation}
Using the identity $\epsilon^{AB} \epsilon^{CD} = \det \Omega \qty(\Omega^{AC} \Omega^{BD} - \Omega^{AD} \Omega^{BC})$, we find for any traceless and symmetric $2\times2$ matrix $X_{AB}$, $\det X = - \frac{1}{2} \det \Omega X^{AB} X_{AB}$.
Hence, we have
\begin{equation}
    \det m = \det \Omega  \qty(r^4 - \frac{1}{2} X^{AB} X_{AB})
\end{equation}

The Hamiltonian constraint consists of two parts. The first one is quadratic in the momenta while the second one is proportional to the Ricci scalar. 
In the following we will analyze them separately.
In addition, we will ignore the factors $\sqrt{\det m}$ in the beginning and reinsert them afterwards.

\subsubsection{Perturbative Expansion of the Momentum Contributions to the Hamiltonian Constraint}

We begin with the momentum contributions and ignore the prefactor involving the square root of the determinant of the metric.
Then, we expand $m$ and $W$ in terms of the symmetric and non-symmetric variables in equation \eqref{eq:DefVariables}. 
We find
\begin{align}
        K&:=\frac{1}{\sqrt{\Omega}} \qty(m_{\mu \rho} m_{\nu \sigma} - \frac{1}{2} m_{\mu \nu} m_{\rho \sigma}) W^{\mu \nu} W^{\rho \sigma} = \sqrt \Omega \Big[\frac{1}{2} (p_v + y_v)^2 + \frac{1}{2} r^2 y_A y^A + \frac{1}{2} X_{AB} y^A y^B\nonumber\\
        &- r^2 (p_v + y_v)(p_h + y_h) - (p_v + y_v) X_{AB}Y^{AB} + r^4 Y^{AB} Y_{AB} + r^2 (p_h + y_h) X_{AB} Y^{AB} + \frac{1}{4}X_{AB}X^{AB} (p_h + y_h)^2\nonumber\\
        & - \frac{1}{2} (X_{AB} Y^{AB})^2 + X_{AB} X_{CD} Y^{AC}Y^{BD} + 2 r^2 Y^{AB}Y^{CD} \Omega_{AC} X_{BD} + (p_h + y_h) Y^{AB} \Omega^{CD} X_{AC} X_{BD}\Big]
\end{align}

The last contribution vanishes because $X_{AB} X_{CD} \Omega^{BD} = \frac{1}{2} \Omega_{AC} X^{BD}X_{BD} \propto \Omega_{AC}$ and $Y^{AC}$ is trace free. 
Similarly, the second to last term vanishes for a similar reason.
Consider a general symmetric and trace-free tensor $T_{AB}$.
Any symmetric tensor can be written in general in the form
\begin{equation}
    T_{AB} = T_1 m_A m_B + T_2 \overline m_A \overline m_B + T_3 (m_A \overline m_B + \overline m_A m_B)
\end{equation}
Imposing the trace-freeness, i.e. $\Omega^{AB} T_{AB} = (m^A \overline m^B + \overline m^A m^B) T_{AB} = 0$, we have the requirement $T_3=0$.
Then, we compute
\begin{equation}
    \Omega^{BD}T_{AB} T_{CD} = T_1 T_2 (m_A \overline m_C + \overline m_A m_C) = T_1 T_2 \Omega_{AC}
\end{equation}
On the other hand, we observe that
\begin{equation}
    T^{AB} T_{AB} = 2 T_1 T_2
\end{equation}
Hence, we conclude that $\Omega^{BD}T_{AB} T_{CD} = \frac{1}{2}\Omega_{AC}T^{BD} T_{BD}$.
In summary, ignoring the factor $\sqrt{\det m}$, the momentum contributions are at most of fourth order in the perturbations.


Expanding $K$ order by order into spherical harmonics, we obtain for the zeroth and first order
\begin{equation}
    {}^{(0)}K_{lm} = \sqrt{4\pi} \delta_{l,0} \qty(\frac{1}{2} p_v^2 - r^2 p_v p_h)
\end{equation}
\begin{equation}
    {}^{(1)}K_{lm} = (p_v - r^2 p_h) y_v^{lm} - r^2 p_v y_h^{lm}
\end{equation}
At second order, we obtain
\begin{align}
    {}^{(2)}K_{lm} &= \sum_{l_1 m_1, l_2 m_2}\Big[C^{lm0}_{l_1 m_1 0, l_2 m_2 0} \qty(\frac{1}{2} y_v^{l_1 m_1} y_v^{l_2 m_2}  - r^2 y_v^{l_1 m_1} y_h^{l_2 m_2}) + \sum_{IJ}\Big(- \frac{1}{2} r^2 C^{lm0}_{l_1 m_1 1, l_2 m_2 -1} q^{IJ} y_I^{l_1 m_1} y_J^{l_2 m_2}\nonumber\\
    &+ C^{lm0}_{l_1 m_1 2, l_2 m_2 -2} q^{IJ}\qty(r^4 Y_I^{l_1 m_1} Y_J^{l_2 m_2} - (p_v - r^2 p_h) X_I^{l_1 m_1}Y_J^{l_2 m_2}  + \frac{1}{4} p_h^2 X_I^{l_1 m_1} X_J^{l_2 m_2}) \Big)\Big]
\end{align}
Next, we have
\begin{align}
    {}^{(3)}K_{lm} &= \sum_{\{l_i,m_i\},IJK,l'm'}\Big[\frac{i^{I+J+K}}{4\sqrt{2}} C^{l m0}_{l_1 m_1 2, l' m' -2} C^{l'm' 2}_{l_2 m_2 -1,l_3 m_3 -1} X^I_{l_1 m_1} y^J_{l_2 m_2}y^K_{l_3 m_3}(-1)^I\qty(1 + (-1)^{I+J+K} (-1)^{l+l_1 +l_2 + l_3})\nonumber \\
    &+ C^{lm 0}_{l_1 m_1 0, l' m' 0} C^{l' m' 0}_{l_2 m_2 2, l_3 m_3 -2} q^{IJ}\Big(- (y_v^{l_1 m_1} - r^2 y_h^{l_1 m_1}) X_I^{l_2 m_2}Y_J^{l_3 m_3} + \frac{1}{2} p_h y_h^{l_1 m_1} X_I^{l_2 m_2} X_J^{l_3 m_3}\Big)\Big]
\end{align}
Finally, the fourth order contributions are
\begin{align}
    {}^{(4)}K_{lm} &= \sum_{\{l_i,m_i\},l',m',l'' m'',I,J}\Big[\frac{1}{4}  C^{l' m' 0}_{l_3 m_3 2, l_4 m_4 -2} C^{l''m''0}_{l_1 m_1 0, l_2 m_2 0}C^{lm0}_{l' m' 0, l'' m'' 0} y_h^{l_1 m_1}y_h^{l_2 m_2} q_{IJ} X^I_{l_3 m_3} X^J_{l_4 m_4}\nonumber\\
    &+ \sum_{KL}\Big(- \frac{1}{2} C^{lm0}_{l'm'0,l'' m''0}C^{l'm'0}_{l_1 m_1 2, l_2 m_2 -2}C^{l'' m''0}_{l_3 m_3 2, l_4 m_4 -2} q_{IJ} X^I_{l_1 m_1}Y^J_{l_2 m_2} q_{KL} X^K_{l_3 m_3} Y^L_{l_4 m_4}\\
    &+\frac{i^{I+J+K+L}}{4} C^{lm0}_{l' m' 0,l'' m'' 0} C^{l' m'0}_{l_1 m_1 2, l_3 m_3 -2} C^{l'' m''0}_{l_2 m_2 2, l_4 m_4 -2}\qty((-1)^{I+J} + (-1)^{K+L} (-1)^{l' + l_1 + l_2 + l'' + l_3 + l_4} )\times\nonumber\\
    &~~~~~~~~~~~~~~~~~\times X^I_{l_1 m_1}X^J_{l_2 m_2} Y^K_{l_3 m_3} Y^L_{l_4 m_4}\Big)\Big]\nonumber
\end{align}
This is the full expansion of the momentum contributions to the Hamiltonian constraint.

\subsubsection{Perturbative Expansion of the Ricci Scalar}
We continue with the Ricci scalar contribution to the Hamiltonian constraint.
For now, we ignore the factor of $\sqrt{\det m}$ and solely focus on the Ricci scalar. 
The scalar curvature is defined in terms of partial derivatives of the Christoffel symbols. 
We will denote by $\overline \Gamma^\mu_{\nu \rho}$ the Christoffel symbol of the 
spherically symmetric part $\overline m_{\mu \nu}$ only and write $\delta \Gamma^\mu_{\nu \rho} = \Gamma^\mu_{\nu \rho}- \overline \Gamma^\mu_{\nu \rho}$.
The correction term is conveniently written in terms of the covariant derivative $\nabla_\mu$ associated to the spherically symmetric metric 
$\overline m_{\mu \nu}$ (i.e. $\nabla_\mu \overline m_{\rho \sigma} = 0$). 
Of course, we have that $\nabla_\mu m_{\nu \rho} = \nabla_\mu \delta m_{\nu\rho}$, where $\delta m_{\mu \nu} = m_{\mu \nu} - \overline m_{\mu \nu}$ 
is the perturbed metric. 
Expanding the inverse metric $m^{\mu \nu}$ to any desired order in the perturbations, we find the corrections to the Christoffel symbol to all orders.

In terms of the Christoffel symbols and its perturbations, the Ricci tensor is given by
\begin{equation}
    R_{\rho \sigma} = \overline R_{\rho \sigma}  + \nabla_\mu \delta \Gamma^\mu_{\rho \sigma} - \nabla_\sigma \delta \Gamma^\mu_{\mu \rho} + \delta \Gamma^\mu_{\mu \tau} \delta \Gamma^\tau_{\rho\sigma} - \delta \Gamma^\mu_{\rho \tau} \delta \Gamma^\tau_{\mu \sigma}
\end{equation}
Here $R_{\rho\sigma}$ is the full Ricci tensor, while $\overline R_{\rho\sigma}$ is the spherically symmetric Ricci tensor.
In the Gullstrand-Painlevé gauge, the metric $\overline m_{\mu \nu}$ is spatially flat and therefore $\overline R_{\mu \nu}$ vanishes. 
We introduced the covariant derivative $\nabla_\mu$ for computational convenience and defined the perturbed Christoffel symbols by 
$\delta \Gamma^\mu_{\nu \rho} = \Gamma^\mu_{\nu \rho}- \overline \Gamma^\mu_{\nu \rho}$ 
In Gullstrand-Painlevé gauge, we are left with
\begin{equation}
    R_{\rho \sigma} = \nabla_\mu \delta \Gamma^\mu_{\rho \sigma} - \nabla_\sigma \delta \Gamma^\mu_{\mu \rho} + \delta \Gamma^\mu_{\mu \tau} \delta \Gamma^\tau_{\rho\sigma} - \delta \Gamma^\mu_{\rho \tau} \delta \Gamma^\tau_{\mu \sigma}
\end{equation}
The non-symmetric contributions to the Christoffel symbol involve the full inverse metric and for the Ricci tensor we have to take its derivative.
For the computation, we would not like to introduce a perturbative expansion of $m^{\mu \nu}$ but keep the discussion general. 
For taking derivatives, we use 
\begin{equation}
    \nabla_\tau m^{\mu \nu} = - m^{\mu \sigma} m^{\nu \rho} \nabla_\tau m_{\sigma \rho}
\end{equation}
which directly follows from the identity $m_{\mu \nu} m^{\nu \rho} = \delta_\mu^\rho$ by taking a covariant derivative.

Inserting the expression for the perturbed Christoffel symbols, the full Ricci tensor is given by
\begin{equation}
\begin{aligned}
        R_{\rho \sigma} &=\frac{1}{2} m^{\mu \alpha} \qty(\nabla_\mu \nabla_\rho m_{\sigma \alpha} +\nabla_\mu \nabla_\sigma m_{\rho \alpha} - \nabla_\mu \nabla_\alpha m_{\rho \sigma} -  \nabla_\sigma \nabla_\rho m_{\mu \alpha})\\
        &- \frac{1}{2}m^{\mu \beta} m^{\alpha \gamma} \qty(\nabla_\mu m_{\beta \gamma}(\nabla_\rho m_{\sigma \alpha} + \nabla_\sigma m_{\rho \alpha} - \nabla_\alpha m_{\rho \sigma}) - \nabla_\sigma m_{\beta \gamma}(\nabla_\mu m_{\rho \alpha} + \nabla_\rho m_{\mu \alpha} - \nabla_\alpha m_{\mu \rho}))\\
        &+ \frac{1}{4}m^{\mu \alpha} m^{\tau \beta} (\nabla_\mu m_{\tau \alpha} +\nabla_\tau m_{\mu \alpha} -\nabla_\alpha m_{\mu \tau})(\nabla_\rho m_{\sigma \beta}+\nabla_\sigma m_{\rho \beta} - \nabla_\beta m_{\rho \sigma})\\
        &- \frac{1}{4} m^{\mu \alpha} m^{\tau \beta} (\nabla_\rho m_{\tau \alpha} + \nabla_\tau m_{\rho \alpha} - \nabla_\alpha m_{\tau \rho})(\nabla_\mu m_{\sigma \beta} + \nabla_\sigma m_{\mu \beta} - \nabla_\beta m_{\mu \sigma})
\end{aligned}
\end{equation}

The scalar curvature is defined by a contraction of the Ricci tensor with the inverse metric.
After some simplifications, we end up with 
\begin{equation}
\begin{aligned}
        R &= (m^{\mu \rho} m^{\nu \sigma} - m^{\mu \nu} m^{\rho \sigma})\nabla_\mu \nabla_\nu m_{\rho \sigma}\\
        &+ (m^{\mu \nu} (m^{\rho \sigma} m^{\alpha \beta} - m^{\rho \alpha} m^{\sigma \beta}) + \frac{1}{4} m^{\mu \sigma} (3 m^{\nu \alpha} m^{\rho \beta} - m^{\nu \rho} m^{\alpha \beta}) - \frac{1}{2} m^{\mu \alpha} m^{\nu \beta} m^{\rho \sigma}) \nabla_\mu m_{\nu \rho} \nabla_\sigma m_{\alpha \beta}
\end{aligned}
\end{equation}

In order to further analyze the expression, we have to split the indices into the radial and angular components. 
This simplifies the expression because in Gullstrand-Painlevé gauge, we have that $m_{33} = 1$, $m_{3A} = 0$ and $m_{AB}= r^2 \Omega_{AB} + X_{AB}$.
For the inverse metric we find $m^{33} = 1$, $m^{3A}=0$ and 
\begin{equation}
\begin{aligned}
    m^{AB} &= \frac{1}{\det m}\epsilon^{AC} \epsilon^{BD}(r^2 \Omega_{CD}  + X_{CD})\\
    &= \frac{\det \Omega }{\det m}\qty(\Omega^{AB}\Omega^{CD} - \Omega^{AD} \Omega^{BC})(r^2 \Omega_{CD} + X_{CD})\\
    &=  \frac{\det \Omega}{\det m}(r^2 \Omega^{AB} - X^{AB})\\
    &=  \frac{1}{r^4 - \frac{1}{2} X^{AB}X_{AB}}(r^2 \Omega^{AB} - X^{AB})
\end{aligned}
\end{equation}
where $X^{AB} := \Omega^{AC} \Omega^{BD} X_{CD}$.

In the expression for the scalar curvature, we split the indices into their radial and angular components and find 
\begin{align}
        R &= m^{AB}(- \nabla_3 \nabla_3 m_{AB} + \nabla_3 \nabla_A m_{3 B} + \nabla_A \nabla_3 m_{B 3} - \nabla_A \nabla_B m_{3 3} ) + (m^{A C} m^{B D} - m^{A B} m^{C D})\nabla_A \nabla_B m_{C D} \nonumber \\
        &+ \frac{1}{2}m^{A B} (\nabla_3 m_{3 3} \nabla_3 m_{A B} - 2 \nabla_3 m_{3 3} \nabla_A m_{3 B} + \nabla_A m_{3 3} \nabla_B m_{3 3})\nonumber\\
        &+ (m^{A B} m^{C D} - 2 m^{A C} m^{B D}) \nabla_3 m_{3 A} \nabla_B m_{C D} + \frac{1}{4} (3 m^{A C} m^{B D} - m^{A B} m^{C D}) \nabla_3 m_{A B} \nabla_3 m_{CD}\\
        &+ (m^{CD} m^{AB} - m^{A D} m^{B C}) \nabla_3 m_{A B} \nabla_C m_{3 D} + (m^{B C} m^{A D} - \frac{1}{2} m^{A B} m^{CD}) \nabla_A m_{3 3} \nabla_B m_{C D}\nonumber\\
        &+ (\frac{3}{2} m^{A C} m^{B D} - \frac{1}{2} m^{A D}m^{B C} - m^{A B}  m^{C D})\nabla_A m_{3 B} \nabla_C m_{3 D}\nonumber\\
        &+ (m^{A B} (m^{C D} m^{EF} - m^{C E} m^{D F}) + \frac{1}{4} m^{A D} (3 m^{B E} m^{C F} - m^{B C} m^{E F}) - \frac{1}{2} m^{A E} m^{B F} m^{C D}) \nabla_A m_{B C} \nabla_D m_{E F}\nonumber
\end{align}

Then, we introduce the unique, torsion-free covariant derivative $D_A$ associated to the metric on the sphere $\Omega_{AB}$. 
The full covariant derivative is then split into partial derivatives in the radial direction $\partial_r$ and covariant derivatives $D_A$ in the angular directions. 
The spherically symmetric Christoffel symbols used in the definition of $\nabla$ read  
\begin{equation}   
    \overline \Gamma^3_{AB} = - r \Omega_{AB}, \quad \overline \Gamma^A_{3B} = r^{-1} \delta^A_B
\end{equation}
Then, the first derivatives of the metric $m$ are given by
\begin{align}
    \begin{split}
    \nabla_3 m_{33} = \nabla_3 m_{3A} &= \nabla_A m_{33} = 0\\
    \nabla_3 m_{AB} = (\partial_r - 2 r^{-1}) X_{AB}, \quad \nabla_A m_{3B} &= - r^{-1} X_{AB}, \quad \nabla_A m_{BC} = D_A X_{BC}
    \end{split}
\end{align}
The second derivatives necessary for the computation are
\begin{align}
    \nabla_3 \nabla_3 m_{AB} &= (\partial_r - 2 r^{-1})(\partial_r - 2 r^{-1}) X_{AB}\nonumber\\
    \nabla_3 \nabla_A m_{3B} &= (\partial_r - 2 r^{-1})(- r^{-1} X_{AB})\nonumber\\
    \nabla_A \nabla_3 m_{B3} &= - r^{-1}(\partial_r - 3 r^{-1}) X_{AB}\\
    \nabla_A \nabla_B m_{33} &= 2 r^{-2} X_{AB}\nonumber\\
    \nabla_A \nabla_B m_{CD} &= D_A D_B X_{CD} + r \Omega_{AB} (\partial_r - 2 r^{-1} ) X_{CD} - 2 \Omega_{A(C}X_{D) B}\nonumber
\end{align}
Inserting the expansion of the covariant derivatives, the Ricci scalar becomes
\begin{align}
\label{eq:RicciCovDsInserted}
        R &= m^{AB}(- \partial_r^2 X_{AB} + 2 r^{-1}\partial_r X_{AB} - 2 r^{-2} X_{AB})\\
        &+ (m^{A C} m^{B D} - m^{A B} m^{C D})(D_A D_B X_{CD} + r \Omega_{AB} (\partial_r - 2 r^{-1} ) X_{CD} - 2 \Omega_{A(C}X_{D) B})\nonumber\\
        &+ \frac{1}{4} (3 m^{A C} m^{B D} - m^{A B} m^{C D}) \partial_3 X_{AB} \partial_3 X_{CD} - 2 r^{-1} m^{AC} m^{BD} \partial_3 X_{AB} X_{CD} + 2 r^{-2} m^{A C} m^{B D} X_{AB} X_{CD}\nonumber\\
        &+ (m^{A B} (m^{C D} m^{EF} - m^{C E} m^{D F}) + \frac{1}{4} m^{A D} (3 m^{B E} m^{C F} - m^{B C} m^{E F}) - \frac{1}{2} m^{A E} m^{B F} m^{C D}) D_A X_{B C} D_D X_{E F}\nonumber
\end{align}

Another simplification occurs when we use a special property of two dimensional metrics.
In two dimensions, we have for any metric $g_{AB}$ that
\begin{equation}
    \epsilon^{AB}\epsilon^{CD} = \det g \qty(g^{AC} g^{BD} - g^{AD} g^{BC})
\end{equation}
This can be used to relate expressions which are written in terms of the full metric $m_{AB}$ to expressions involving the background metric $\Omega_{AB}$:
\begin{equation}
    m^{AC}m^{BD} - m^{AB} m^{CD} = \frac{1}{\det m} \epsilon^{AD}\epsilon^{BC} = \frac{\det \Omega}{\det m}\qty(\Omega^{AC} \Omega^{BD} - \Omega^{AB} \Omega^{CD})
    \label{eq:IdentityReplmToOmega}
\end{equation}
The above identity allows us to rewrite combinations of inverses of the full two dimensional metric $m^{AB}$ in terms of the spherically symmetric metric $\Omega^{AB}$. 
Using the identity and the trace-freeness of $X$, the entire second line in equation \eqref{eq:RicciCovDsInserted} reduces to 
\begin{equation}
    \frac{\det \Omega}{\det m} \Omega^{AC} \Omega^{BD} D_A D_B X_{CD}
\end{equation}
The last line in equation \eqref{eq:RicciCovDsInserted} can also be rewritten and we obtain
\begin{equation}
    \begin{aligned}
        &m^{AB}(m^{CD} m^{EF} - m^{CE} m^{DF}) + \frac{1}{2} m^{BE}(m^{AD} m^{CF} - m^{AF} m^{CD}) + \frac{1}{4} m^{AD}(m^{BE} m^{CF} - m^{BC} m^{EF})\\
        &= \qty(\frac{\det \Omega}{\det m})^2\Big[(r^2 \Omega^{AB} - X^{AB})(\Omega^{CD} \Omega^{EF} - \Omega^{CE} \Omega^{DF}) + \frac{1}{2}(r^2\Omega^{BE} - X^{BE})(\Omega^{AD} \Omega^{CF} - \Omega^{AF} \Omega^{CD})\\
        &~~~~~~~~~~~~~~~~~~+ \frac{1}{4}(r^2 \Omega^{AD}- X^{AD}) (\Omega^{BE} \Omega^{CF} - \Omega^{BC} \Omega^{EF})\Big]
    \end{aligned}
\end{equation}
Then contracting with $D_A X_{BC} D_D X_{EF}$, we have
\begin{equation}
    \begin{aligned}
        &\qty(\frac{\det \Omega}{\det m})^2\Big[- (r^2 \Omega^{AB} - X^{AB})\Omega^{CE} \Omega^{DF} + \frac{1}{2}(r^2\Omega^{BE} - X^{BE})(\Omega^{AD} \Omega^{CF} - \Omega^{AF} \Omega^{CD})\\
        &~~~~~~~~~~~~~~~~+ \frac{1}{4}(r^2 \Omega^{AD}- X^{AD}) \Omega^{BE} \Omega^{CF}\Big] D_A X_{BC} D_D X_{EF}
    \end{aligned}
\end{equation}

Combining all the results above, the Ricci tensor is written as
\begin{equation}
    R = \qty(\frac{\det \Omega}{\det m})^2 \frac{1}{\sqrt{\Omega}}\Big[{}^{(1)}R + {}^{(2)}R + {}^{(3)}R + {}^{(4)}R\Big]
\end{equation}
where the subscript in ${}^{(i)} R$ denotes the order in the perturbations of the contribution. 
The first and second order are given by
\begin{equation}
    {}^{(1)}R = \sqrt{\Omega} r^4 D^A D^B X_{AB}
\end{equation}
\begin{equation}
\begin{aligned}
    {}^{(2)}R &=  \sqrt{\Omega}\Big[ r^4X^{AB}(\partial_r^2  - 4 r^{-1} + 4 r^{-2}) X_{AB} + \frac{3r^4}{4} \partial_r X^{AB} \partial_r X_{AB}\\
    & + r^2 \qty( \frac{3}{4}\Omega^{AD} \Omega^{BE}\Omega^{CF} - \frac{1}{2}\Omega^{AE}\Omega^{BF} \Omega^{CD} - \Omega^{AB}\Omega^{CE} \Omega^{DF}) D_A  X_{BC} D_D X_{EF}\Big]
\end{aligned}
\end{equation}
The mode expansion of the first order contribution is
\begin{equation}
    {}^{(1)}R_{lm} = \sqrt{\frac{(l+2)(l+1)l(l-1)}{2}} r^4 X^e_{lm}
\end{equation}
For the second order contribution, the first line is given by the contraction of the tensor perturbation $X_{AB}$ and is easily expanded into modes. 
For the second line we use the spin-weighted harmonic formalism to handle the covariant derivatives.
We find the following expression for the second line:
\begin{align}
    &\sum_{l_2 m_2 I, l_3 m_3 J} \frac{i^{I+J}}{16}r^2(-1)^I\Big[\eth L^2_{l_2 m_2} \overline \eth L^{-2}_{l_3 m_3} - \overline \eth L^2_{l_2 m_2} \eth L^{-2}_{l_3 m_3} + (-1)^{I+J}\qty(\overline \eth L^{-2}_{l_2 m_2} \eth L^2_{l_3 m_3} - \eth L^{-2}_{l_2 m_2} \overline \eth L^2_{l_3 m_3})\Big] X^I_{l_1 m_1} X^J_{l_2 m_2}\nonumber\\
    &=\sum_{l_2 m_2 I, l_3 m_3 J} \frac{i^{I+J}}{16}r^2(-1)^I\Big[- \sqrt{(l_2+3)(l_2-2)}\sqrt{(l_3+3)(l_3-2)} (L^3_{l_2 m_2} L^{-3}_{l_3 m_3} + (-1)^{I+J} L^{-3}_{l_2 m_2} L^3_{l_3 m_3})\nonumber\\
    &+ \sqrt{(l_2+2)(l_2-1)}\sqrt{(l_3+2)(l_3-1)} (L^1_{l_2 m_2} L^{-1}_{l_3 m_3} + (-1)^{I+J} L^{-1}_{l_2 m_2} L^1_{l_3 m_3})\Big]X^I_{l_1 m_1} X^J_{l_2 m_2}
\end{align}
Decomposing into harmonics, the second order perturbation becomes
\begin{align}
\label{eq:2ndRicci}
    {}^{(2)}R_{lm} &= \sum_{l_2 m_2 I, l_3 m_3 J} \Big[C^{l m 0}_{l_2 m_2 2, l_3 m_3 -2} q_{IJ} \qty(X^I_{l_2 m_2}(r^4 \partial_r^2 - 4 r^3\partial_r + 4 r^2 )X^J_{l_3 m_3} + \frac{3 r^4}{4} \partial_r X^I_{l_2 m_2} \partial_r  X^J_{l_3 m_3})\\
    &+\frac{r^2}{8} \qty(\frac{\lambda_{l_2,2} \lambda_{l_3,2}}{\lambda_{l_2,1} \lambda_{l_3,1}} C^{lm0}_{l_2 m_2 1, l_3 m_3 -1} - \frac{\lambda_{l_2,3} \lambda_{l_3,3}}{\lambda_{l_2,2} \lambda_{l_3,2}}C^{lm0}_{l_2 m_2 3, l_3 m_3 -3}) q_{IJ} X^I_{l_2 m_2} X^J_{l_3 m_3}\Big]\nonumber
\end{align}

At third order we find the contribution
\begin{align}
    {}^{(3)}R &= \sqrt{\Omega} \Big[- \frac{3r^2}{2}\Omega^{AC} X^{BD}\partial_r X_{AB}\ \partial_r X_{CD}  - \frac{1}{2} X^{CD}X_{CD} D^A D^B X_{AB}\\
    & + \Big[X^{AB}\Omega^{CE} \Omega^{DF} - \frac{1}{2}X^{BE} (\Omega^{AD} \Omega^{CF} - \Omega^{AF} \Omega^{CD}) - \frac{1}{4}X^{AD} \Omega^{BE} \Omega^{CF}\Big] D_A X_{BC} D_D X_{EF}\Big]\nonumber
\end{align}
All the contributions are now expanded in terms of the $\eth$-formalism. 
The first term involving the radial partial derivatives vanishes because $\Omega^{AC} \partial_r X_{AB} \partial_r X_{CD}$ is proportional to $\Omega_{BD}$ and $X^{BD}$ is trace free.
Expanding the remaining contributions into spin-weighted spherical harmonics gives
\begin{align}
    {}^{(3)}R = \sqrt{\Omega} \sum_{l_1 m_1 I. l_2 m_2 J, l_3 m_3 K}& X^I_{l_1 m_1} X^J_{l_2 m_2} X^K_{l_3 m_3}  \frac{i^{I+J+K}}{4\sqrt{2}}\times\nonumber\\
    &\times\Big[- \frac{1}{2}((-1)^I L_{l_1 m_1}^2 L_{l_2 m_2}^{-2} + (-1)^J L_{l_1 m_1}^{-2} L_{l_2 m_2}^{2}) (\eth \eth L^{-2}_{l_3 m_3} + (-1)^K \overline \eth \overline \eth L^{2}_{l_3 m_3})\nonumber\\
    &+ \frac{3}{4}\qty((-1)^J L^{-2}_{l_1 m_1}\eth L^2_{l_2 m_2} \eth L^{-2}_{l_3 m_3} + (-1)^{I+K} L^{2}_{l_1 m_1}\overline \eth L^{-2}_{l_2 m_2} \overline \eth L^2_{l_3 m_3})\\
    &- \frac{1}{4}\qty((-1)^K L^{-2}_{l_1 m_1}\eth L^{-2}_{l_2m_2}\eth L^{2}_{l_3 m_3} + (-1)^{I+J} L^2_{l_1 m_1} \overline \eth L^2_{l_2m_2}\overline\eth L^{-2}_{l_3 m_3})\nonumber\\
    &+ \frac{1}{2}\qty((-1)^{J+K}L^{-2}_{l_1 m_1} \overline \eth L^2_{l_2 m_2} \overline \eth L^2_{l_3 m_3}+ (-1)^I L^2_{l_1 m_1} \eth L^{-2}_{l_2 m_2} \eth L^{-2}_{l_3 m_3})\Big]\nonumber
\end{align}
In the fourth line, we can exploit the total symmetry of the three $X$'s to make the replacement $(l_2 m_2 J) \leftrightarrow (l_3 m_3 K)$.
We obtain 
\begin{align}
    {}^{(3)}R = \sqrt{\Omega} \sum_{l_1 m_1 I. l_2 m_2 J, l_3 m_3 K}& X^I_{l_1 m_1} X^J_{l_2 m_2} X^K_{l_3 m_3}  \frac{i^{I+J+K}}{4\sqrt{2}}\times\nonumber\\
    &\times\Big[- \frac{1}{2}((-1)^I L_{l_1 m_1}^2 L_{l_2 m_2}^{-2} + (-1)^J L_{l_1 m_1}^{-2} L_{l_2 m_2}^{2}) (\eth \eth L^{-2}_{l_3 m_3} + (-1)^K \overline \eth \overline \eth L^{2}_{l_3 m_3})\nonumber\\
    &+ \frac{1}{2}\qty((-1)^J L^{-2}_{l_1 m_1}\eth L^2_{l_2 m_2} \eth L^{-2}_{l_3 m_3} + (-1)^{I+K} L^{2}_{l_1 m_1}\overline \eth L^{-2}_{l_2 m_2} \overline \eth L^2_{l_3 m_3})\\
    &+ \frac{1}{2}\qty((-1)^{J+K}L^{-2}_{l_1 m_1} \overline \eth L^2_{l_2 m_2} \overline \eth L^2_{l_3 m_3}+ (-1)^I L^2_{l_1 m_1} \eth L^{-2}_{l_2 m_2} \eth L^{-2}_{l_3 m_3})\Big]\nonumber
\end{align}
Evaluating the $\eth$ terms, we find
\begin{align}
    {}^{(3)}R &= \sqrt{\Omega} \sum_{l_1 m_1 I, l_2 m_2 J, l_3 m_3 K} X^I_{l_1 m_1} X^J_{l_2 m_2} X^K_{l_3 m_3} \frac{i^{I+J+K}}{4\sqrt{2}}\times\nonumber\\
    &\times\Big[- ((-1)^I L_{l_1 m_1}^2 L_{l_2 m_2}^{-2} + (-1)^J L_{l_1 m_1}^{-2} L_{l_2 m_2}^{2}) \lambda_{l_3,2} (1 + (-1)^K ) L^{0}_{l_3 m_3}\nonumber\\
    &+ \sqrt{(l_2+3)(l_2-2)} \sqrt{(l_3 + 2)(l_3 -1)}\qty((-1)^J L^{-2}_{l_1 m_1} L^3_{l_2 m_2} L^{-1}_{l_3 m_3} + (-1)^{I+K} L^{2}_{l_1 m_1} L^{-3}_{l_2 m_2} L^1_{l_3 m_3})\\
    &+ \sqrt{(l_2+2)(l_2-1)}\sqrt{(l_3+2)(l_3-1)}\qty((-1)^{J+K}L^{-2}_{l_1 m_1} L^1_{l_2 m_2} L^1_{l_3 m_3}+ (-1)^I L^2_{l_1 m_1} L^{-1}_{l_2 m_2} L^{-1}_{l_3 m_3})\Big]\nonumber
\end{align}
We couple the harmonics in the following way:
\begin{align}
    L^2_{l_1 m_1} L^{-2}_{l_2 m_2} &= \sum_{l' m'}C^{l' m'0}_{l_1 m_1 2, l_2 m_2 -2} L_{l' m'}\nonumber\\
    L^{2}_{l_1 m_1} L^{-3}_{l_2 m_2} &= - \sum_{l' m'}C^{l' m' 1}_{l_1 m_1 2, l_2 m_2 -3} L^{-1}_{l'm'}\\
    L^{-2}_{l_1 m_1} L^{1}_{l_2 m_2} &= - \sum_{l' m'} C^{l'm' 1}_{l_1 m_1 -2, l_2 m_2 1} L^{-1}_{l' m'}\nonumber
\end{align}
Then, we have
\begin{align}
    {}^{(3)}R_{lm} &= \sum_{\{l_i m_i\} I,J, K, l' m'} X^I_{l_1 m_1} X^J_{l_2 m_2} X^K_{l_3 m_3} \frac{i^{I+J+K}}{4\sqrt{2}}\times\\
    &\times\Big[- C^{l'm'0}_{l_1 m_1 2, l_2 m_2 -2} C^{lm0}_{l' m' 0, l_3 m_3 0} ((-1)^I + (-1)^J (-1)^{l' + l_1 + l_2}) \lambda_{l_3,2} (1 + (-1)^K )\nonumber\\
    &+ \sqrt{(l_2+3)(l_2-2)} \sqrt{(l_3 + 2)(l_3 -1)} C^{l' m' 1}_{l_1 m_1 2, l_2 m_2 -3} C^{lm0}_{l' m' - 1, l_3 m_3 1} \qty((-1)^J  (-1)^{l + l_1 + l_2 + l_3} + (-1)^{I+K})\nonumber\\
    &+ \sqrt{(l_2+2)(l_2-1)}\sqrt{(l_3+2)(l_3-1)}C^{l' m' 1}_{l_1 m_1 -2, l_2 m_2 1} C^{lm0}_{l' m' -1, l_3 m_3 1}  \qty((-1)^I (-1)^{l + l_1 + l_2 + l_3}  + (-1)^{J+K})\Big]\nonumber
\end{align}
In the third line, we use the identity
\begin{equation}
    \sqrt{(l_2+3)(l_2-2)} C^{l' m' 1}_{l_1 m_1 2, l_2 m_2 -3} = - \sqrt{l'(l' + 1)} C^{l' m' 0}_{l_1 m_1 2, l_2 m_2 -2} - \sqrt{(l_1 +2)(l_1 - 1)}C^{l'm'1}_{l_1 m_1 1,l_2 m_2 -2}
\end{equation}
and otain
\begin{align}
    {}^{(3)}R_{lm} &= \sum_{l_1 m_1 I, l_2 m_2 J, l_3 m_3 K, l' m'} X^I_{l_1 m_1} X^J_{l_2 m_2} X^K_{l_3 m_3} \frac{i^{I+J+K}}{4\sqrt{2}}\times\\
    &\times\Big[- C^{l'm'0}_{l_1 m_1 2, l_2 m_2 -2} C^{lm0}_{l' m' 0, l_3 m_3 0} ((-1)^I + (-1)^J (-1)^{l' + l_1 + l_2}) \lambda_{l_3,2} (1 + (-1)^K )\nonumber\\
    &- \sqrt{(l_3 + 2)(l_3 -1)} \sqrt{l'(l'+1)}C^{l' m'0}_{l_1 m_1 2, l_2 m_2 -2} C^{lm0}_{l' m' - 1, l_3 m_3 1}  \qty((-1)^J (-1)^{ l + l_1 + l_2 + l_3}+ (-1)^{I+K})\Big]\nonumber
\end{align}
In the computation some cancellation occurred when exchanging the indices $(l_1 m_1 I) \leftrightarrow (l_2 m_2 J)$ due to the symmetry of the $X$-terms.

Finally, at fourth order we have
\begin{align}
    \frac{{}^{(4)}R}{\sqrt{\Omega}} &= \frac{1}{4}(3 X^{AC} X^{BD} - X^{AB} X^{CD}) \partial_r X_{AB} \partial_r X_{CD} - \frac{1}{2}X^{CD}X_{CD} X^{AB} \partial_r^2 X_{AB}\\
    &= \frac{3}{4}\qty(\partial_r (X_{AB} X^{AC})-X_{AB}\partial_r X^{AC})X^{BD}\partial_r X_{CD} - \frac{1}{4} \qty(X^{AB} \partial_r X_{AB})^2 - \frac{1}{2} X^{CD} X_{CD} X^{AB} \partial_r^2 X_{AB}\nonumber\\
    &=\frac{1}{2}(X^{AB} \partial_r X_{AB})^2 - \frac{3}{8} X^{AB}X_{AB} \partial_r X^{CD}\partial_r X_{CD}  - \frac{1}{2} X^{CD} X_{CD} X^{AB} \partial_r^2 X_{AB}\nonumber
\end{align}
where we used the identity $X_{AB}X^{AC} = \frac{1}{2} X^{AD} X_{AD} \delta^C_B$.
Expanding into spherical harmonics, we find
\begin{align}
    {}^{(4)}R_{lm} &= \sum_{l_i m_i I_i} \sum_{l' m', l'' m''} \frac{1}{2} q_{I_1 I_2} q_{I_3 I_4} C^{lm0}_{l' m' 0,l''m''0}C^{l' m' 0}_{l_1 m_1 2, l_2 m_2 -2} C^{l'' m'' 0}_{l_3 m_3 2, l_4 m_4 -2}  \times\\
    &\times \qty[X^{I_1}_{l_1 m_1}\partial_r X^{I_2}_{l_2 m_2}X^{I_3}_{l_3 m_3} \partial_r X^{I_4}_{l_4 m_4} - \frac{3}{4} X^{I_1}_{l_1 m_1}X^{I_2}_{l_2 m_2}\partial_rX^{I_3}_{l_3 m_3} \partial_r X^{I_4}_{l_4 m_4} - X^{I_1}_{l_1 m_1}X^{I_2}_{l_2 m_2} X^{I_3}_{l_3 m_3} \partial_r^2 X^{I_4}_{l_4 m_4}]\nonumber
\end{align}

Therefore, we successfully computed the expansion of the Ricci tensor to all orders.
However the Hamiltonian constraint $V_0$ depends non-polynomially on the Ricci tensor and the momentum contributions due to the factors of $\sqrt{\det m}$.
The momentum contributions come with a factor of $\qty(\det m)^{-\frac{1}{2}}$ and the Ricci scalar contributions come with a non-polynomial factor of $\qty(\det m)^{-\frac{3}{2}}$.
We observe, that multiplying $V_0$ by $\qty(\det m)^{\frac{3}{2} + n}$ with $n \in \mathbb{N}_0$, the non-polynomial contributions will turn into a polynomial 
expression in the non-symmetric variables.
For the minimal choice $n=0$, the resulting polynomial will be of degree $6$.
Note that the procedure for obtaining a polynomial version of the Hamiltonian constraint works in general and not only for the specific case discussed here.
In $3+1$ dimensions, the minimal degree of the polynomial version of the Hamiltonian constraint is $10$ because the Ricci scalar involves three 
factors of the inverse metric times $\partial^2 m, (\partial m)^2$ hence to make it polynomial one needs to multiply the constraint 
by $\det(m)^{5/2}$ and $\det(m) m^{-1}$ is a polynomial of order two. Then the kinetic term becomes $\det(m)^2 K$ which is of order $10$ 
while the curvature term becomes $\det(m)^3 R$ which is of order $8$. 
The identity \eqref{eq:IdentityReplmToOmega} is only reducing that order from $10$ to $6$ only for our our specific gauge condition which requires 
to multiply by $\det(m)^{3/2}$ only and because $\det(m)$ itself is only of order $2$ so that $\det(m) K$ which is of order $6$ 
while $\det(m)^3 R$ which is of order $4$. This is a tremendous simplification as the number of algebraically independent 
terms grows in general factorially with the 
polynomial order.

\subsubsection{Mode Expansion of the Constraints}

In order that the perturbative expansion of the constraint terminates, we define an equivalent Hamiltonian constraint with modified density weight. 
As discussed in the previous section, the lowest degree of $6$ is obtained by using
\begin{equation}
    \tilde V^0 = \frac{\sqrt{\det m}^{3}}{\sqrt{\det \Omega}^{3}} V_0 = \frac{\det m}{\det \Omega} K - \qty (\frac{\det m}{\det \Omega})^2 \sqrt \Omega R
\end{equation}
In the section we already computed the mode expansion of the contributions $K$ and $R$. 
For the multiplication by $\det m$, we use that
\begin{equation}
    \qty(\frac{\det m}{\det\Omega})_{lm} = r^4 \delta_{l,0} - \frac{1}{2} (X^2)_{lm}\,,
\end{equation}
where
\begin{equation}
    (X^2)_{l m} := \sum_{l_1 m_1 I, l_2 m_2 J} C^{l m 0}_{l_1 m_1 2, l_2 m_2 -2} q_{IJ} X^I_{l_1 m_1}X^J_{l_2 m_2}
\end{equation}
Then, we obtain
\begin{align}
    \begin{split}
    {}^{(0)}\tilde V^0 &= r^4 {}^{(0)}K_{lm}\\
    {}^{(1)}\tilde V^0 &= r^4 {}^{(1)}K_{lm} - {}^{(1)}R_{lm}\\
    {}^{(2)}\tilde V^0 &= r^4 {}^{(2)}K_{lm} - \frac{1}{2} \sum_{l_1m_1,l_2 m_2}C^{lm0}_{l_1m_10,l_2 m_20} (X^2)_{l_1m_1} {}^{(0)}K_{l_2 m_2} - {}^{(2)}R_{lm}\\
    {}^{(3)}\tilde V^0 &= r^4 {}^{(3)}K_{lm} - \frac{1}{2}\sum_{l_1m_1,l_2 m_2}C^{lm0}_{l_1m_10,l_2 m_20} (X^2)_{l_1m_1} {}^{(1)}K_{l_2m_2} - {}^{(3)}R_{lm}\\
    {}^{(4)}\tilde V^0 &= r^4 {}^{(4)}K_{lm}  - \frac{1}{2}\sum_{l_1m_1,l_2 m_2}C^{lm0}_{l_1m_10,l_2 m_20} (X^2)_{l_1m_1} {}^{(2)}K_{l_2m_2} - {}^{(4)}R_{lm}\\
    {}^{(5)}\tilde V^0 &=  - \frac{1}{2}\sum_{l_1m_1,l_2 m_2}C^{lm0}_{l_1m_10,l_2 m_20} (X^2)_{l_1m_1} {}^{(3)}K_{l_2m_2}\\
    {}^{(6)}\tilde V^0 &=  - \frac{1}{2}\sum_{l_1m_1,l_2 m_2}C^{lm0}_{l_1m_10,l_2 m_20} (X^2)_{l_1m_1} {}^{(4)}K_{l_2m_2}
    \end{split}
\end{align}

This gives the full expansion of the modified Hamiltonian constraint to all orders. 
We now explicitly decompose the above expressions into the spherically symmetric and non-symmetric contributions up to third order. 
At zeroth order we only have a symmetric contribution:
\begin{equation}
    {}^{(0)}C_v = 2\pi r^4 (p_v^2 - 2 r^2 p_v p_h)
\end{equation}
At first order, we only have the non-symmetric constraint
\begin{equation}
    {}^{(1)}Z^v_{lm} = r^4 \qty[(p_v -r^2 p_h) y_v^{lm} - r^2 p_v y_h^{lm} - \sqrt{\frac{(l+2)(l+1)l(l-1)}{2}} X^e_{lm}] 
\end{equation}
The symmetric second order constraints are
\begin{align}
\begin{split}
    {}^{(2)}C_v &= r^4\Big[\sum_I\Big(r^4 Y_I \cdot Y_I - (p_v - r^2 p_h) X^I \cdot Y_I - \frac{3}{4} \partial_r X^I \cdot \partial_r X_I  - X^I \cdot (\partial_r^2 - 4 r^{-1} \partial_r + \frac{7}{2} r^{-2} ) X_I\\
    &- \frac{1}{4} (r^{-4} p_v^2 - 2 r^{-2} p_v p_h - p_h^2) X^I \cdot X_I + \frac{r^2}{2}  y_I \cdot y_I \Big)  +\frac{1}{2} y_v \cdot y_v - r^2 y_v \cdot y_h \Big]
\end{split}
\end{align}
For two functions $f$ and $g$ with expansion coefficients $f_{lm}$ and $g_{lm}$ we defined $f\cdot g := \sum_{l m} f_{lm} g_{lm}$.
The non-symmetric second order constraints are
\begin{align}
    {}^{(2)}Z^v_{lm} &= r^4 \sum_{l_1 m_1, l_2 m_2}\Big[C^{lm0}_{l_1 m_1 0, l_2 m_2 0} \qty(\frac{1}{2} y_v^{l_1 m_1} y_v^{l_2 m_2}  - r^2 y_v^{l_1 m_1} y_h^{l_2 m_2}) + \sum_{IJ}\Big(- \frac{1}{2} r^2 C^{lm0}_{l_1 m_1 1, l_2 m_2 -1} q^{IJ} y_I^{l_1 m_1} y_J^{l_2 m_2}\nonumber\\
    &+ C^{lm0}_{l_1 m_1 2, l_2 m_2 -2} q^{IJ}\qty(r^4 Y_I^{l_1 m_1} Y_J^{l_2 m_2} - (p_v - r^2 p_h) X_I^{l_1 m_1}Y_J^{l_2 m_2} - \frac{1}{4 r^4} \qty( p_v^2 - 2 r^2 p_v p_h - r^4p_h^2) X_I^{l_1 m_1} X_J^{l_2 m_2}) \Big)\Big]\nonumber\\
    &- \sum_{l_2 m_2 I, l_3 m_3 J} \Big[C^{l m 0}_{l_2 m_2 2, l_3 m_3 -2} q_{IJ} \qty(X^I_{l_2 m_2}(r^4 \partial_r^2 - 4 r^3\partial_r + 4 r^2 )X^J_{l_3 m_3} + \frac{3 r^4}{4} \partial_r X^I_{l_2 m_2} \partial_r  X^J_{l_3 m_3})\\
    &+\frac{r^2}{8} \qty(\frac{\lambda_{l_2,2} \lambda_{l_3,2}}{\lambda_{l_2,1} \lambda_{l_3,1}} C^{lm0}_{l_2 m_2 1, l_3 m_3 -1} - \frac{\lambda_{l_2,3} \lambda_{l_3,3}}{\lambda_{l_2,2} \lambda_{l_3,2}}C^{lm0}_{l_2 m_2 3, l_3 m_3 -3}) q_{IJ} X^I_{l_2 m_2} X^J_{l_3 m_3}\Big]\nonumber
\end{align}
At third order, we have the symmetric constraint
\begin{align}
    {}^{(3)}C_v &= r^4 \sum_{\{l_i,m_i\},IJK}\Big[\frac{i^{I+J+K}}{4\sqrt{2}} C^{l_1 m_1 2}_{l_2 m_2 -1,l_3 m_3 -1} X^I_{l_1 m_1} y^J_{l_2 m_2}y^K_{l_3 m_3} (-1)^I \qty(1 + (-1)^{I+J+K} (-1)^{l_1 +l_2 + l_3}) \nonumber\\
    &+ C^{l_1 m_1 0}_{l_2 m_2 2, l_3 m_3 -2} q^{IJ}\Big(- (y_v^{l_1 m_1} - r^2 y_h^{l_1 m_1}) X_I^{l_2 m_2}Y_J^{l_3 m_3}\Big)\\
    &+ \frac{1}{2}C^{l_1 m_1 0}_{l_2 m_2 2, l_3 m_3 -2} \Big((p_h + r^{-2} p_v) y_h^{l_1 m_1}  - r^{-4}(p_v - r^{2} p_h) y_v^{l_1 m_1}\Big) q^{IJ} X_I^{l_2 m_2} X_J^{l_3 m_3}\Big]\nonumber\\
    & + \sum_{\{l_i m_i\}, IJK} X^I_{l_1 m_1} X^J_{l_2 m_2} X^K_{l_3 m_3} \frac{i^{I+J+K}}{4\sqrt{2}}C^{l_3 m_3 0}_{l_1 m_1 2, l_2 m_2 -2} \lambda_{l_3,2} \Big[ ((-1)^I + (-1)^J (-1)^{l_1 + l_2 + l_3}) (1 + (-1)^K )\nonumber\\
    &~~~~~~~~~- \qty((-1)^J (-1)^{l_1 + l_2 + l_3}+ (-1)^{I+K})\Big]\nonumber
    \label{eq:SolCv3}
\end{align}

\section{Symplectic Reduction I -- Solution of the Constraints}
\label{sec:SolConstraints}

The next two sections apply the symplectic reduction formalism of \cite{I} to the terms of third order .
In the first step, we need to solve the Hamiltonian and diffeomorphism constraints for the gauge momenta $p,y$.
In the second step, we compute the physical (reduced) Hamiltonian for the physical degrees of freedom $X,Y$. 
In \cite{II}, we explicitly constructed the solution of the constraints up to second order for the pure gravity case. 
In the following, we go beyond this and continue with the solution to third order. 
For completeness, we also review the solution of the constraints up to second order.

In  the computation we assume a perturbative expansion of $p$ and $y$ of the form
\begin{align}
    p = p^{(0)} + p^{(1)} + p^{(2)} + p^{(3)} + O(p^4), \quad y = y^{(0)} + y^{(1)} + y^{(2)} + y^{(3)} + O(y^4)
\end{align}
Then, we iterativley solve the symmetric constraints $C$ for $p$ and the non-symmetric constraints $Z$ for $y$. 
Notice that by construction, there is no first order symmetric and no zeroth order non-symmetric constraint.
Thus, we have that $p^{(1)}$ and $y^{(0)}$ vanish.

The construction of the solution to the constraints is performed as follows:
We first solve the zeroth order symmetric constraints ${}^{(0)} C$ and obtain $p^{(0)}$. 
Then, using this result, we determine $y^{(1)}$ from the equation ${}^{(1)}Z=0$.
After that, we use the second order constraint equations and the zeroth and first order solutions for $y$ and $p$ to find the second order solutions $p^{(2)}$ and $y^{(2)}$.
The procedure then continues iteratively to all higher orders.
In \cite{I}, it was proven that this method gives a solution to the constraints to any desired order.

In the following, we perform the computation to third order.
However, we proceed in a different order:
First, we consider only the symmetric constraints $C_{v/h}$ and then compute their solution up to third order for $p_{v/h}$.
For this computation, we assume that we already found the solution for the non-symmetric momenta $y$.
Then, we solve the non-symmetric constraints $Z_{v/h/e/o}$ for the momenta $y_{v/h/e/o}$ up to second order.
This result can then be plugged back into the solution of the symmetric constraints to find their full solution.

\subsection{Solution of the Symmetric Constraints}

There are two symmetric constraints, the symmetric radial diffeomorphism constraint $C_h$ and the symmetric Hamiltonian constraint $C_v$. 
The solution of the zeroth order constraints is straight forward and we obtain
\begin{equation}
    p_v^{(0)} = 2 \sqrt{r r_s}, \quad p_h^{(0)} = \sqrt{\frac{r_s}{r^3}}
\end{equation}
where $r_s=2M$ is an integration constant.
At second order, the spherically symmetric diffeomorphism constraint implies that
\begin{equation}
    p_h^{(2)} = \frac{1}{r} \partial_r p_v^{(2)} - \frac{1}{8\pi} \sum_I Y_I \cdot \partial_r X^I
\end{equation}
We insert this in the second order, symmetric Hamiltonian constraint and obtain
\begin{equation}
    2 \pi r^4 \qty( 2 p_v^{(0)} p_v^{(2)} - 2 r p_v^{(0)} \partial_r p_v^{(2)} + 2 r^2 p_v^{(0)} \frac{1}{8\pi} \sum_I Y_I \cdot \partial_r X^I - 2 r^2 p_v^{(2)} p_h^{(0)}) + {}^{(2)}C_v(p = p^{(0)}, y = y^{(1)}) = 0
\end{equation}
In the equation, we assumed the solution of the non-symmetric first-order constraints has been achieved.
The result needs to be inserted into the expression for ${}^{(2)}C_v$.
Dividing this equation by $4\pi r^6$ and rearranging some terms we find
\begin{equation}
    \frac{1}{r} p_v^{(0)} \partial_r p_v^{(2)} - \frac{1}{r^2} p_v^{(0)} p_v^{(2)} + p_h^{(0)} p_v^{(2)} = \frac{p_v^{(0)}}{8\pi}\sum_I Y_I \cdot \partial_r X^I +  \frac{1}{4\pi r^6} {}^{(2)}C_v(p = p^{(0)}, y = y^{(1)})
\end{equation}
From the zeroth order radial diffeomorphism constraint we have that $p_h^{(0)} = r^{-1} \partial_r p_v^{(0)}$.
This identity allows us to rewrite the left-hand side as a total radial derivative
\begin{equation}
    \partial_r \qty(\frac{1}{r} p_v^{(0)} p_v^{(2)}) = \frac{p_v^{(0)}}{8\pi}\sum_I Y_I \cdot \partial_r X^I +  \frac{1}{4\pi r^6} {}^{(2)}C_v(p = p^{(0)}, y = y^{(1)})
\end{equation}
Integrating the equation, we obtain the solution
\begin{equation}
    p_v^{(2)} = \frac{r}{4 \pi p_v^{(0)}} \int \dd r \qty(\frac{1}{2} p_v^{(0)} \sum_I Y_I \cdot \partial_r X^I + \frac{1}{r^6}{}^{(2)}C_v(p = p^{(0)}, y = y^{(1)}))
\end{equation}
The second order, symmetric Hamiltonian constraint gives a solution for $p_h^{(2)}$:
\begin{equation}
\begin{aligned}
        p_h^{(2)} &= \frac{1}{2r^2} p_v^{(2)} + \frac{1}{4 \pi r^6 p_v^{(0)}} {}^{(2)}C_v(p = p^{(0)}, y = y^{(1)})
\end{aligned}
\end{equation}

At third order we have the following equations from the radial diffeomorphism and the Hamiltonian constraint:
\begin{align}
    8 \pi(r p_h^{(3)} - \partial_r p_v^{(3)} ) =& 0\\
    4\pi r^4 \qty(p_v^{(3)}( r^2 p_h^{(0)} - p_v^{(0)} ) + r^2 p_h^{(3)}p_v^{(0)})=& A^{(2)} +  {}^{(3)}C_v(p=p^{(0)}, y = y^{(1)})
\end{align}
The term $A^{(2)}$ comes from the second order Hamiltonian constraint.
We have to insert the first and second order solution $y = y^{(1)} + y^{(2)}$ into the  Hamiltonian constraint.
The second order contribution $y^{(2)}$ leads to an increase of the perturbative order of $C_v^{(2)}$.
This contribution is given by
\begin{equation}
    A^{(2)} = r^4 \qty[y_v^{(2)}\cdot \qty(y_v^{(1)} - r^2 y_h^{(1)}) - r^2 y_h^{(2)}\cdot y_v^{(1)} + r^2 \sum_I y_I^{(1)}\cdot y_I^{(2)}]
\end{equation}
Due to the absense of any first order symmetric constraints, the solutions $p^{(2)}$ do not play any role in the equations for $p^{(3)}$.
The third order, symmetric, radial diffeomorphism constraint implies
\begin{equation}
    p_h^{(3)} = \frac{1}{r} \partial_r p_v^{(3)}
\end{equation}
We insert this into the third order, symmetric Hamiltonian constraint.
In addition, we use again the relation between the zeroth order momenta $p_h^{(0)} = r^{-1}\partial_r p_v^{(0)}$ and find
\begin{equation}
    \partial_r \qty(\frac{1}{r} p_v^{(0)} p_v^{(3)}) = \frac{1}{4 \pi r^6}\qty( A^{(2)} + {}^{(3)}C_v(p=p^{(0)}))
\end{equation}
A direct integration of this differential equation gives
\begin{equation}
    p_v^{(3)} = \frac{r}{4 \pi p_v^{(0)}}\int \frac{\dd r}{r^6} \qty( A^{(2)} + {}^{(3)}C_v(p = p^{(0)}, y = y^{(1)}))
\end{equation}
Finally, from the third order, symmetric Hamiltonian constraint we obtain the remaining momentum
\begin{equation}
    p_h^{(3)} = \frac{1}{2r^2} p_v^{(3)} + \frac{1}{4\pi r^6 p_v^{(0)}} \qty(A^{(2)} + {}^{(3)}C_v(p = p^{(0)}, y = y^{(1)}))
\end{equation}

\subsection{Solution of the Non-symmetric constraints}

We continue with the solution of the non-symmetric constraints, which we already assumed in the computations of the previous subsection.
As there are no zeroth order, non-symmetric constraints, we start with the first order corrections. 
The solution can be found separately for the odd and even polarity sectors. 

In the odd polarity sector, we only have an angular diffeomorphism constraint ${}^{(1)}Z^o_{lm}$ and it has the solution
\begin{equation}
    [y_o^{(1)}]_{lm} = \frac{\sqrt{2(l+2)(l-1)}}{r^2} \int \dd r \qty(r^2 Y_o^{lm} + \frac{p_h^{(0)}}{2} X^o_{lm})
\end{equation}
For the special case $l=1$, the tensor modes $X,Y$ are undefined and we obtain $[y_o^{(1)}]_{lm} \propto r^{-2}$.
The proportionality constant is an integration constant and related to the angular momentum of a slowly rotating black hole (see \cite{II}).

The even polarity sector is more complex. 
The non-symmetric Hamiltonian constraint implies 
\begin{equation}
    [y_h^{(1)}]_{lm} = \frac{1}{2r^2} [y_v^{(1)}]_{lm} - \sqrt{\frac{(l+2)(l+1)l(l-1)}{2}} \frac{1}{r^2 p_v^{(0)}} X^e_{lm}
\end{equation}
Then, the non-symmetric radial diffeomorphism constraint gives
\begin{equation}
    [y_e^{(1)}]_{lm} = \frac{2}{\sqrt{l(l+1)}}(\partial_r [y_v^{(1)}]_{lm} - r [y_h^{(1)}]_{lm})
\end{equation}
Finally, we obtain a differential equation for $y_v^{(1)}$ using the even polarity, angular diffeomorphism constraint:
\begin{equation}
    \partial_r (r^2 [y_e^{(1)}]_{lm}) + r^2 \sqrt{l(l+1)} [y_h^{(1)}]_{lm} = \sqrt{2(l+2)(l-1)}\qty(r^2 Y_e^{lm} + \frac{p_h^{(0)}}{2} X^e_{lm})
\end{equation}
Inserting the expressions for $y_e^{(1)}$, we find
\begin{equation}
    2 r^2 \partial_r^2 [y_v^{(1)}]_{lm} + 3 r \partial_r [y_v^{(1)}]_{lm} +  \frac{1}{2} (l+2)(l-1) [y_v^{(1)}]_{lm} = s_1(r)
\end{equation}
where
\begin{equation}
    s_1(r) = \sqrt{2(l+2)(l+1)l(l-1)}\Big[(r^2 Y_e^{lm} + \frac{p_h^{(0)}}{2} X^e_{lm}) + (l(l+1)-1) \frac{1}{2 p_v^{(0)}} X^e_{lm} - \frac{r}{p_v^{(0)}} \partial_r X^e_{lm}\Big] 
\end{equation}
The solution of this differential equation is given by an integral which was derived in \cite{II}.
\begin{equation}
    y^{(1)}_v = r^{\alpha_-} \int \dd \tilde r\qty(\frac{2 i \tilde r^{\alpha_+-1/2}}{\sqrt{4l(l+1)-9}} s_1(\tilde r)) - r^{\alpha_+} \int \dd \tilde r\qty(\frac{2 i \tilde r^{\alpha_--1/2}}{\sqrt{4l(l+1)-9}} s_1(\tilde r))
    \label{eq:yv1}
\end{equation}
with
\begin{align}
    \alpha_\pm = \frac{1}{4}\qty(-1 \pm i \sqrt{4 l(l+1) - 9})
\end{align}
For the special case $l=1$, we have that $s_1(r)$ vanishes. Using that $\alpha_\pm |_{l=1} = - \frac{1}{4} \pm \frac{1}{4}$, we find 
\begin{equation}
    [y^{(1)}_v]_{1m} = A_m + B_m r^{-1/2}
\end{equation}
with integration constants $A_m$ and $B_m$.

At second order, the solution for the odd polarity momentum $y_o^{(2)}$ is given by solving the second order, non-symmetric, angular diffeomorphism constraint with odd polarity:
\begin{equation}
    [y_o^{(2)}]_{lm} = \frac{1}{r^2} \int \dd r {}^{(2)}Z^o_{lm}\Big|_{p = p^{(0)},y = y^{(1)}}
\end{equation}

For the second order even polarity momenta, we solve the non-symmetric Hamiltonian constraint and obtain
\begin{equation}
    [y_h^{(2)}]_{lm} = \frac{1}{2r^2} [y_v^{(2)}]_{lm} + \frac{1}{r^6 p_v^{(0)}} {}^{(2)} Z^v_{lm}\Big|_{p = p^{(0)},y = y^{(1)}}
\end{equation}
Next, the radial diffeomorphism constraint implies 
\begin{align}
    [y_e^{(2)}]_{lm} &= \frac{1}{\sqrt{l(l+1)}}\qty( 2 \partial_r [y_v^{(2)}]_{lm} - 2 r [y_h^{(2)}]_{lm} - {}^{(2)}Z^h_{lm}\Big|_{p = p^{(0)},y = y^{(1)}})\\
    &=\frac{1}{\sqrt{l(l+1)}}\qty( 2 \partial_r [y_v^{(2)}]_{lm} - \frac{1}{r} [y_v^{(2)}]_{lm} - \Bigg[\frac{2}{r^5 p_v^{(0)}} {}^{(2)} Z^v_{lm}+ {}^{(2)}Z^h_{lm}\Bigg]_{p = p^{(0)},y = y^{(1)}})
\end{align}
The remaining even polarity, angular diffeomorphism constraint gives a differential equation for $y_v^{(2)}$:
\begin{equation}
    \partial_r (r^2 [y_e^{(2)}]_{lm}) + r^2 \sqrt{l(l+1)} [y_h^{(2)}]_{lm} = {}^{(2)}Z^e_{lm}\Big|_{p = p^{(0)},y = y^{(1)}}
\end{equation}
Inserting the expressions found above, we have
\begin{equation}
     2 r^2 \partial_r^2 [y_v^{(2)}]_{lm} + 3 r \partial_r [y_v^{(2)}]_{lm} + \frac{1}{2} (l+2)(l-1) [y_v^{(2)}]_{lm}= s_2(r)
\end{equation}
where
\begin{equation}
    s_2(r) = \Big[\partial_r (r^2 {}^{(2)}Z^h_{lm}) + \sqrt{l(l+1)} {}^{(2)}Z^e_{lm} + \frac{l(l+1)-1}{r^4 p_v^{(0)}}{}^{(2)} Z^v_{lm}  - \frac{2r}{p_v^{(0)}} \partial_r \qty(r^{-4}{}^{(2)}Z^v_{lm}) \Big]_{p = p^{(0)},y = y^{(1)}}
\end{equation}
This is precisely the same differential equation that appeared in the solution of the first order non-symmetric constraints.
Therefore, the solution is given by
\begin{equation}
    y^{(2)}_v = r^{\alpha_-} \int \dd \tilde r\qty(\frac{2 i \tilde r^{\alpha_+-1/2}}{\sqrt{4l(l+1)-9}} s_2(\tilde r)) - r^{\alpha_+} \int \dd \tilde r\qty(\frac{2 i \tilde r^{\alpha_--1/2}}{\sqrt{4l(l+1)-9}} s_2(\tilde r))
\end{equation}

For the second order solutions for the momenta, we do not have to take special care of the case $l=1$.
Due to the presence of the Clebsch-Gordan coefficients, the variables $X,Y$ with $l\geq2$ couple to contribute to terms with $l=1$.

As a consistency check, we computed the asymptotic behaviour of the second order non-symmetric constraints based on the decay conditions of $X,Y$.
Then, the solutions above imply the asymptotic behaviour of the variables $y^{(2)}$ and this behaviour is compatible with the assumed behaviour of the $y$ variables.

\section{Symplectic Reduction II -- The Physical Hamiltonian}
\label{sec:PhysHam}

After all the preparatory steps, we now derive the physical Hamiltonian up to third order in the perturbations. 
In \cite{I}, an implicit formula was derived which gives the physical Hamiltonian for one asymptotic end in terms of a boundary term at infinity of the 
spherically symmetric momentum $p_v$. 
In particular, we have
\begin{equation}
    H = \lim_{r\to\infty} \frac{2\pi}{\kappa r} p_v^2 = \lim_{r\to \infty} \frac{2\pi}{\kappa r}((p_v^{(0)})^2 + 2 p_v^{(0)}p_v^{(2)} + 2 p_v^{(0)} p_v^{(3)} + O(4))
\end{equation}
Inserting the solution of the previous section for $p_v$ we find
\begin{equation}
    H = M + \frac{1}{\kappa} \int_{\mathbb{R}^+} \dd r\qty[\frac{1}{2} p_v^{(0)} \sum_I Y_I \cdot \partial_r X^I + \frac{1}{r^6} \qty( A^{(2)} + [{}^{(2)}C_v + {}^{(3)}C_v]_{p= p^{(0)},y = y^{(1)}})]
    \label{eq:PhysHamBeforeTrafo}
\end{equation}
The final step is to replace the first and second order non-symmetric momenta $y$ by their solutions above.
This will in general give a very complicated expression. 
In the second order case, we observed that we can introduce new canonical variables $(Q,P)$ through a canonical transformation (see \cite{II}).
The transformation is conveniently written in terms of variables $\tilde X, \tilde Y$ defined as $X^I = \sqrt{2} r \tilde X^I$ and $\tilde Y_I = \sqrt{2} r Y_I$.
Up to second order this transformation is given by a type 1 generating functional:
\begin{align}
    G &= \int \dd r\Bigg[\frac{1}{r} (r Q^o)' \cdot \tilde X^o - \frac{\sqrt{r r_s}}{2 r^2}(\tilde X^o \cdot \tilde X^o + Q^o \cdot Q^o) +\frac{r \Lambda}{(l+2)(l-1)} \sqrt{\frac{r}{r_s}} \qty(\partial_r \tilde X^e \cdot \partial_r \tilde X^e  + \partial_r Q^e \cdot \partial_r Q^e)\nonumber\\
    &- \frac{1}{8 r \sqrt{r r_s}}\qty(3(l+2)(l-1)r + 7 r_s) \tilde X^e \cdot \tilde X^e+ \frac{(l+2)(l+1)l(l-1)r^2 - 6 r r_s + 3 r_s^2}{16 r^2 \sqrt{r r_s} \Lambda} Q^e \cdot Q^e\\
    &- \frac{2 r^2 \Lambda^2}{(l+2)(l-1)}\sqrt{\frac{r}{r_s}}(Q^e)' \cdot \pdv{r}\qty(\frac{\tilde X^e}{r \Lambda}) + \frac{1}{4 r \sqrt{r r_s}}\qty(l(l+1)r + 3 r_s)Q^e \cdot \tilde X^e \Bigg]\nonumber
\end{align}
The canonical transformation is then given by the relations
\begin{equation}
    \tilde Y^I_{lm} = \fdv{G}{\tilde X^I_{lm}}, \quad P^I_{lm} = - \fdv{G}{Q^I_{lm}}
\end{equation}

In terms of these variables the second order contributions to the integral in \eqref{eq:PhysHamBeforeTrafo} reduces to
\begin{equation}
    H^{(2)} = \sum_I \frac{1}{\kappa} \int_{\mathbb{R}^+} \dd{r} \qty[\sqrt{\frac{r_s}{r}} P_I \cdot \partial_r Q^I + \frac{1}{2} \qty(P_I \cdot P_I + \partial_r Q^I \cdot \partial_r Q^I + V_I Q^I \cdot Q^I)]
\end{equation}
where $V_I$ are the Regge-Wheeler Zerilli potentials.
In \cite{II}, it was shown that a boundary term that appears in the computation vanishes due to the fall-off conditions on the fields.

At third order the situation becomes significantly more complicated.
Now, the modes with even and odd polarity start to couple to each other through the Clebsch-Gordan coefficients.
In the remaining part of this subsection, we will briefly sketch the necessary steps for the computation of the full physical Hamiltonian to third order.
After that, we specialize to the pure odd polarity subsector in the next subsection, where we assume all the even polarity modes to vanish.
Strictly speaking this is not justified due to the mode coupling at third order, where two first order odd polarity perturbations can couple to an even 
polarity one. 
Nevertheless, it gives some insights into the general structure of the full results.

The third order contributions to the physical Hamiltonian are
\begin{equation}
    H^{(3)} = \frac{1}{\kappa} \int_{\mathbb{R}^+} \frac{\dd r}{r^6} \qty( A^{(2)} + {}^{(3)}C_v|_{p= p^{(0)},y = y^{(1)}})
\end{equation}
The second contribution can be directly copied from equation \eqref{eq:SolCv3}, where the non-symmetric momenta $y$
need to be replaced by their first order solutions.

For $A^{(2)}$, we start by replacing the second order solutions for the non-symmetric momenta by their solution:
\begin{align}
    A^{(2)} &= r^4 \Big[\frac{2 r^2}{\sqrt{l(l+1)}} y_e^{(1)}\cdot \partial_r y_v^{(2)} + y_v^{(2)}\cdot \qty(\frac{1}{2} y_v^{(1)} - r^2 y_h^{(1)} - \frac{r}{\sqrt{l(l+1)}} y_e^{(1)}) - \frac{1}{r^4 p_v^{(0)}} {}^{(2)}Z^v \cdot y_v^{(1)} \nonumber\\
    &+ y_o^{(1)}\cdot \int \dd r {}^{(2)}Z^o - \frac{r^2}{\sqrt{l(l+1)}} y_e^{(1)}\cdot \Bigg[\frac{2}{r^5 p_v^{(0)}} {}^{(2)} Z^v + {}^{(2)}Z^h\Bigg]_{p = p^{(0)},y = y^{(1)}}\Big]
\end{align}
Then using the first order solutions, we obtain

\begin{align}
    A^{(2)} &= r^4 \Big[\frac{2 r^2}{\sqrt{l(l+1)}} y_e^{(1)}\cdot \partial_r y_v^{(2)} + y_v^{(2)}\cdot \qty(\sqrt{\frac{(l+2)!}{2(l-2)!}} \frac{1}{p_v^{(0)}} X^e - \frac{r}{\sqrt{l(l+1)}} y_e^{(1)}) - \frac{1}{r^4 p_v^{(0)}} {}^{(2)}Z^v \cdot y_v^{(1)} \nonumber\\
    &+ y_o^{(1)}\cdot \int \dd r {}^{(2)}Z^o - \frac{r^2}{\sqrt{l(l+1)}} y_e^{(1)}\cdot \Bigg[\frac{2}{r^5 p_v^{(0)}} {}^{(2)} Z^v_{lm}+ {}^{(2)}Z^h_{lm}\Bigg]_{p = p^{(0)},y = y^{(1)}}\Big]
\end{align}

In the above equation we have to substitute $y_e^{(1)}$ in terms of $y_v^{(1)}$ and $X^e$.
Then, using the integral solutions for $y_v$, we obtain $A^{(2)}$ purely in terms of the physical degrees of freedom $X,Y$.
Finally, we combine the expressions for ${}^{(3)} C_v$ and $A^{(2)}$ to get the reduced Hamiltonian to third order in terms of $X,Y$ variables.

At second order, we found that introducing the quantities $Q,P$ through a canonical transformation significantly simplifies the reduced Hamiltonian. 
In particular, the complicated solution for $y_v$ in terms of integrals can be avoided. 
At third order, we can similarly replace $X,Y$ in favour of the variables $Q,P$ introduced before through the generating functional. 
This does however not remove the integral dependence of $y_v^{(2)}$ on $X,Y$ and furhter investigations are necessary to find a generalized 
canonical transformation which simplifies the third order reduced Hamiltonian.

\subsection{Detailed Analysis of the odd polarity subsector}

The third order symmetric Hamiltonian constraint is given by
\begin{align}
    {}^{(3)}C_v &= \frac{i r^4}{2\sqrt{2}}  \sum_{\{l_i,m_i\}} \sigma_-  \Big[ C^{l_1 m_1 2}_{l_2 m_2 -1,l_3 m_3 -1} X^o_{l_1 m_1} y^o_{l_2 m_2}y^o_{l_3 m_3} + C^{l_3 m_3 0}_{l_1 m_1 2, l_2 m_2 -2} \lambda_{l_3,2} X^o_{l_1 m_1} X^o_{l_2 m_2} X^o_{l_3 m_3} \Big]
\end{align}
In the above equation we neglected all even polarity contributions $(X^e,Y_e)$ as well as $y^{(1)}_{v/h/e}$.
The reason for this is that given $(X^e,Y_e) = 0$, we automatically find that the solutions $y^{(1)}_{v/h/e}$ need to vanish (no mode coupling at first order).

Consider the second contribution proportional to $(X^o)^3$. 
This product is completely symmetric in the pair of indices $(l_i m_i)$. 
However, we have that
\begin{equation}
    \sigma_- C^{l_3 m_3 0}_{l_1 m_1 2, l_2 m_2 -2} = \sigma_- C^{l_3 m_3 0}_{l_2 m_2 -2, l_1 m_1 2} = \sigma_- (-1)^{l_1 + l_2 + l_3} C^{l_3 m_3 0}_{l_2 m_2 2, l_1 m_1 -2} = - \sigma_- C^{l_3 m_3 0}_{l_2 m_2 2, l_1 m_1 -2}
\end{equation}
Comparing the left and the right terms, we observe that this contribution is antisymmetric in exchanging $(l_1 m_1)$ and $(l_2 m_2)$.
Therefore, the last term propotional to $(X^o)^3$ vanishes and we have
\begin{align}
    {}^{(3)}C_v &= \frac{i r^4}{2\sqrt{2}}  \sum_{\{l_i,m_i\}} \sigma_- C^{l_1 m_1 2}_{l_2 m_2 -1,l_3 m_3 -1} X^o_{l_1 m_1} y^o_{l_2 m_2}y^o_{l_3 m_3}
\end{align}

The other contribution of third order to the physical Hamiltonian comes from $A^{(2)}$.
It depends both on $y^{(1)}$ and $y^{(2)}$.
Since only $y_o^{(1)}$ is non-vanishing in our scenario, we only need to compute $y_o^{(2)}$ to find $A^{(2)}$.
We have
\begin{align}
    {}^{(2)} Z^o_{lm} &= - \sum_{\{l_i,m_i\}}\partial_r(X^o_{l_2 m_2} y_o^{l_3 m_3}) \frac{i}{\sqrt{2}} \sigma_- C^{l m -1}_{l_2 m_2 2, l_3 m_3 -1}\\
    &+ \sum_{\{l_im_i\}} \frac{i}{2} \sigma_- \Big[\frac{\lambda_{l_2,2}}{\lambda_{l_2,1}}C^{lm1}_{l_2 m_2 1, l_3 m_3 -2} - \frac{\lambda_{l_3,2}}{\lambda_{l_3,1}}C^{l m 1}_{l_2 m_2 -2, l_3 m_3 1} + \lambda_{l,1}C^{l m0}_{l_2 m_2 -2, l_3 m_3 2}\Big]X^o_{l_2 m_2} Y_o^{l_3 m_3}\nonumber
\end{align}
This gives
\begin{align}
    [y_o^{(2)}]_{lm} &= r^{-2} \int \dd r {}^{(2)} Z^o =  - \frac{1}{\sqrt{2} r^2} \sum_{\{l_i,m_i\}} i\sigma_- C^{l m -1}_{l_2 m_2 2, l_3 m_3 -1} X^o_{l_2 m_2} y_o^{l_3 m_3} \\
    &+ \frac{1}{2 r^2} \sum_{\{l_im_i\}} i \sigma_- \Big[\frac{\lambda_{l_2,2}}{\lambda_{l_2,1}}C^{lm1}_{l_2 m_2 1, l_3 m_3 -2} - \frac{\lambda_{l_3,2}}{\lambda_{l_3,1}}C^{l m 1}_{l_2 m_2 -2, l_3 m_3 1} + \lambda_{l,1}C^{l m0}_{l_2 m_2 -2, l_3 m_3 2}\Big] \int \dd r X^o_{l_2 m_2} Y_o^{l_3 m_3}\nonumber
\end{align}
We obtain for the remaining third order contribution to the physical Hamiltonian:
\begin{align}
    A^{(2)} &= r^6 y_o^{(2)} \cdot y_o^{(1)}\nonumber\\
    &= - \frac{r^4}{\sqrt{2}} \sum_{\{l_i,m_i\}} i\sigma_- C^{l_1 m_1 2}_{l_2 m_2 -1, l_3 m_3 -1} X^o_{l_1 m_1} y_o^{l_2 m_2} y_o^{l_3 m_3} \\
    &+ \frac{r^4}{2} \sum_{\{l_im_i\}} i \sigma_- \Big[\frac{\lambda_{l_2,2}}{\lambda_{l_2,1}}C^{lm1}_{l_2 m_2 1, l_3 m_3 -2} - \frac{\lambda_{l_3,2}}{\lambda_{l_3,1}}C^{l m 1}_{l_2 m_2 -2, l_3 m_3 1} + \lambda_{l,1}C^{l m0}_{l_2 m_2 -2, l_3 m_3 2}\Big] y_o^{l_1 m_1} \int \dd r X^o_{l_2 m_2} Y_o^{l_3 m_3}\nonumber
\end{align}

Summing the two contributions we find
\begin{align}
    H^{(3)}_{oo \rightarrow o} &= \frac{1}{2 \kappa} \sum_{\{l_i,m_i\}} i \sigma_-\int \frac{\dd{r}}{r^2} \Big[-\frac{1}{\sqrt{2}} C^{l_1 m_1 2}_{l_2 m_2 -1,l_3 m_3 -1} X^o_{l_1 m_1} y^o_{l_2 m_2}y^o_{l_3 m_3}\\
    &+ \Big(\frac{\lambda_{l_2,2}}{\lambda_{l_2,1}}C^{l_1 m_1 1}_{l_2 m_2 1, l_3 m_3 -2} - \frac{\lambda_{l_3,2}}{\lambda_{l_3,1}}C^{l_1 m_1 1}_{l_2 m_2 -2, l_3 m_3 1} + \lambda_{l_1,1}C^{l_1 m_1 0}_{l_2 m_2 -2, l_3 m_3 2}\Big) y_o^{l_1 m_1} \int \dd r X^o_{l_2 m_2} Y_o^{l_3 m_3}\nonumber\Big]
\end{align}
Using that $X^o = \sqrt{2} r \tilde X^o$ and $Y_o = 1/(\sqrt{2}r) \tilde Y_o$:
\begin{align}
    H^{(3)}_{oo \rightarrow o} &= \frac{1}{2 \kappa} \sum_{\{l_i,m_i\}} i \sigma_-\int \frac{\dd{r}}{r^2} \Big[- r C^{l_1 m_1 2}_{l_2 m_2 -1,l_3 m_3 -1} \tilde X^o_{l_1 m_1} y^o_{l_2 m_2}y^o_{l_3 m_3}\\
    &+ \Big(\frac{\lambda_{l_2,2}}{\lambda_{l_2,1}}C^{l_1 m_1 1}_{l_2 m_2 1, l_3 m_3 -2} - \frac{\lambda_{l_3,2}}{\lambda_{l_3,1}}C^{l_1 m_1 1}_{l_2 m_2 -2, l_3 m_3 1} + \lambda_{l_1,1}C^{l_1 m_1 0}_{l_2 m_2 -2, l_3 m_3 2}\Big) y_o^{l_1 m_1} \int \dd r \tilde X^o_{l_2 m_2} \tilde Y_o^{l_3 m_3}\nonumber\Big]
\end{align}
In this equation, we have to replace $y_o = \sqrt{(l+2)(l-1)}/r Q^o$ and
\begin{align}
    \tilde X^o &= r \int \qty(\frac{P^o}{r} - \frac{\sqrt{r r_s}}{r^3}Q^o)\dd r\\
    \tilde Y_o &= \frac{1}{r} (r Q^o)' - \frac{\sqrt{r r_s}}{r^2}X^o(Q^o, P^o)
\end{align}
We start by replacing $y_o$:
\begin{align}
    H^{(3)}_{oo \rightarrow o} &= \frac{1}{2 \kappa} \sum_{\{l_i,m_i\}} i \sigma_-\int \dd{r} \Big[- r^{-3} \frac{\lambda_{l_2,2}\lambda_{l_3,2}}{\lambda_{l_2,1}\lambda_{l_3,1}} C^{l_1 m_1 2}_{l_2 m_2 -1,l_3 m_3 -1} \tilde X^o_{l_1 m_1} Q^o_{l_2 m_2} Q^o_{l_3 m_3}\\
    &+ r^{-3} \frac{\lambda_{l_1,2}}{\lambda_{l_1,1}}\Big(\frac{\lambda_{l_2,2}}{\lambda_{l_2,1}}C^{l_1 m_1 1}_{l_2 m_2 1, l_3 m_3 -2} - \frac{\lambda_{l_3,2}}{\lambda_{l_3,1}}C^{l_1 m_1 1}_{l_2 m_2 -2, l_3 m_3 1} + \lambda_{l_1,1}C^{l_1 m_1 0}_{l_2 m_2 -2, l_3 m_3 2}\Big) Q^o_{l_1 m_1} \int \dd r \tilde X^o_{l_2 m_2} \tilde Y_o^{l_3 m_3}\nonumber\Big]
\end{align}
Then, replacing $Y_o$ we find
\begin{align}
    H^{(3)}_{oo \rightarrow o} &= \frac{1}{2 \kappa} \sum_{\{l_i,m_i\}} i \sigma_-\int \dd{r} \Big[- r^{-3} \frac{\lambda_{l_2,2}\lambda_{l_3,2}}{\lambda_{l_2,1}\lambda_{l_3,1}} C^{l_1 m_1 2}_{l_2 m_2 -1,l_3 m_3 -1} \tilde X^o_{l_1 m_1} Q^o_{l_2 m_2} Q^o_{l_3 m_3}\nonumber\\
    &+ r^{-3} \frac{\lambda_{l_1,2}}{\lambda_{l_1,1}}\Big(\frac{\lambda_{l_2,2}}{\lambda_{l_2,1}}C^{l_1 m_1 1}_{l_2 m_2 1, l_3 m_3 -2} - \frac{\lambda_{l_3,2}}{\lambda_{l_3,1}}C^{l_1 m_1 1}_{l_2 m_2 -2, l_3 m_3 1} + \lambda_{l_1,1}C^{l_1 m_1 0}_{l_2 m_2 -2, l_3 m_3 2}\Big) Q^o_{l_1 m_1} \times\\
    &~~~~~~~~~~~~~~~\qty(X^o_{l_2 m_2} Q^o_{l_3 m_3} - \int \dd r P^o_{l_2 m_2} Q^o_{l_3 m_3} + \int \dd r \frac{\sqrt{ r r_s}}{r^2}(Q^o_{l_2 m_2} Q^o_{l_3 m_3} - X^o_{l_2 m_2} X^o_{l_3 m_3}))\nonumber\Big]
\end{align}
Due to the symmetry of exchanging $(l_2 m_2)$ with $(l_3 m_3)$ of the $(Q^o)^2$ and the $(X^o)^2$ terms in the last line,
we find that these terms vanish when summed with the bracket of Clebsch-Gordan coefficients.
The first term in the last line can be combined with the first line after exchanging $(l_1 m_1)$ with $(l_2 m_2)$.
\begin{align}
    H^{(3)}_{oo \rightarrow o} &= \frac{1}{2 \kappa} \sum_{\{l_i,m_i\}} i \sigma_- \int \dd{r} \Big[r^{-3} \frac{\lambda_{l_2,2}}{\lambda_{l_2,1}}\Big(\frac{\lambda_{l_1,2}}{\lambda_{l_1,1}}C^{l_1 m_1 1}_{l_2 m_2 1, l_3 m_3 -2} - \lambda_{l_2,1}C^{l_1 m_1 2}_{l_2 m_2 0, l_3 m_3 -2}\Big) X^o_{l_1 m_1} Q^o_{l_2 m_2} Q^o_{l_3 m_3}\\
    &- r^{-3} \frac{\lambda_{l_1,2}}{\lambda_{l_1,1}}\Big(\frac{\lambda_{l_2,2}}{\lambda_{l_2,1}}C^{l_1 m_1 1}_{l_2 m_2 1, l_3 m_3 -2} - \frac{\lambda_{l_3,2}}{\lambda_{l_3,1}}C^{l_1 m_1 1}_{l_2 m_2 -2, l_3 m_3 1} + \lambda_{l_1,1}C^{l_1 m_1 0}_{l_2 m_2 -2, l_3 m_3 2}\Big) Q^o_{l_1 m_1} \int \dd r P^o_{l_2 m_2} Q^o_{l_3 m_3}\nonumber\Big]
\end{align}
Here $X^o$ needs to be expressed in terms of the integral involving $Q^o$ and $P^o$.

In appendix \ref{sec:RWG}, we perform the derivation of the odd polarity interaction Hamiltonian in the Regge-Wheeler gauge. 
This gauge is the typical gauge in the treatment of black hole perturbation theory using the Einstein equations. 
For odd polarity perturbations, it is characterized by the gauge condition $m_{33} = 1$, $m_{3A} = \sum_{lm} x^o_{lm} [L^o_{lm}]_A$ and $m_{AB} = r^2 \Omega_{AB}$ where $x_A$ is now the physical degree of freedom.
Not surprisingly, the result is different from the one above.
In the end of the appendix, we show how the master variables in the two gauges are related and we leave it for future investigations to study how the 
third order physical Hamiltonians are related.
We do not expect the third order Hamiltonians to be equivalent in the Regge-Wheeler and Gullstrand-Painlevé gauge fixing due to the reasons 
discussed in more detail in \cite{TTNew1}.

\subsection{Comparison with Lagrangian Perturbation Theory}

In this section we derive the equations of motion of the physical Hamiltonian and discuss its relation to the non-linear perturbation theory
of the equations of motion in the literature \cite{Madrid1,Madrid2,Madrid3,Italy1,UK1,Japan1}. 
In the beginning of this section, we observed that the physical Hamiltonian can be written as
\begin{equation}
    H = M + \frac{1}{\kappa} \int_{\mathbb{R}^+}  \qty[\sqrt{\frac{r_s}{r}}P \cdot \partial_r Q + \frac{1}{2}\qty(P\cdot P + \partial_r Q \cdot \partial_r Q + V Q \cdot Q)]\dd r + H_{\mathrm{int}}(Q,P)
\end{equation}
where we used the explicit result for the quadratic order and introduced $H_\mathrm{int}$ for the third order interactions.
Then, the equations of motion of the full physical Hamiltonian give
\begin{align}
    \dot Q_{lm} &= \fdv{H}{P_{lm}} = \sqrt{\frac{r_s}{r}} \partial_r Q_{lm} + P_{lm} + \fdv{H_{\mathrm{int}}}{P_{lm}}\\
    \dot P_{lm} &= - \fdv{H}{Q_{lm}} = \partial_r\qty(\sqrt{\frac{r_s}{r}}P_{lm}) + \partial_r^2 Q_{lm} - V Q_{lm} - \fdv{H_{\mathrm{int}}}{Q_{lm}}
\end{align}
Notice that this is a coupled first-order differential equation in the Gullstrand-Painlevé time $\tau$ for the master variables $(Q,P)$.
In order to compare this to the Lagrangian perturbation theory, we have to combine the two equations to a second order differential equation for $Q_{lm}$.
For generic interaction Hamiltonians depending on both $Q$ and $P$ this is not straight forward. 
Since, we are working to third order only, we can use the following strategy:
The first equation can be solved for $P_{lm}$ up to second order in $(Q,P)$:
\begin{equation}
    P_{lm} = \dot Q_{lm} - \sqrt{\frac{r_s}{r}} \partial_r Q_{lm} - \fdv{H_{\mathrm{int}}}{P_{lm}}\Big|_{P_{lm} = \dot Q_{lm} - \sqrt{\frac{r_s}{r}} \partial_r Q_{lm}}
\end{equation}
Then, the second equation gives
\begin{align}
    \begin{split}
    &\partial_t \qty(\dot Q_{lm} - \sqrt{\frac{r_s}{r}} \partial_r Q_{lm}) -\partial_r\qty(\sqrt{\frac{r_s}{r}}(\dot Q_{lm} - \sqrt{\frac{r_s}{r}} \partial_r Q_{lm})) - \partial_r^2 Q_{lm} + V Q_{lm} =\\
    &=\partial_t \qty(\fdv{H_{\mathrm{int}}}{P_{lm}}\Big|_{P_{lm} = \dot Q_{lm} - \sqrt{\frac{r_s}{r}} \partial_r Q_{lm}}) - \partial_r\qty(\sqrt{\frac{r_s}{r}}\fdv{H_{\mathrm{int}}}{P_{lm}}\Big|_{P_{lm} = \dot Q_{lm} - \sqrt{\frac{r_s}{r}} \partial_r Q_{lm}})  - \fdv{H_{\mathrm{int}}}{Q_{lm}}
    \end{split}
\end{align}
This is now a second-order differential equation in time for $Q$. 
It is convenient to rewrite the equation in a more covariant form introducing a two-dimensional metric $g$ on the $(t,r)$ subspace of the 
Schwarzschild spacetime in Gullstrand-Painlevé coordinates:
\begin{equation}
    g_{tt} = -\qty(1-\frac{r_s}{r}), \quad g_{t3} = g_{3t}= \sqrt{\frac{r_s}{r}}, \quad g_{33} =1, \quad \quad g^{tt} = -1, \quad g^{3t} = g^{t3} = \sqrt{\frac{r_s}{r}}, \quad g^{33} =  1-\frac{r_s}{r}
\end{equation}
We define the operator $\square \cdot := \sqrt{-g}^{-1} \partial_\mu (\sqrt{-g} g^{\mu \nu} \partial_\nu \cdot)$ and introduce the timelike normal $n^\mu$ 
to the spacelike hypersurfaces of the foliation of spacetime with respect to the Gullstrand-Painlevé time coordinate.
The normal is given by $n^t = 1$, $n^3 = -\sqrt{\flatfrac{r_s}{r}}$ ($n_\mu n^\mu=-1$) and we find
\begin{align}
    &(\square - V )Q_{lm} =-\nabla_\mu\qty(n^\mu \fdv{H_{\mathrm{int}}}{P_{lm}}\Big|_{P_{lm} = n^\nu \nabla_\nu Q_{lm}}) + \fdv{H_{\mathrm{int}}}{Q_{lm}}
    \label{eq:HamEqnOfMotion}
\end{align}

In the Lagrangian perturbation theory one starts from the vacuum Einstein equations and inserts the expansion $g = g^{(0)} + g^{(1)} + g^{(2)}+ \dots$.
Then, $g^{(0)}$ is assumed to be a solution of the spherically symmetric Einstein equations, i.e. the Schwarzschild solution. 
Up to third order the remaining equations are
\begin{equation}
    G^{(1)}_{\mu \nu}[g^{(1)}] = 0, \quad \quad \quad G^{(1)}_{\mu \nu}[g^{(2)}] = - G^{(2)}_{\mu \nu}[g^{(1)},g^{(1)}]
\end{equation}
where $G_{\mu \nu}$ is the Einstein tensor. 
It is expanded as
\begin{align}
    G^{(1)}_{\mu \nu}[h](x) &:= \int \fdv{G_{\mu \nu}(x)}{g_{\mu \nu}(y)}\Big|_{g = g^{(0)}} h_{\mu \nu}(y)\dd y\\
    G^{(2)}_{\mu \nu}[h^1,h^2](x) &:= \int \frac{\delta^2 G_{\mu \nu}(x)}{\delta g_{\mu \nu}(y_1) \delta g_{\mu \nu}(y_2)}\Big|_{g = g^{(0)}} h^1_{\mu \nu}(y_1) h^2_{\mu \nu}\dd y_1 \dd y_2
\end{align}
Then, the solution proceeds iteratively. 
First, the first order equations are solved for $g^{(1)}$ and this solution then acts as a source for the second order solutions $g^{(2)}$.

The metric perturbations are not gauge invariant, i.e. they are varying under changes of coordinates. 
For this reason, there are two options:
(i) Define combinations of the components of the metric perturbations such that they are gauge invariant.
(ii) Fix the gauge by setting some components of the metric perturbations to zero. 
In the end, the two approaches lead to equivalent results.
Given the gauge invariant viewpoint, one can fix the gauge and connect to the gauge fixed theory. 
From the gauge fixed theory, one may extend the gauge fixed quantity to a gauge invariant one by adding suitable combinations of gauge degrees of freedom (relational observables).
At second order in perturbation theory, it is most convenient to work in the gauge fixed framework. 
In order to define the gauge invariant variables one has to find gauge invariant expressions which are linear in $g^{(2)}$ and
quadratic in $g^{(1)}$ and the expressions are quite long and complicated \cite{Madrid2}.

Let us therefore adopt the gauge fixed viewpoint. A convenient gauge fixing is the so-called Regge-Wheeler gauge.
In this gauge, one can define master variables $\psi^{(1/2)}_\pm$ depending on $g^{(1)}$ and $g^{(2)}$ respectively.
The subscript $\pm$ stands for the axial and polar subsector in an expansion into spherical harmonics.
The master variables are scalar quantities such that $G^{(1)}_{\mu \nu}$ reduces to a wave operator, i.e.
\begin{equation}
    G^{(1)}_{\mu \nu}[\psi^{i}_\pm] = (\square - V_\pm) \psi_\pm^{(i)}
\end{equation}
where $i=1,2$.
In this notation, the first order Einstein equations then simply reduce to the famous Regge-Wheeler-Zerilli equation $(\square - V_\pm) \psi_\pm^{(1)} = 0$.
The second order equation is the Regge-Wheeler-Zerilli equation with a non-trivial source term depending on $\psi_\pm^{(1)}$, i.e. 
\begin{equation}
    (\square - V_\pm) \psi_\pm^{(2)} = S_\pm(\psi_\pm^{(1)},\psi_\pm^{(1)})
    \label{eq:LagrangianEqnOfMotion}
\end{equation}
Note that in this source term, the axial and polar perturbations as well as the modes with different $l,m$ will be coupled.

Comparing \eqref{eq:HamEqnOfMotion} with \eqref{eq:LagrangianEqnOfMotion}, we observe that the two expressions both involve the operator $\square - V$ on the 
left-hand side.
In order to match the two approaches, we have to expand the master variable $Q = Q^{(1)} + Q^{(2)} + \dots$ and collect terms of equal order in the perturbations.
At first order, we obtain the Regge-Wheeler equation without source and for the second order we find a non-trivial source term given by
\begin{equation}
    \qty[-\nabla_\mu\qty(n^\mu \fdv{H_{\mathrm{int}}}{P_{lm}}\Big|_{P_{lm} = n^\nu \nabla_\nu Q_{lm}}) + \fdv{H_{\mathrm{int}}}{Q_{lm}}]_{Q = Q^{(1)}}
    \label{eq:SourceHamInteraction}
\end{equation}
Thus, the two approaches match provided the two source terms are equivalent.

For the comparison, we now focus on the purely odd polarity sector, i.e. ignoring all the even polarity modes. 
The corresponding master equation including the source term is given by (see \cite{II})
\begin{equation}
    (\square - V_-) \psi_-^{lm} = - 2 r \epsilon^{ab} \nabla_a t^{lm}_b
\end{equation}
where $t^{lm}_b$ is the axial contribution to the Einstein tensor $G_{Ab}$. 
Using Mathematica the Einstein tensor to second order in the perturbations reads
\begin{align}
    \begin{split}
    G_{Ab} &= \frac{1}{r^3}\Big[r D_A h^{Cd} \nabla_b h_{Cd} - r D^C h_{Ad} \nabla^d h_{bC} - r D_A h_{bC} \nabla_d h^{Cd} + r D_C h_{Ab} \nabla_d h^{Cd}\\
    &+ h^{Cd} (2 r D_A \nabla_b h_{Cd} + 2 r D_C \nabla_d h_{Ab} - r D_A \nabla_d h_{bC} - r D_C \nabla_b h_{Ad}- 4 \gamma_b D_A h_{Cd} + 2 \gamma_b D_C h_{Ad} \\\
    &~~~~~~~~~~~~~+ 2 \gamma_d D_A h_{bC} - 2 \gamma_d D_C h_{Ab})\Big]
    \end{split}
\end{align}
Then, we expand the metric perturbation according to $h_{Ab} = \sum_{lm} h^{lm}_b [L^o_{lm}]_A$.
Using the definition of the vector and tensor spherical harmonics, we obtain
\begin{equation}
    D_A [L^o_{lm}]_{B} = \sqrt{\frac{(l+2)(l-1)}{2}} [L^o_{lm}]_{AB} + \frac{1}{2} \sqrt{l(l+1)} \epsilon_{AB} L_{lm}
\end{equation}
This gives
\begin{align}
    \begin{split}
    G_{Ab} &= \frac{1}{2 r^3}\Big[\sqrt{2(l_1+2)(l_1-1)} [L^1_o]_{AF} [L^2_o]^F\qty((r \nabla_b h^1_d + r \nabla_d h^1_b  - 2 \gamma_b h^1_d)h_2^d + h_1^d\qty(r \nabla_b h^2_d - r \nabla_d h^2_b))\\
    &- \sqrt{l_1(l_1+1)} L^1 [L_2^e]_A\big( h_1^d (r \nabla_b h^2_d + r \nabla_d h^2_b) + h^1_b(- 2 r \nabla_d h_2^d + 4 \gamma_d h_2^d)\\
    &~~~~~~~~~~~~~+ 3 h_2^d (r \nabla_b h^1_{d} - r \nabla_d h^1_b - 2 \gamma_b h^1_{d})\big)\Big]
    \end{split}
\end{align}
In the above equation we abbreviated the mode expansion where $1$, $2$ stand for the indices $l_1 m_1$, $l_2 m_2$ respectively.
In vacuum, the first order Einstein equations imply that $\nabla_d h^d = 0$.
Defining $t_b^{lm} = \int G_{Ab} [L^o_{lm}]^A\dd \Omega$, we find
\begin{align}
    t_{b}^{lm} &= \sum_{\{l_i, m_i\}}\frac{i \sigma_- }{2 r^3}\Big[\sqrt{(l_1+2)(l_1-1)} C^{l_1 m_1 2}_{l_2 m_2 -1, l m -1} \qty((r \nabla_b h^1_d + r \nabla_d h^1_b  - 2 \gamma_b h^1_d)h_2^d + h_1^d\qty(r \nabla_b h^2_d - r \nabla_d h^2_b))\nonumber\\
    &+ \sqrt{l_1(l_1+1)} C^{l_1 m_1 0}_{l_2 m_2 1, l m -1} \big( h_1^d (r \nabla_b h^2_d + r \nabla_d h^2_b) + 4 h^1_b \gamma_d h_2^d + 3 h_2^d (r \nabla_b h^1_{d} - r \nabla_d h^1_b - 2 \gamma_b h^1_{d})\big)\Big]
\end{align}
In these equations one has to replace $h_a$ by the odd polarity master variable defined by:
\begin{equation}
    \psi_- = \gamma^3 \epsilon^{ab}\nabla_a (\gamma^{-2} h_b)
\end{equation}
The inverse relation is
\begin{equation}
    h_a = \frac{1}{(l+2)(l-1)} \epsilon_{ab} \nabla^b (r \psi_-)
\end{equation}
Thus, the full source term is expressed in terms of the odd polarity master variable and its covariant derivatives.

It remains to compute the source term using equation \eqref{eq:SourceHamInteraction} based on the third order interaction Hamiltonian 
computed in the Regge-Wheeler gauge in appendix \ref{sec:RWG} (see \eqref{eq:3rdAxialRW}). 
The result then needs to be compared to the source term found above in terms of the odd polarity master variable.
Furthermore, for a complete picture, one has to check, whether the master variable $Q$ and the Regge-Wheeler master variable $\psi_-$
defined above correspond to the same physical quantity. 
Both variables arise in a different way:
The Hamiltonian master variable is introduced through a canonical transformation which simplifies the Hamiltonian. 
The Lagrangian master variable is capturing the dynamics of the gauge fixed variable $h_a$.
To first order, it was shown that both variables agree (see \cite{II}) but this does not necessarily generalise to higher orders.\\ 
\\
\\
Why this is nevertheless physically correct is discussed in more detail in \cite{TTNew1}. It has to do with the correspondence between the 
gauge fixed or reduced frameworks and the relational observable frameworks induced by different gauge fixing conditions. Roughly speaking,
there is a Poisson isomorphism between the true degrees of of freedom and the relational observables induced by the same gauge fixing 
condition. One can therefore identify these two sets of variables as one possible description of the observables of the theory. Those 
observables evolve by the corresponding reduced Hamiltonian in the reduced framework or equivalently the physical Hamiltonian in the 
relational framework. This can be done for different gauges. Since the dimension of the reduced phase space does not depend 
on the choice of gauge fixing, it is possible to write the relational Dirac observables corresponding to second gauge in terms of the 
relational Dirac observables of the first which is a canonical transformation by construction. By using the identifications with the 
reduced phase spaces there is a corresponding canonical transformation at the reduced phase space level. However, there is no 
reason to expect that the reduced Hamiltonians of the second gauge when written in terms of the true degrees of freedom of the first gauge
using this canonical transformation must equal the reduced Hamiltonian of the first gauge. As trivial example, one may consider gauge fixing 
conditions on reference clock variables of different physical dimension. 
Then the corresponding reduced Hamiltonians also have different dimension because 
the product of clock and Hamiltonian dimension must be that of an action. They therefore can never equal each other.

\section{Conclusion}
\label{sec:Conclusion}

In this manuscript, we investigated beyond second order black hole perturbation theory in the Hamiltonian framework for vacuum General Relativity.
It generalises earlier work on the subject in \cite{II,III,IV} where second order perturbation theory was analysed in detail for electromagnetic 
and gravitational fields within the gravitational sector.
The computation is based on the reduced phase space approach to Hamiltonian perturbation theory, which includes backreaction and addresses 
the issue of gauge invariance in a systematic, non-perturbative way to all orders.
 
In the first half of this manuscript, we carefully expanded all vacuum constraints of general relativity in the ADM formulation into their symmetric and 
non-symmetric contributions
with respect to spherical symmetry to all orders. This is possible by working with polynomial versions of the constraints which in the Gullstrand 
Painleve gauge is possible by using the density weight four rather than density one Hamiltonian constraint. One obtains a polynomial of degree
six upon multiplying the Hamiltonian constraint by the 
determinant of the intrinsic metric raised the power $3/2$. This is a remarkable simplification because in generic gauges the power $5/2$ is required
which generates a polynomial of degree ten.

In the second half of the work, we investigated the solutions of these constraints to third order and obtained the reduced third order interaction Hamiltonian
of the system.
The Hamiltonian only depends on the non-symmetric true degrees of freedom $X,Y$ (and the ADM mass) 
which coordinatise the reduced phase space of polar and axial gravitational waves and generates their dynamics on a black hole spacetime 
parametrised by the ADM mass. If one would add matter such as fermions or scalar fields there would also be 
symmetric true degrees of freedom and the reduced Hamiltonian would govern their dynamics and their interaction or backreaction with 
$X,Y$ as well \cite{I}.
   
In the previous papers of the series, we used a canonical transformation to introduce new variables $Q,P$ such that the physical Hamiltonian simplifies.
In these new variables, we recovered the famous Regge-Wheeler-Zerilli equations for $Q$ after computing the Hamiltonian equations of motion. 
At third order, the situation is more complicated and it is unclear whether new variables $Q,P$ exist, such that the equations of motion for $Q$ match 
the results in the literature. In fact, on generic grounds the existence of such a canonical transformation is not granted because already for simple 
mechanical systems the reduced phase space dynamics encoded by the reduced Hamiltonian does depend on the chosen gauge in the following sense:
While there exists a natural canonical transformation between two reduced phase descriptions induced by two different gauge conditions which follows 
from the Poisson isomorphism between the reduced and relational description, the reduced 
Hamiltonians are not images of each other under this canonical transformation.
This appears to be alarming at first, apparently indicating a loss of gauge invariance. However, this is not the case. For instance, the kinematical 
phase space of a constrained sytem may be the Cartesian product of two cotangent bundles one of which is over a compact configuration space, 
the other over a non-compact configuration 
space. Picking a compact configuration clock induces a reduced Hamiltonian that generically depends on 
the non-compact configuration true degrees of of freedom and vice versa. 
While it is possible 
have a canonical transformation between two such reduced phase spaces (e.g. $(-1,1)$ and $\mathbb{R}$ are diffeomorphic) it is never possible 
to change the range of a function by pulling it back by the canonical transformation. Yet, both descriptions fully describe the same physical system. 
See \cite{TTNew1} for more details. 

Whether in the present situation the natural canonical transformation between the GP and RZ gauge descriptions still
maps between the corresponding reduced Hamiltonians was beyond the scope of the present paper and a full analysis is left for future investigations.  
In appendix \ref{sec:RWG} we started this analysis and imposed the Regge-Wheeler gauge on the non-symmetric degrees of freedom and obtained the 
reduced Hamiltonian for the corresponding true degrees of freedom. 
To second order, the reduced Hamiltonian expressed in the Regge-Wheeler master variables $Q,P$ appears to be 
identical to the one in Gullstrand-Painlevé gauge. 
This is motivated but not strictly proved by the results of \cite{IV}, which showed that at second order, the choice of gauge for the symmetric variables only changes
the physical Hamiltonian by a boundary term.
At third order, the results in the Regge-Wheeler and Gullstrand-Painlevé gauge differ and it is unclear whether new master variables exist 
such that their form is identical. 

We have also started to investigate the Hamiltonian equations of motion for self-interacting gravitational waves. While we have computed all axial and 
polar terms and their interaction to third order, in this paper we restricted those equations of motion to the purely axial terms in order to gain 
first insights into the phenomenon of self-interactions. In the future we need to incorporate of course the self interacting polar 
modes and their interaction with the axial modes. 

The results in this manuscript are the basis for gauge invariant higher black hole perturbation theory, both classically and quantum mechanically. 
Classically, the theory is important for the investigation of non-linear effects in black hole perturbation theory,
such as quasi-normal modes and the simulation of merging black holes.
For the quantum theory, the second order physical Hamiltonian motivates a Fock quantisation of the non-symmetric degrees of freedom.
The new, third order contributions will then lead to a non-trivial S-matrix.

Finally, also inclusion of matter to third order is required. This extension is not difficult with the tools developed in this paper.
For instance for electromagnetic matter and third order there are interaction terms quadratic in the non-symmetric modes of the electromagnetic 
field and linear in the non-symmetric modes of the gravitational field, respectively of both polarisations.

\appendix

\section{Axial Third Order Interaction in Regge-Wheeler Gauge}
\label{sec:RWG}

In section \ref{sec:PhysHam} of the main text, we derived the physical Hamiltonian in the GP gauge.
For comparison we compute the pure axial third order interaction term in this appendix using the Regge-Wheeler gauge. 
In this gauge the spatial metric and conjugate momentum read
\begin{gather}
    m_{33} = 1, \quad m_{3A} = x_A, \quad m_{AB} = r^2 \Omega_{AB}\\
    W^{33} = \sqrt{\Omega} p_v, \quad W^{3A} = \frac{1}{2}\sqrt{\Omega} y^A, \quad W^{AB} =\sqrt\Omega\frac{p_h}{2} \Omega^{AB} + Y^{AB}
\end{gather}
In this subsection we are only interested in computing the perturbation theory up to third order.
Therefore, in contrast to the main text, we only work up to third order and truncate all higher order contributions.
Inverting the metric we find 
\begin{align}
\begin{aligned}
    m^{33} &= 1 + \frac{1}{r^2} x^A x_A + O(x^4)\\
    m^{3A} &= - \frac{1}{r^2} x^A - \frac{1}{r^4} x^A x^B x_B+ O(x^4)\\
    m^{AB} &= \frac{1}{r^2}\Omega^{AB} + \frac{1}{r^4} x^A x^B+ O(x^4)
\end{aligned}
\end{align}

Then, we have to use the expansion of the metric to find the diffeomorphsim and Hamiltonian constraints up to second order.
For the axial perturbation, we need to compute the angular diffeomorphism constraint which is given by
\begin{equation}
    V_A = \sqrt\Omega \qty(- 2 r^2 D^B Y_{AB} - \partial_r (r^2 y_A + 2 p_v x_A) + y^B D_{A} x_B - D_B (x_A y^B))
\end{equation}

We split the Hamiltonian constraint again into the momentum contributions and the Ricci scalar contributions.
For the momentum contributions $K$, we find
\begin{align}
\begin{aligned}
    K = \sqrt{\Omega}\Big[&\frac{1}{2} p_v^2 - r^2 p_v p_h +  r^4 Y^{AB} Y_{AB} + \frac{1}{2} r^2 y_A y^A + p_v y^A x_A + p_v p_h x^A x_A\\
    &+ 2 p_v Y^{AB} x_A x_B + 2 r^2 Y^{AB} y_A x_B\Big]
\end{aligned}
\end{align}
For the Ricci scalar we have
\begin{align}
    R &= R^{(2)} +  \nabla_3 m_{3A} \nabla_B m_{3C} (- 2 m^{3A} m^{BC}) + \nabla_A m_{33} \nabla_B m_{3C} (2 m^{AB} m^{3C} - m^{BC} m^{3A})\nonumber\\
    &+ \nabla_A m_{3B} \nabla_C m_{DE}( m^{3D}(3 m^{AC} m^{BE} - 2 m^{AB} m^{CE} - m^{AE} m^{BC}) + m^{3C}(m^{AB} m^{DE} - m^{AD} m^{BE})\\
    &+ m^{3B}(2m^{AD} m^{CE} - m^{AC} m^{DE}) + m^{3A}(m^{BC} m^{DE} - 2 m^{BE} m^{CD})) \nonumber
\end{align}
where $R^{(2)}$ is the second order contribution computed in \cite{II}.
The first derivative of the metric perturbations are given by
\begin{align}
    \nabla_3 m_{3A} = \qty(\partial_r - \frac{1}{r})x_A, \quad \nabla_A m_{33} = - \frac{2}{r} x_A, \quad \nabla_A m_{3 B} = D_A x_B, \quad \nabla_A m_{BC} = 2 r \Omega_{A(B}x_{C)}
\end{align}
Hence, we find for the third order contribution to the Ricci scalar
\begin{equation}
    {}^{(3)}R = \frac{4}{r^5} x^A x^B D_A x_B
\end{equation}
This completes the derivation of the third order contributions to the constraints.

The next step is the expansion of the constraints in terms of spherical harmonics.
We use the convention
\begin{equation}
    x_A = \sum_{lm} x^o_{lm} [L^o_{lm}]_A, \quad y^A = \sum_{lm} y_o^{lm} [L^o_{lm}]^A
\end{equation}
The odd polarity angular diffeomorphism constraint in terms of modes $l,m$ is given by
\begin{equation}
    Z^o_{lm} = \sqrt{2(l+2)(l-1)} r^2 Y_o^{lm} - \partial_r (r^2 y_o^{lm} + 2 p_v x^o_{lm}) + \int_{S^2} [L^o_{lm}]^A \qty(y^B D_A x_B- D_B (x_A y^B))
\end{equation}
The integral can be evaluated explicitly using the $\eth$-formulation and we have
\begin{align}
    \int_{S^2} &[L^o_{lm}]^A \qty(y^B D_A x_B- D_B (x_A y^B)) = -  \sum_{\{l_i,m_i\}}  i \sigma_- \sqrt{l_2(l_2 + 1)} C^{lm 1}_{l_1 m_1 -1, l_2 m_2 0} y_o^{l_1 m_1} x^o_{l_2 m_2}
\end{align}
The solution of the odd polarity diffeomorphism constraint for $Y^o$ is straight-forward.
In contrast to the Gullstrand-Painlevé gauge, it does not involve any differential equations. 
This is an advantages of the Regge-Wheeler gauge over the Gullstrand-Painlevé gauge and the reason for its usage in the literature.
The solution is
\begin{equation}
    Y_o^{lm} = \frac{1}{\sqrt{2(l+2)(l-1)} r^2}\qty(\partial_r (r^2 y_o^{lm} + 2 p_v^{(0)} x^o_{lm})  + \sum_{\{l_i,m_i\}} i \sigma_- \sqrt{l_2(l_2 + 1)} C^{lm1}_{l_1 m_1 -1, l_2 m_2 0} y_o^{l_1 m_1} x^o_{l_2 m_2})
\end{equation}

For the third order Hamiltonian constraint, we only need the spherically symmetric contribution. 
Explicitly it is given by
\begin{align}
    \begin{split}
    {}^{(3)}C_v &= \int_{S^2} \qty(r^4 {}^{(3)} K - r^8 {}^{(3)} R ) \dd \Omega\\
    &= \sum_{\{l_i,m_i\}} i \sigma_- \Big[\sqrt{2} r^4 C^{l_1 m_1 2}_{l_2 m_2 -1, l_3 m_3 -1} Y_o^{l_1 m_1}( p_v^{(0)}x^o_{l_2 m_2} x^o_{l_3 m_3} + r^2 x^o_{l_2 m_2} y^o_{l_3 m_3})\\
    &~~~~~~~~~~~~~~~~~~~- 2 \sqrt{(l_1 + 2)(l_1 - 1)} C^{l_1 m_1 2}_{l_2 m_2 -1, l_3 m_3 -1} r^3 x^o_{l_1 m_1} x^o_{l_2 m_2} x^o_{l_3 m_3} \Big]
    \end{split}
\end{align}

The third order contribution to the physical Hamiltonian is given by ${}^{(3)}C_v$ where we need to replace $Y_o$ by its solution. 
In addition, we have to use ${}^{(2)}C_v$ where we insert the expansion of $Y_o$ to second order and keep the third order term.
In total the third order Hamiltonian is given by
\begin{align}
    \begin{aligned}
    H^{(3)} &=  \int_{\mathbb{R}^+} \sum_{\{l_i,m_i\}} i \sigma_-  \sqrt{2} \qty(\frac{\sqrt{l_3(l_3 + 1)}}{\sqrt{(l_1+2)(l_1-1)}} C^{l_1 m_1 1}_{l_2 m_2 -1, l_3 m_3 0} + C^{l_1 m_1 2}_{l_2m_2 -1, l_3 m_3 -1}) Y_o^{l_1 m_1} y_o^{l_2 m_2} x^o_{l_3 m_3}\\
    &~~~~~~~~~~~+ C^{l_1 m_1 2}_{l_2 m_2 -1, l_3 m_3 -1} \qty(\sqrt{2} r^{-2}  p_v^{(0)} Y_o^{l_1 m_1} - 2 \sqrt{(l_1 + 2)(l_1 - 1)} r^{-3} x^o_{l_1 m_1}) x^o_{l_2 m_2} x^o_{l_3 m_3} \dd r 
    \end{aligned}
\end{align}
In the first line, we use the recursion relation for the Clebsch-Gordan coefficients $C$ and find
\begin{align}
    \begin{aligned}
    H^{(3)} &=  \int_{\mathbb{R}^+} \sum_{\{l_i,m_i\}} - i \sigma_-  \sqrt{2} \frac{\sqrt{l_2(l_2 + 1)}}{\sqrt{(l_1+2)(l_1-1)}} C^{l_1 m_1 1}_{l_2 m_2 0, l_3 m_3 -1} Y_o^{l_1 m_1} y_o^{l_2 m_2} x^o_{l_3 m_3}\\
    &~~~~~~~~~~~+ i \sigma_- C^{l_1 m_1 2}_{l_2 m_2 -1, l_3 m_3 -1} \qty(\sqrt{2} r^{-2}  p_v^{(0)} Y_o^{l_1 m_1} - 2 \sqrt{(l_1 + 2)(l_1 - 1)} r^{-3} x^o_{l_1 m_1}) x^o_{l_2 m_2} x^o_{l_3 m_3} \dd r
    \end{aligned}
\end{align}

In \cite{II} it turned out that the second order physical Hamiltonian simplifies when expressed in master variables $(Q^o, P^o)$ which 
are related to $(X^o,Y^o)$ through a canonical transformation. 
The new variables are defined as
\begin{align}
    Q^o_{lm} &:= \frac{1}{\sqrt{(l+2)(l-1)}r}(r^2 y_o^{lm} + 2 p_v^{(0)} x^o_{lm})\\
    P^o_{lm} &:= - \frac{\sqrt{(l+2)(l-1)}}{r}x^o_{lm} + \frac{p_v}{2r^2} Q^o_{lm}
\end{align}
In terms of these variables the first order solution for the angular diffeomorphism constraint reads $Y_o = \partial_r (r Q^o)/(\sqrt{2}r^2)$.
Furthermore, from the canonical transformation we find
\begin{align}
    y_o^{lm} &= \frac{\sqrt{(l+2)(l-1)}}{r} Q^o_{lm} - \frac{2 p_v^{(0)}}{r^2} x^o_{lm}\\
    P_o^{lm} &= - \frac{\sqrt{(l+2)(l-1)}}{r} x^o_{lm} + \frac{p_v^{(0)}}{2 r^2} Q^o_{lm}
\end{align}
Hence, this transformation can also be rewritten in terms of a type 1 generating functional:
\begin{align}
    G= \int \frac{\sqrt{(l+2)(l-1)}}{r} Q^o \cdot x^o  - \frac{p_v^{(0)}}{4r^2} (Q^o \cdot Q^o + 4 x_o \cdot x_o ) \dd{r} 
\end{align}

Let us briefly review how this transformation simplifies the second order physical Hamiltonian.
Recall the second order physical Hamiltonian in terms of $x^o,y_o$, and $Y_o$ from \cite{II}:
\begin{align}
    H^{(2)}= \int_{\mathbb{R}^+} - \sqrt{\frac{r_s}{r}} x^o \partial_r y_o + \frac{1}{2} y_o \cdot y_o + \frac{2 \sqrt{r r_s}}{r^2} y_o \cdot x^o + \frac{1}{r^2}x^o \cdot \qty(4 r \partial_r + \frac{l(l+1)}{2} - 3 + 2 \frac{r_s}{r})x^o + r^2 Y_o \cdot Y_o \dd r
\end{align}
After inserting $Y_o$ and replacing $y_o$ in terms of $x^o$ and $Q^o$, we find
\begin{align}
    \begin{aligned}
    H^{(2)}=&\int_{\mathbb{R}^+}- \sqrt{\frac{r_s}{r}}  \frac{\sqrt{(l+2)(l-1)}}{r^2} (x^o \cdot Q^o + r x^o  \cdot \partial_r Q^o) + \frac{l(l+1)}{2r^2} Q^o \cdot Q^o + \frac{(l+2)(l-1)}{2r^2} x^o \cdot x^o\\
    &+ \frac{1}{2} (\partial_r Q^o) \cdot (\partial_r Q^o) + \dv{r}\qty(\frac{2(r + r_s)}{r^2} x^o \cdot x^o + \frac{1}{2r} Q^o \cdot Q^o)\dd r
    \end{aligned}
\end{align}
Then, after inserting $x^o$, we find the fimilar expression for the odd polarity contribution to the second order physical Hamiltonian
\begin{align}
    \begin{aligned}
    H^{(2)}=&\int_{\mathbb{R}^+}\sqrt{\frac{r_s}{r}}P^o  \cdot \partial_r Q^o + \frac{1}{2} \qty(P^o \cdot P^o + (\partial_r Q^o) \cdot (\partial_r Q^o) + \frac{l(l+1) r - 3 r_s}{r^3} Q^o \cdot Q^o)\\
    &+ \dv{r}\qty(2 \frac{r + r_s}{r^2} x^o \cdot x^o + \frac{r-r_s}{2 r^2} Q^o \cdot Q^o)\dd r
    \end{aligned}
\end{align}
The total derivative term in the second line leads to a boundary term and using the fall-off conditions on the canonical variables, one can show that
this contribution to the physical Hamiltonian vanishes.

At third order, we have
\begin{align}
    H^{(3)} &= \int_{\mathbb{R}^+}\sum_{\{l_i,m_i\}} -  i \sigma_- r^{-3} \frac{\lambda_{l_2,2}}{\sqrt{(l_1+2)(l_1-1)}} C^{l_1 m_1 1}_{l_2 m_2 0, l_3 m_3 -1} \partial_r (r Q^o_{l_1 m_1}) Q^o_{l_2m_2} x^o_{l_3 m_3}\\
    &~~~~~~~~~~~+ i \sigma_-  r^{-4}  p_v^{(0)}\qty(C^{l_1 m_1 2}_{l_2 m_2 -1, l_3 m_3 -1} +  2  \frac{\sqrt{l_2(l_2 + 1)}}{\sqrt{(l_1+2)(l_1-1)}} C^{l_1 m_1 1}_{l_2 m_2 0, l_3 m_3 -1})\partial_r (r Q^o_{l_1 m_1}) x^o_{l_2m_2} x^o_{l_3 m_3}\nonumber\\
    &~~~~~~~~~~~- 2  i \sigma_- r^{-3}\sqrt{(l_1 + 2)(l_1 - 1)} C^{l_1 m_1 2}_{l_2 m_2 -1, l_3 m_3 -1} x^o_{l_1 m_1} x^o_{l_2 m_2} x^o_{l_3 m_3} \dd r\nonumber
\end{align}
We use the recursion relation of the Clebsch-Gordan coefficients in the second and third line in the following form:
\begin{equation}
    C^{l_1 m_1 2}_{l_2 m_2 -1, l_3 m_3 -1} = - \frac{1}{\sqrt{(l_1+2)(l_1 -1)}}\qty( \sqrt{l_2 (l_2+1)} C^{l_1 m_1 1}_{l_2 m_2 0, l_3 m_3 -1}+\sqrt{l_3 (l_3+1)} C^{l_1 m_1 1}_{l_2 m_2 -1, l_3 m_3 0})
\end{equation}
Additionally, we exploit the symmetry of $x^o_{l_2 m_2} x^o_{l_3 m_3}$ under exchanging $(l_2 m_2) \leftrightarrow (l_3 m_3)$. 
We find
\begin{align}
    \begin{split}
    H^{(3)} &= \int_{\mathbb{R}^+} \sum_{\{l_i,m_i\}} - i \sigma_- r^{-3} \frac{\lambda_{l_2,2}}{\sqrt{(l_1+2)(l_1-1)}} C^{l_1 m_1 1}_{l_2 m_2 0, l_3 m_3 -1} \partial_r (r Q^o_{l_1 m_1}) Q^o_{l_2m_2} x^o_{l_3 m_3}\\
    &~~~~~~~~~~~+ 4 i \sigma_- r^{-3} \sqrt{l_2 (l_2+1)} C^{l_1 m_1 1}_{l_2 m_2 0, l_3 m_3 -1} x^o_{l_1 m_1} x^o_{l_2 m_2} x^o_{l_3 m_3} \dd r
    \end{split}
\end{align}
The term proportional to the product of the three $x^o$'s vanishes because it is symmetric while the remaining sum is antisymmetric when exchanging the $(l_1 m_1)$
with $(l_3 m_3)$:
\begin{align}
    \sigma_- C^{l_1 m_1 1}_{l_2 m_2 0, l_3 m_3 -1} = \sigma_- (-1)^{l_1 + l_2 + l_3} C^{l_3 m_3 1}_{l_2 m_2 0, l_1 m_1 -1} = - \sigma_- C^{l_3 m_3 1}_{l_2 m_2 0, l_1 m_1 -1}
\end{align}
Thus,
\begin{align}
    H^{(3)} &= \int_{\mathbb{R}^+} - \sum_{\{l_i,m_i\}} i \sigma_- r^{-3} \frac{\lambda_{l_2,2}}{\sqrt{(l_1+2)(l_1-1)}} C^{l_1 m_1 1}_{l_2 m_2 0, l_3 m_3 -1} \partial_r (r Q^o_{l_1 m_1}) Q^o_{l_2m_2} x^o_{l_3 m_3}\dd r
\end{align}
Finally, replacing $x_o$ in terms $Q^o$ and $P^o$, we find
\begin{align}
    H^{(3)} &= \int_{\mathbb{R}^+}  \sum_{\{l_i,m_i\}} i \sigma_- r^{-2} \lambda_{l_2,2} \frac{\lambda_{l_1,1} \lambda_{l_3,1}}{\lambda_{l_1,2} \lambda_{l_3,2}} C^{l_1 m_1 1}_{l_2 m_2 0, l_3 m_3 -1} \partial_r (r Q^o_{l_1 m_1}) Q^o_{l_2m_2} \qty(P^o_{l_3 m_3} - \frac{\sqrt{r r_s}}{r^2} Q^o_{l_3 m_3})\dd r
    \label{eq:3rdAxialRW}
\end{align}

In the end of this appendix, we show how the master variable in the Regge-Wheeler and the Gullstrand-Painlevé gauge are related. 
For the Regge-Wheeler gauge, we have the observable degrees of freedom $x^o,y_o$ while for the Gullstrand-Painlevé gauge, we have $X^o,Y_o$.
Starting from the Regge-Wheeler variables $x^o,y_o$ or the corresponding master variables $Q^o_\mathrm{RW},P^o_\mathrm{RW}$, we can compute the gauge invariant extension of these variables using the observable map.
Then, restricting the gauge invariant variables to the GP gauge, we can compare them with the variables $X^o,Y_o$ and master variables $Q^o_\mathrm{GP},P^o_\mathrm{GP}$ used in the GP gauge. 

Up to third order, the relational observable \cite{Rovelli2, Dittrich1, Dittrich2, Dittrich3, ThiemannReduced} for the axial perturbations is given by
\begin{align}
    &O_F = F + \sum_{lm} \int \dd r' X^o_{lm}(r') \qty{Y_o^{lm}(r') - [Y_o^{\mathrm{sol}}]^{l m}(r'),F}\\
    &+ \sum_{l_1 m_1, l_2 m_2}  \int \dd r_1 \dd r_2 X^o_{l_1 m_1}(r_1)X^o_{l_2 m_2}(r_2)  \qty{Y_o^{l_1m_1}(r_1) - [Y_o^{\mathrm{sol}}]^{l_1m_1}(r_1),\qty{Y_o^{l_2m_2}(r_2) - [Y_o^{\mathrm{sol}}]^{l_2m_2}(r_2),F}}\nonumber
\end{align}
where $Y_o^{\mathrm{sol}} = [Y_o^{(1)}]^{lm} + [Y_o^{(2)}]^{lm}$ is the solution of the constraints.
Using this map we find
\begin{align}
    O_{x^o}^{lm} = x^o_{lm} &- \frac{r^2}{\sqrt{2 (l+2)(l-1)}}\partial_r \qty(r^{-2}X^o_{lm}) + \sum_{\{l_i,m_i\}}\frac{i \sigma_-}{r^2} \frac{\sqrt{l_2(l_2+1)}}{\sqrt{2 (l_1+2)(l_1-1)}} C^{lm-1}_{l_1 m_1 1, l_2 m_2 0} X^o_{l_1 m_1}x^o_{l_2 m_2}\\
    &- \sum_{\{l_i,m_i\}} i \sigma_- \frac{\sqrt{l_1(l_1+1)}}{\sqrt{2(l_1+2)(l_1-1)}\sqrt{2(l_2+2)(l_2-1)}} C^{l_2 m_2 1}_{l m -1, l_1 m_1 0} \partial_r \qty(r^{-2} X^o_{l_1 m_1}) X^o_{l_2 m_2}\nonumber
\end{align}
Similarly, for $y_o$ we have
\begin{align}
    O_{y_o}^{lm} = y_o^{lm} &+ \frac{2 p_v}{\sqrt{2 (l+2)(l-1)}}\partial_r \qty(r^{-2}X^o_{lm}) - \sum_{\{l_i,m_i\}}\frac{i \sigma_-}{r^2} \frac{\sqrt{l(l+1)}}{\sqrt{2 (l_1+2)(l_1-1)}} C^{l_1m_1 1}_{l_2 m_2 -1, l m 0} X^o_{l_1 m_1}y_o^{l_2 m_2}\\
    &- \sum_{\{l_i,m_i\}} i \sigma_- \frac{2 p_v}{r^2} \frac{\sqrt{l(l+1)}}{\sqrt{2(l_1+2)(l_1-1)}\sqrt{2(l_2+2)(l_2-1)}} C^{l_2 m_2 1}_{l_1 m_1 -1,lm0}\partial_r \qty(r^{-2} X^o_{l_1 m_1}) X^o_{l_2 m_2}\nonumber
\end{align}
The gauge invariant extension of the Regge-Wheeler master variable $Q^o_{\mathrm{RW}}$ restricted to the GP gauge is 
\begin{align}
    [Q^o_{\mathrm{RW}}] &= \frac{r}{\sqrt{(l+2)(l-1)}}\qty[O_{y_o} + \frac{2 p_v}{r^2}O_{x^o}]_{x_o=0,y_o=y_o^{(1)}+y_o^{(2)}}\\
    &=\frac{r}{\sqrt{(l+2)(l-1)}}\Big[y_o^{(1)}+ y_o^{(2)} - \sum_{\{l_i,m_i\}}\frac{i \sigma_-}{r^2} \frac{\sqrt{l(l+1)}}{\sqrt{2 (l_1+2)(l_1-1)}} C^{l_1m_1 1}_{l_2 m_2 -1, l m 0} X^o_{l_1 m_1}[y_o^{(1)}]_{l_2 m_2}\nonumber\\
    &~~~~- \sum_{\{l_i,m_i\}} i \sigma_- \frac{2 p_v}{r^2} \frac{\sqrt{l(l+1)} C^{l_2 m_2 1}_{l_1 m_1 -1,lm0} + \sqrt{l_1(l_1+1)}C^{l_2 m_2 1}_{l_1 m_1 0, l m -1} }{\sqrt{2(l_1+2)(l_1-1)}\sqrt{2(l_2+2)(l_2-1)}} \partial_r \qty(r^{-2} X^o_{l_1 m_1}) X^o_{l_2 m_2}\Big]\nonumber\\
    &= Q^o_{\mathrm{GP}} + \frac{r}{\sqrt{(l+2)(l-1)}}\Big[y_o^{(2)} - \sum_{\{l_i,m_i\}}\frac{i \sigma_-}{r^2} \frac{\sqrt{l(l+1)}}{\sqrt{2 (l_1+2)(l_1-1)}} C^{l_1m_1 1}_{l_2 m_2 -1, l m 0} X^o_{l_1 m_1}[y_o^{(1)}]_{l_2 m_2}\nonumber\\
    &~~~~+ \sum_{\{l_i,m_i\}} i \sigma_- \frac{p_v}{r^2} \frac{C^{l_2 m_2 2}_{l_1 m_1 -1,lm -1}}{\sqrt{(l_1+2)(l_1-1)}} \partial_r \qty(r^{-2} X^o_{l_1 m_1}) X^o_{l_2 m_2}\Big]\nonumber
\end{align}
In the last step, we introduced the GP master variable $Q^o_\mathrm{GP}$ defined in the main text.
For the gauge invariant extension of $P^o_\mathrm{RW}$, we obtain
\begin{align}
    [P^o_\mathrm{RW}]_{lm} &= \frac{p_v}{2 r^2}[Q^o_\mathrm{RW}]_{lm} - \frac{\sqrt{(l+2)(l-1)}}{r}[O_{x^o}]_{lm}\Big|_{x_o=0,y_o=y_o^{(1)}+y_o^{(2)}}\\
    &= P^o_\mathrm{GP} + \frac{p_v}{2 r\sqrt{(l+2)(l-1)}}\Big[y_o^{(2)} - \sum_{\{l_i,m_i\}}\frac{i \sigma_-}{r^2} \frac{\sqrt{l(l+1)}}{\sqrt{2 (l_1+2)(l_1-1)}} C^{l_1m_1 1}_{l_2 m_2 -1, l m 0} X^o_{l_1 m_1}[y_o^{(1)}]_{l_2 m_2}\nonumber\\
    &~~~~~~~~~~~~~~~~~~+ \sum_{\{l_i,m_i\}} i \sigma_- \frac{p_v}{r^2} \frac{C^{l_2 m_2 2}_{l_1 m_1 -1,lm -1}}{\sqrt{(l_1+2)(l_1-1)}} \partial_r \qty(r^{-2} X^o_{l_1 m_1}) X^o_{l_2 m_2}\Big]\nonumber\\
    &~~~~~~~~~~+ \sum_{\{l_i,m_i\}} i \sigma_- \frac{\sqrt{(l+2)(l-1)} \sqrt{l_1(l_1+1)}}{r \sqrt{2(l_1+2)(l_1-1)}\sqrt{2(l_2+2)(l_2-1)}} C^{l_2 m_2 1}_{l m -1, l_1 m_1 0} \partial_r \qty(r^{-2} X^o_{l_1 m_1}) X^o_{l_2 m_2}\nonumber
\end{align}
Here $P^o_\mathrm{GP}$ is the Gullstrand-Painlevé master variable.
In both cases, to first order, we have $Q^o_\mathrm{RW} = Q^o_\mathrm{GP}$ and $P^o_\mathrm{RW}  = P^o_\mathrm{GP}$.
At second order they do not correspond to the same gauge invariant variable.
This aligns well with the fact that we found two different expressions for the third order physical Hamiltonian in the Regge-Wheeler and GP gauge. 
Due to the difference at second order, it would have been surprising if they were identical.

\section{Boundary Term Modified Gauge Fixing Conditions}
\label{sec:BTModifiedGauge}

The boundary term is given by the sum of the original boundary term $B_e$ and new contributions depending on $X^e_\infty$:
\begin{align}
    \tilde B_e &= B_e -\sqrt{\frac{(l+2)(l-1)}{2l(l+1)}} \frac{2(r(r - 2r_s) l(l+1) - r_s(r_s - r))}{l(l+1)r p_v^{(0)} \Lambda}X^e_\infty \partial_r y_v^{(0)}\\
    &- \frac{(l+2)(l-1)}{2l(l+1)}\frac{1}{r^4 (p_v^{(0)})^2 \Lambda^2} \Big[-\left(l^2+l-1\right) (2 l (l+1)+5) r^2 r_s^2+(l-1) l (l+1) (l+2) r^4\nonumber\\
    &~~~~~~~~~-((l-2) l (l+1) (l+3)+2) r^3 r_s-3 l (l+1) r r_s^3-3 r_s^4\Big]X^e_\infty X^e\nonumber\\
    &-\sqrt{\frac{(l+2)(l-1)}{2l(l+1)}}\frac{1}{4 l(l+1)r^4 p_v^{(0)} \Lambda^3}\Big[l (l+1) \left(l^2+l-2\right)^2 \left(l^2+l-1\right) r^4\nonumber\\
    &~~~~~~~~~+(l-1) (l+2) \left(l (l+1) \left(l (l+1) \left(l^2+l+5\right)-9\right)+2\right) r^3 r_s\nonumber\\
    &~~~~~~~~~+\left(l (l+1) \left(4 l^2 (l+1)^2-15\right)+16\right) r^2 r_s^2+3 (l (l+1) (5 l (l+1)-3)-11) r r_s^3\nonumber\\
    &~~~~~~~~~+3 (4 l (l+1)+7) r_s^4\Big]X^e_\infty y_v^{(1)}\nonumber\\
    &-\frac{(l+2)(l-1)}{8 l^2(l+1)^2 r^4 (p_v^{(0)})^2 \Lambda^3}\Big[-(l-1)^2 l^2 (l+1)^2 (l+2)^2 r^5+(l-1) l^2 (l+1)^2 (l+2) \left(l^2+l-8\right) r^4 r_s\nonumber\\
    &~~~~~~~~~+2 \left(l^2+l+1\right) \left(l^2 (l+1)^2 (2 l(l+1)-5)-2\right) r^3 r_s^2+\left(17 l^6+51 l^5+11 l^4-63 l^3+40 l+16\right) r^2 r_s^3\nonumber\\
    &~~~~~~~~~+3 (2 l (l+1) (7 l (l+1)-11)-9) r r_s^4+3 (11 l (l+1)+5)r_s^5\Big] (X^e_\infty)^2\nonumber
\end{align}

\providecommand{\href}[2]{#2}\begingroup\raggedright\endgroup

\end{document}